\documentclass[aps,prd,showkeys,preprint,nofootinbib,floatfix]{revtex4-1}
\pdfoutput=1

\usepackage{yfonts}
\usepackage{comment}
\usepackage{color}
\usepackage{amsmath}
\usepackage{amssymb}
\usepackage{amsthm}
\usepackage{mathrsfs}
\usepackage{graphicx}
\usepackage{fancyhdr}
\usepackage{array}
\usepackage{latexsym}
\usepackage[all]{xy}
\usepackage{eufrak}
\usepackage{euscript}
\usepackage{enumerate}
\usepackage{dsfont}
\usepackage{slashed}
\usepackage{hyperref}
\usepackage{caption}

\hypersetup{pdftex,colorlinks=true,linkcolor=blue,citecolor=blue,menucolor=black,urlcolor=blue,filecolor=blue}

\def\lsim{\mathop{\hbox{${\lower 3.8pt\hbox{$<$}}\atop{\raise 0.2pt\hbox{$\sim$}}$}}}
\def\gsim{\mathop{\hbox{${\lower 3.8pt\hbox{$>$}}\atop{\raise 0.2pt\hbox{$\sim$}}$}}}

\begin{document}

\title{Momentum transport in strongly coupled anisotropic plasmas in the presence of strong magnetic fields}

\author{Stefano Ivo Finazzo}
\email{stefano@ift.unesp.br}
\affiliation{Instituto de F\'{i}sica Te\'orica, Universidade do Estado de S\~{a}o Paulo, Rua Dr. Bento T. Ferraz, 271, CEP 01140-070, S\~{a}o Paulo, SP, Brazil}

\author{Renato Critelli}
\email{renato.critelli@usp.br}
\affiliation{Instituto de F\'{i}sica, Universidade de S\~{a}o Paulo, Rua do Mat\~{a}o, 1371, Butant\~{a}, CEP 05508-090, S\~{a}o Paulo, SP, Brazil}

\author{Romulo Rougemont}
\email{romulo@if.usp.br}
\affiliation{Instituto de F\'{i}sica, Universidade de S\~{a}o Paulo, Rua do Mat\~{a}o, 1371, Butant\~{a}, CEP 05508-090, S\~{a}o Paulo, SP, Brazil}

\author{Jorge Noronha}
\email{noronha@if.usp.br}
\affiliation{Instituto de F\'{i}sica, Universidade de S\~{a}o Paulo, Rua do Mat\~{a}o, 1371, Butant\~{a}, CEP 05508-090, S\~{a}o Paulo, SP, Brazil}

\begin{abstract}
We present a holographic perspective on momentum transport in strongly coupled, anisotropic non-Abelian plasmas in the presence of strong magnetic fields. We compute the anisotropic heavy quark drag forces and Langevin diffusion coefficients and also the anisotropic shear viscosities for two different holographic models, namely, a top-down deformation of strongly coupled $\mathcal{N} = 4$ Super-Yang-Mills (SYM) theory triggered by an external Abelian magnetic field, and a bottom-up Einstein-Maxwell-dilaton (EMD) model which is able to provide a quantitative description of lattice QCD thermodynamics with $(2+1)$-flavors at both zero and nonzero magnetic fields. We find that, in general, energy loss and momentum diffusion through strongly coupled anisotropic plasmas are enhanced by a magnetic field being larger in transverse directions than in the direction parallel to the magnetic field. Moreover, the anisotropic shear viscosity coefficient is smaller in the direction of the magnetic field than in the plane perpendicular to the field, which indicates that strongly coupled anisotropic plasmas become closer to the perfect fluid limit along the magnetic field. We also present, in the context of the EMD model, holographic predictions for the entropy density and the crossover critical temperature in a wider region of the $(T,B)$ phase diagram that has not yet been covered by lattice simulations. Our results for the transport coefficients in the phenomenologically realistic magnetic EMD model could be readily used as inputs in numerical codes for magnetohydrodynamics.
\end{abstract}



\maketitle
\tableofcontents


\section{Introduction}
\label{intro}

The study of the behavior of QCD matter under extreme conditions is a very active area of research regarding the physics of the strong interactions. Ultrarelativistic heavy ion collisions \cite{expQGP1,expQGP2,expQGP3,expQGP4,expQGP5} are currently probing matter in a region of the QCD phase diagram close to the crossover transition \cite{Aoki:2006we}, where the system behaves as a strongly coupled quark-gluon plasma (QGP) \cite{QGP} (for recent reviews, see \cite{reviewQGP1,reviewQGP2}). One of the most striking features of this strongly coupled QGP is its nearly perfect fluid behavior characterized by a very small value (when compared to weak coupling QCD calculations \cite{Arnold:2000dr,Arnold:2003zc}) for the shear viscosity to entropy density ratio, which according to recent hydrodynamic simulations \cite{Ryu:2015vwa} simultaneously matching experimental data for different physical observables, is given by the value $\eta/ s\approx 0.095$ (at least near the crossover region). This small value is remarkably close to the estimate $\eta/ s=1/4\pi$ valid for a broad class of strongly coupled holographic plasmas with spatially isotropic and translationally invariant gravity duals characterized by actions with at most two derivatives \cite{Policastro:2001yc,Buchel:2003tz,Kovtun:2004de}. This observation suggested that the holographic gauge/gravity correspondence \cite{adscft1,adscft2,adscft3,adscft4} could be useful to obtain insight on the non-equilibrium transport properties of strongly coupled non-Abelian plasmas such as the QGP (for recent reviews on applications of the holographic correspondence to the physics of the QGP, see \cite{solana,adams}). The fact that the gauge/gravity duality may be employed to calculate real time non-equilibrium observables \cite{Son:2002sd,Herzog:2002pc,Gubser:2008sz,Skenderis:2008dg} is particularly interesting since weak coupling QCD calculations cannot reliably describe the strongly coupled region close to the crossover transition, while lattice QCD simulations, though very successful in handling calculations of equilibrium quantities such as the equation of state (at least at zero baryon density), suffer from severe technical difficulties to perform real time calculations \cite{Meyer:2011gj}. Therefore, one may resort to the holographic duality as a non-perturbative tool to compute observables which are very difficult to calculate using first principle QCD techniques as it is the case of real time transport coefficients  near the crossover region.

\newpage

Indeed, the gauge/gravity duality has already been used to compute several transport coefficients of different strongly coupled non-Abelian plasmas - see for instance Refs. \cite{Policastro:2001yc,Buchel:2003tz,Kovtun:2004de,GN2,Noronha:2009ud,Ficnar:2011yj,Ficnar:2012yu,hydro,gubser2,finitemu,Rougemont:2015ona,CaronHuot:2006te,conductivity,Finazzo:2015xwa,Rougemont:2016nyr}.

One of the many branches of applications of holographic techniques to the physics of strongly coupled systems, which is the one we are particularly interested in exploring in the present work, regards the influence of strong external Abelian magnetic fields on the equilibrium and transport properties of strongly interacting non-Abelian plasmas. In fact, very intense magnetic fields ranging from\footnote{Note: $eB=1$ GeV$^2\Rightarrow B \simeq 1.69\times 10^{20}$ G.} $eB\sim m_\pi^2\sim 0.02 \, \mathrm{GeV^2}$ at the Relativistic Heavy Ion Collider (RHIC) to $eB\sim 15 m_\pi^2\sim 0.3 \, \mathrm{GeV^2}$ at the Large Hadron Collider (LHC) may be produced at the earliest stages of ultrarelativistic peripheral heavy ion collisions \cite{noncentralB1,noncentralB2,noncentralB3,noncentralB4,noncentralB5,noncentralB6}.\footnote{At first, one may expect that such strong magnetic fields rapidly decrease in intensity in the later stages when the QGP is formed (after $\sim 1$ fm/c) due to the departure of the spectators from the collision region. However, the electric conductivity of the QGP may sensitively slow down the decay of the magnetic field in the medium \cite{Tuchin:2013apa,Gursoy:2014aka} and the quantum nature of the sources \cite{Holliday:2016lbx} may delay this decay even further. However, it remains unclear whether the large magnetic fields produced at the earliest stages of peripheral collisions remain strong enough to affect transport and equilibrium properties of the QGP.} Less intense, but still very strong magnetic fields (at least when compared to fields of terrestrial, non-astrophysical origins) up to $eB\sim 5\times 10^{-5} m_\pi^2\sim 1\, \textrm{MeV}^2$ are expected to be present inside magnetars \cite{magnetar}, while much stronger fields of order $eB\sim 200 m_\pi^2\sim 4\, \textrm{GeV}^2$ are believed to have been generated in the primordial Universe \cite{universe1,universe2,latticedata0}. Due to this wide range of scenarios where strong magnetic fields may play a relevant role in the properties and the evolution of different physical systems, a large amount of research on related topics has been carried out in the last years, see for instance Refs. \cite{Gusynin:1994re,Gusynin:1995nb,Miransky:2002rp,Fukushima:2016vix,Magdy:2015eda,Agasian:2008tb,Ferrer:2005vd,Fukushima:2007fc,Noronha:2007wg,Mizher:2010zb,Fukushima:2012xw,Fukushima:2012kc,Bali:2012zg,Blaizot:2012sd,Bali:2013esa,Bonati:2014ksa,Fukushima:2013zga,Machado:2013rta,Fraga:2012ev,Fraga:2013ova,Andersen:2013swa,Bali:2013owa,Ruggieri:2014bqa,Ferreira:2013oda,Ferreira:2014kpa,Farias:2014eca,Farias:2016gmy,Ayala:2014iba,Ayala:2014gwa,Ferrer:2014qka,Kamikado:2014bua,Yu:2014xoa,Braun:2014fua,Mueller:2015fka,Endrodi:2015oba,cohen} and also \cite{reviewfiniteB1,reviewfiniteB2,reviewfiniteB3,reviewfiniteB4} for recent reviews.

In the holographic scenario, some calculations of physical observables in the presence of strong magnetic fields in different gauge/gravity models were discussed, for instance, in Refs. \cite{Wu:2013qja,Evans:2010xs,Preis:2010cq,Mamo:2013efa,Mamo:2015dea,Mamo:2015aia,Li:2016bbh,Ballon-Bayona:2013cta,Callebaut:2013ria,Dudal:2015wfn,Drwenski:2015sha,Evans:2016jzo,DK1,DK2,DK3,DK-applications1,DK-applications2,DK-applications3,DK-applications4,Rougemont:2015oea,Kiritsis:2011ha,Hartnoll:2007ai,Hartnoll:2009sz,McInnes:2015kec,McInnes:2016dwk,Rajagopal:2015roa,Sadofyev:2015hxa}. In the present work, we are specifically interested in analyzing anisotropic momentum transport coefficients of strongly interacting magnetized plasmas, namely, heavy quark drag forces, Langevin diffusion coefficients, and also anisotropic shear viscosities.

\newpage

For hard probes plowing through the strongly coupled plasma, the momentum loss of the probe to the medium may be described by its drag force \cite{Gubser:2006bz,Herzog:2006gh,Herzog:2006se,CasalderreySolana:2006rq,Gubser:2006qh,Gursoy:2009kk}. If one considers the influence of thermal fluctuations, it may be further characterized by the momentum diffusion of the probe along (and transversely) to its initial velocity via the diffusion coefficients associated to the Brownian motion of the probe described by a local Langevin equation \cite{Gubser:2006nz,CasalderreySolana:2007qw,Gursoy:2010aa}. Another important transport coefficient associated with the hydrodynamic evolution of the energy-momentum tensor of the system is the shear viscosity. As we shall review in this work, the presence of an external Abelian magnetic field explicitly breaks $SO(3)$ rotational symmetry down to $SO(2)$ rotations in the plane transverse to the magnetic field direction inducing an anisotropy in the system which, in turn, implies in a splitting of these observables into several new transport coefficients. For instance, while in the isotropic case at zero magnetic field there is one drag force, one shear viscosity, and two Langevin diffusion coefficients (one  corresponding to fluctuations transverse and the other to fluctuations parallel to the heavy quark velocity), the anisotropy induced by a nonzero magnetic field (as in the models considered in the present work) causes the appearance of two different nontrivial shear viscosities, two different drag forces, and five different Langevin diffusion coefficients depending on the orientation of the momentum diffusion relative to the directions of the magnetic field and the velocity of the probe.

In the context of a top-down anisotropic deformation of a strongly coupled $\mathcal{N} = 4$ Super-Yang-Mills (SYM) plasma driven by a nontrivial profile for a bulk axion field \cite{Mateos:2011ix,Mateos:2011tv}, the corresponding anisotropic drag forces \cite{Chernicoff:2012iq,Giataganas:2012zy,Misobuchi:2015ioa}, Langevin diffusion coefficients \cite{Giataganas:2013zaa,Giataganas:2013hwa,Chakrabortty:2013kra}, and shear viscosities \cite{Rebhan:2011vd}, have been already computed but a detailed holographic study of anisotropic momentum transport driven by an external magnetic field has not yet been done.

Recently, in Refs. \cite{Fukushima:2015wck,Li:2016bbh}, some of the weakly coupled perturbative QCD Langevin diffusion coefficients at leading order in the strong coupling constant, $\alpha_s$, and strong magnetic fields were computed in the $\alpha_s eB \ll T^2 \ll eB$ limit. In Ref. \cite{Li:2016bbh}, the anisotropic drag forces and some of the Langevin diffusion coefficients were also computed strictly in the particular limit of strong magnetic fields, $eB/T^2\gg 1$, for the top-down anisotropic deformation of a strongly coupled SYM plasma driven by an external magnetic field, called the ``magnetic brane model'' \cite{DK1,DK2,DK3}. In those works, it was found that the heavy quark diffusion for a probe moving perpendicularly to the magnetic field is larger than in the case of parallel motion suggesting that this may contribute to heavy quark elliptic flow \cite{Fukushima:2015wck}.

In the present work, as a warm-up calculation in a top-down holographic model, we go beyond the analytical limit $eB/T^2\gg 1$ worked out in Ref. \cite{Li:2016bbh} and derive full numerical results for the anisotropic momentum transport coefficients of the magnetic brane model, which are valid for any value of the ratio $eB/T^2$. We compute for the first time the full results for the two drag forces, the five Langevin diffusion coefficients (two of them were not discussed in Ref. \cite{Li:2016bbh} in any limit) and, for completeness, we also review the main result of Ref. \cite{DK-applications2} concerning the calculation of the anisotropic shear viscosities for this magnetized SYM plasma.

However, when thinking about possible applications to real world QCD at finite temperature, it is desirable to work with a holographic model which is able to emulate at least some of the effects of the dynamical infrared breaking of conformal symmetry associated to the emergence of the dimensional transmutation scale $\Lambda_{\textrm{QCD}}$. This is clearly not the case of the top-down magnetic brane model proposed in Refs. \cite{DK1,DK2,DK3} since, for instance, all the observables in this model are functions of the dimensionless ratio $eB/T^2$ instead of $eB$ and $T$, separately. The reason for that is the fact that the SYM plasma is a conformal system at zero magnetic field and, in this case, if $B=0$ the temperature is the only scale of the system and one is only able to physically distinguish the zero temperature from the finite temperature case with $T\neq 0$ being a fixed scale of the system. When this theory is deformed by the introduction of an external magnetic field (which explicitly breaks conformal symmetry \cite{DK-applications4}), the value of this field is then naturally measured in terms of the fixed temperature scale.

The situation in QCD is completely different since due to the dynamical breaking of conformal symmetry in the infrared regime, $\Lambda_{\textrm{QCD}}$ emerges as the natural (quantum) scale of the theory, even in the vacuum. By turning on the temperature in QCD, $T$ is in fact a variable (differently from what happens in the magnetic brane setup), which is naturally measured in terms of the fixed scale $\Lambda_{\textrm{QCD}}$. In the very same way, by applying an external magnetic field to QCD matter, the magnetic field is naturally measured in terms of $\Lambda_{\textrm{QCD}}$ and, therefore, both $T$ and $eB$ may be independently varied.

In order to induce a dynamical breaking of conformal symmetry in holographic settings, one may consider a bottom-up Einstein-dilaton model with a nontrivial dilaton potential responsible for emulating the effects of the $\Lambda_{\textrm{QCD}}$ scale, as originally proposed in \cite{Gubser:2008ny} (see also \cite{GN2,conductivity,Noronha:2009ud,Ficnar:2011yj,Ficnar:2012yu,hydro,Rougemont:2016nyr} for further applications). In Ref. \cite{DeWolfe:2010he} (see also \cite{gubser2,finitemu,Rougemont:2015ona,Finazzo:2015xwa} for further applications), an extension of the holographic setup proposed in \cite{Gubser:2008ny} encompassed the construction of an Einstein-Maxwell-dilaton (EMD) model describing a QCD-like theory at finite temperature, nonzero baryon chemical potential, and zero magnetic field at the boundary of isotropic, asymptotically AdS$_5$ spaces. More recently, some of us proposed a different EMD model \cite{Rougemont:2015oea} describing a QCD-like theory at finite temperature and nonzero magnetic field (at zero chemical potential) at the boundary of spatially anisotropic, asymptotically AdS$_5$ spaces. The fundamental reasoning involved in these phenomenological bottom-up approaches may be properly dubbed as some type of ``black hole engineering'', which consists in adequately ``teaching'' the dilatonic black hole model how to behave in a QCD-like manner, on phenomenologically interesting regions of the QCD phase diagram. More precisely, one seeds the holographic model with adequate lattice and/or experimental/observational data, which are used to dynamically fix the free parameters of the bottom-up setup considered. Once these parameters are fixed, further calculations of different observables (usually related to real time transport coefficient) provide true predictions of the holographic setup. 

In the magnetic EMD setting proposed in Ref. \cite{Rougemont:2015oea}, the free parameters of the model were dynamically fixed by matching the holographic equation of state and magnetic susceptibility at zero magnetic field with the corresponding $(2+1)$-flavor lattice QCD data with physical quark masses presented in Refs. \cite{latticedata1,latticedata2}, respectively. Then, the equation of state at finite magnetic field follows as a prediction of the holographic model. In \cite{Rougemont:2015oea} we obtained a reasonable agreement with the equation of state at finite $T$ and $B$ calculated recently on the lattice \cite{latticedata3} for magnetic fields up to $eB\sim 0.3$ GeV$^2$. In the present work, we update the model proposed in \cite{Rougemont:2015oea} by seeding it with more recent lattice data for the equation of state at $B=0$ \cite{Borsanyi:2013bia} and by performing a global matching to different observables characterizing the equation of state at zero magnetic field. The updated magnetic EMD model to be discussed in the present work greatly improves the quantitative agreement with the finite $T$ and $B$ lattice QCD equation of state of Ref. \cite{latticedata3}, also extending it to higher values of the magnetic field. As the main results of the present work, we employ this updated magnetic EMD model to compute for the first time the $T$ and $B$ dependence of the anisotropic drag forces, Langevin diffusion coefficients and shear viscosities in a realistic magnetized QCD-like holographic dual.

We finish this introductory section by providing an overview of the paper in order to guide the reader:
\begin{enumerate}[(i)]
 
\item The thermodynamics, drag forces, Langevin diffusion coefficients, and shear viscosities for the top-down magnetic brane model are discussed in Section \ref{sec3.0};

\item The thermodynamics, drag forces, Langevin diffusion coefficients, and shear viscosities for the phenomenological QCD-like bottom-up magnetic EMD model are presented in Section \ref{sec4.0}. The results presented in this section will be of more interest to the phenomenologically oriented reader;

\item In the concluding Section \ref{conclusion} we discuss the most important implications of our calculations and outline future projects that may be pursued;

\item For completeness, we provide in Appendix \ref{transport} a review of the general holographic formalism for the computation of drag forces, Langevin diffusion coefficients and shear viscosities, in both isotropic and anisotropic settings. In Appendix \ref{apb}, we present a comprehensive derivation of the anisotropic Kubo formulas for the several shear (and bulk) viscosities appearing in first order viscous magnetohydrodynamics.

\end{enumerate}

In this work, we use natural units $c=\hbar=k_B=1$ and a mostly plus (Lorentzian) metric signature.

\section{The magnetic brane model}
\label{sec3.0}

\subsection{The model and its thermodynamics}
\label{sec3.1}

\textit{Action.} The first class of magnetized backgrounds we will use in the present work corresponds to a top-down model of magnetic branes dual to a deformation of strongly coupled SYM theory triggered by an external magnetic field \cite{DK1,DK2,DK3}. The model is described by the Einstein-Maxwell action,
\begin{equation}
\label{eq:magbranesaction}
S = \frac{1}{16\pi G_5}\int_{\mathcal{M}_5}d^5x\sqrt{-g}\left[R + \frac{12}{L^2}- F_{\mu\nu}^2\right] +S_{\textrm{CS}}+S_{\textrm{GHY}}+S_{\textrm{CT}},
\end{equation}
where $L$ is the asymptotic $\mathrm{AdS}_5$ radius, which we set to unity, $S_{\textrm{CS}}$ is the topological $(4+1)$-dimensional Abelian Chern-Simons term (which vanishes on-shell for the backgrounds considered here though it is useful \cite{DK1} to fix the relation between the bulk magnetic field - which we denote in this section by $B$ - and the physically observable magnetic field at the boundary - which we denote in this section by $\mathcal{B}$), $S_{\textrm{GHY}}$ is the Gibbons-Hawking-York term \cite{ghy1,ghy2} needed in order to give a well posed initial value problem, and $S_{\textrm{CT}}$ is the counterterm action \cite{ren1,ren2,ren3,ren4,ren5} needed in order to render the complete on-shell action finite. However, as we will not need to compute here the on-shell action, we do not need to specify the explicit form of $S_{\textrm{CS}}$, $S_{\textrm{GHY}}$, and $S_{\textrm{CT}}$.

\vspace{8pt}

\textit{Ansatz and equations of motion.} The magnetic brane background is described by the following ansatz in coordinates which we call the standard coordinates, denoted by a tilde,
\begin{equation}
\label{eq:magbranatz}
ds^2=-\tilde{U}(\tilde{r})d\tilde{t}^2+\frac{d\tilde{r}^2}{\tilde{U}(\tilde{r})} + e^{2\tilde{V}(\tilde{r})}(d\tilde{x}^2 + d\tilde{y}^2) + e^{2\tilde{W}(\tilde{r})} d\tilde{z}^2, \quad F = B d\tilde{x} \wedge d\tilde{y}.
\end{equation}
One can check that Maxwell's equations following from Eq. \eqref{eq:magbranesaction} are trivially satisfied by the ansatz \eqref{eq:magbranatz}, while Einstein's equations reduce to
\begin{align}
\tilde{U}(\tilde{V}''-\tilde{W}'') + (\tilde{U}' + \tilde{U}(2\tilde{V}'+\tilde{W}'))(\tilde{V}'-\tilde{W}') & = -2B^2 e^{-4\tilde{V}}, \\
2 \tilde{V}''+ \tilde{W}'' + 2(\tilde{V}')^2 + (\tilde{W}')^2 & = 0, \\
\frac{1}{2} \tilde{U}'' + \frac{1}{2} \tilde{U}' (2 \tilde{V}' + \tilde{W}') & = 4 + \frac{2}{3} B^2 e^{-4 \tilde{V}}.
\end{align}
The fourth equation follows from the above and it is a constraint on initial data:
\begin{equation}
2 \tilde{U}' \tilde{V}' + \tilde{U}' \tilde{W}' + 2 \tilde{U} (\tilde{V}')^2 + 4 \tilde{U} \tilde{V}' \tilde{W}' = 12 - 2 B^2 e^{-4 \tilde{V}}.
\end{equation}

\vspace{8pt}

\textit{Asymptotics.} The magnetic brane solution corresponds to a holographic renormalization group flow interpolating between a $\mathrm{BTZ} \times \mathrm{R}^2$ near horizon solution given by ($r = r_H$ is the BTZ black hole \cite{Banados:1992wn} horizon)
\begin{equation}
\label{eq:BTZ}
ds^2 = \left[ -3 (\tilde{r}^2 - \tilde{r}_H^2) dt^2 + 3 \tilde{r}^2 d\tilde{z}^2 + \frac{d\tilde{r}^2}{3(\tilde{r}^2-\tilde{r}_H^2)} \right] + \frac{B}{\sqrt{3}} (d\tilde{x}^2 + d\tilde{y}^2)
\end{equation}
describing the deep infrared, and a near boundary $\mathrm{AdS}_5$ asymptotic solution at $\tilde{r} \to \infty$,
\begin{equation}
ds^2 = \tilde{r}^2 (- d\tilde{t}^2 + d\tilde{x}^2+d\tilde{y}^2+ d\tilde{z}^2) + \frac{d\tilde{r}^2}{\tilde{r}^2},
\end{equation}
describing the ultraviolet.

As discussed in Ref. \cite{DK1}, the gauge variation of the Chern-Simons term in Eq. \eqref{eq:magbranesaction} may be used to  compute a 3-point function in the gauge theory and compare it with the SYM chiral anomaly, which gives the following relation between the physically observable magnetic field at the boundary and the bulk magnetic field: $\mathcal{B} = \sqrt{3} B$.

\vspace{8pt}

\textit{Numerical solutions.} In order to numerically solve the equations of motion we introduce new coordinates, which we call numerical coordinates, represented without the tildes
\begin{align}
\tilde{t} = t, \,\,\, \tilde{r} = r, \,\,\,\tilde{x} = \frac{x}{\sqrt{v(b)}}, \,\,\,\tilde{y} = \frac{y}{\sqrt{v(b)}}, \,\,\,\tilde{z} = \frac{z}{\sqrt{w(b)}};\,\,\, \left(\Rightarrow B=\frac{b}{v(b)}\right),
\label{eq:NumCoord}
\end{align}
where $b$ is the rescaled magnetic field in the numerical coordinates, which is taken as an initial condition (each value of this initial condition will generate a numerical background corresponding to some definite physical state at the gauge theory), and $v(b)$, $w(b)$ are functions extracted from the numerical solutions $V(r)$, $W(r)$ by fitting their near-boundary behavior as $e^{2V(r\to\infty)}\sim v(b)r^2$, $e^{2W(r\to\infty)}\sim w(b) r^2$, respectively. These numerical coordinates are chosen such that the horizon is located at $r_H = 1$ and that the rescaled metric functions $U(r)$, $V(r)$, and $W(r)$ satisfy the boundary conditions $U'(1) = 1\, (\Rightarrow T=1/4\pi)$, $V(1) = W(1) = 0$. Moreover, $U(1)=0$. One can then show that, on-shell,
\begin{align}
V'(1) = 4-\frac{4}{3} b^2 \quad \mathrm{and} \quad W'(1) = 4+\frac{2}{3} b^2,
\end{align}
and one can then numerically integrate the equations of motion from the horizon to the boundary.

\begin{figure}[h]
\begin{center}
\includegraphics[width=0.6\textwidth]{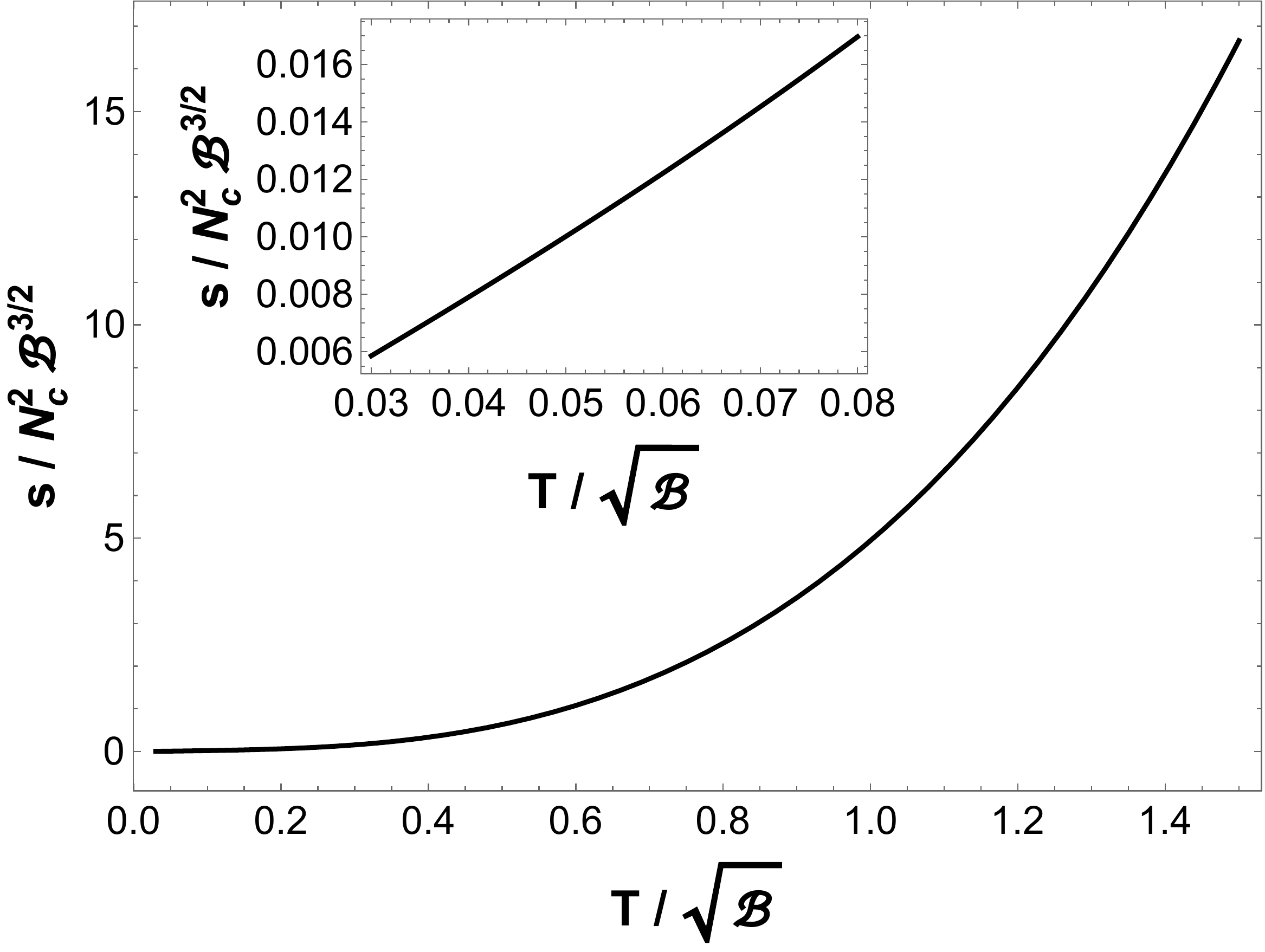} 
\end{center}
\caption{Normalized entropy density for the magnetic brane setup as a function of the dimensionless ratio $T/\sqrt{\mathcal{B}}$. In the inset we show the corresponding behavior for large magnetic fields.}
\label{fig:Fpar}
\end{figure}

One can also compute the temperature and the entropy density normalized by the gauge theory magnetic field $\mathcal{B}$ in terms of the scaling functions $v(b)$ and $w(b)$ \cite{DK1},
\begin{align}
\frac{T}{\sqrt{\mathcal{B}}} = \frac{1}{4\pi\, 3^{1/4}} \sqrt{\frac{v(b)}{b}} \quad \mathrm{and} \quad
\frac{s}{N_c^2 \mathcal{B}^{3/2}} = \frac{1}{2\pi\, 3^{3/4}} \sqrt{\frac{v(b)}{b^3 w(b)}}.
\end{align}
The corresponding equation of state is plotted in Fig. \ref{fig:Fpar}, where one can notice that at large (small) magnetic fields (compared to the fixed temperature scale) the behavior of the dimensionless ratio $s/N_c^2\mathcal{B}^{3/2}$ is linear (cubic) in the dimensionless ratio $T/\sqrt{\mathcal{B}}$. Therefore, one indeed recovers the AdS$_5$-Schwarzschild result for D3-branes in the ultraviolet, $s\sim T^3$.

\subsection{Drag force}
\label{sec3.2}

\begin{figure}[h]
\begin{center}
\begin{tabular}{c}
\includegraphics[width=0.45\textwidth]{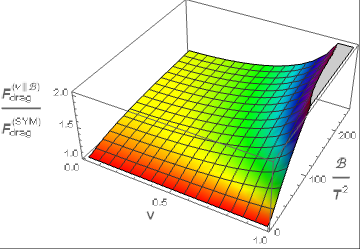} 
\end{tabular}
\begin{tabular}{c}
\includegraphics[width=0.45\textwidth]{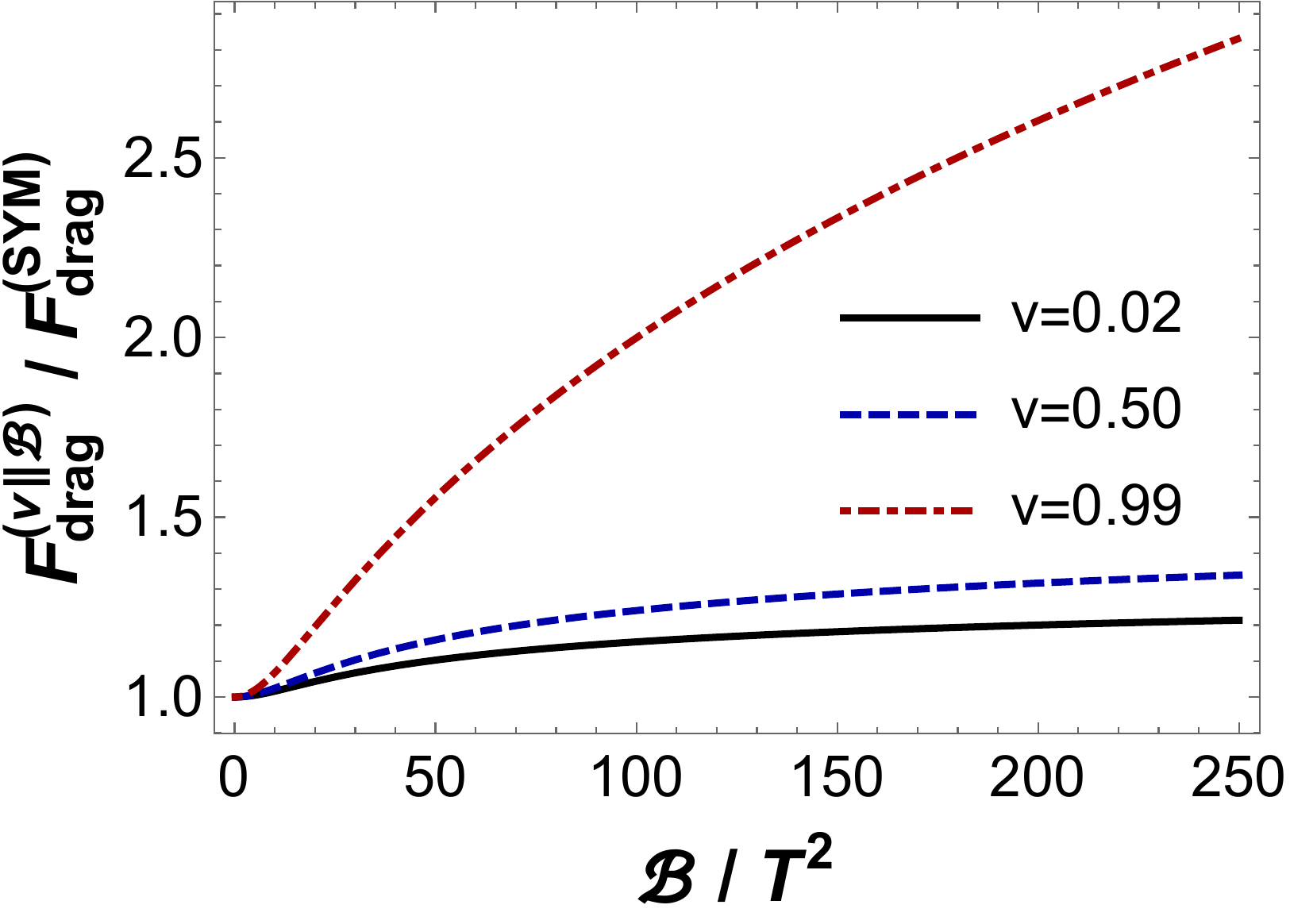} 
\end{tabular}
\end{center}
\caption{(Color online) Anisotropic drag force $F_{\textrm{drag}}^{(v\parallel \mathcal{B})}$ in the magnetic brane model normalized by the isotropic SYM result at zero magnetic field. \emph{Left:} 3D plot as function of $\mathcal{B}/T^2$ and $v$. \emph{Right:} as a function of $\mathcal{B}/T^2$ for some fixed values of $v$.}
\label{fig:DKdrag1}
\end{figure}

\begin{figure}[h]
\begin{center}
\begin{tabular}{c}
\includegraphics[width=0.45\textwidth]{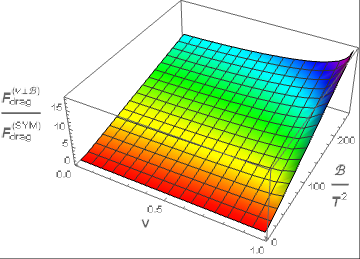} 
\end{tabular}
\begin{tabular}{c}
\includegraphics[width=0.45\textwidth]{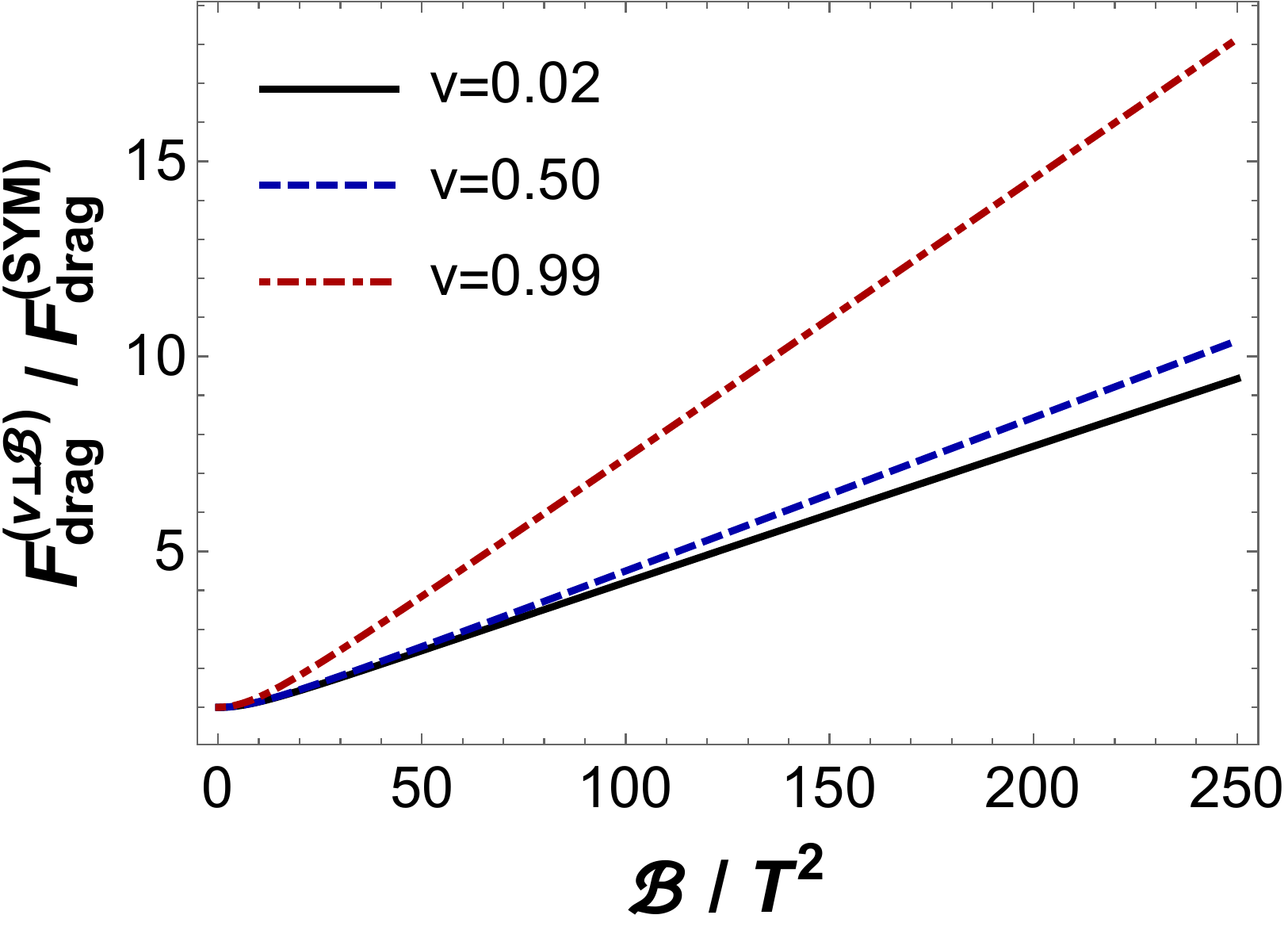} 
\end{tabular}
\end{center}
\caption{(Color online) Anisotropic drag force $F_{\textrm{drag}}^{(v\perp \mathcal{B})}$ in the magnetic brane model normalized by the isotropic SYM result at zero magnetic field. \emph{Left:} 3D plot as function of $\mathcal{B}/T^2$ and $v$. \emph{Right:} as a function of $\mathcal{B}/T^2$ for some fixed values of $v$.}
\label{fig:DKdrag2}
\end{figure}

The anisotropic drag forces described by Eqs. \eqref{eq:dragpal} to \eqref{eq:rperp} may be computed in the magnetic brane backgrounds by considering,\footnote{Since the background dilaton field is zero in the magnetic brane model it follows that $g_{\mu\nu}^{(s)}=g_{\mu\nu}$.}
\begin{align}
g_{tt}^{(s)}=-\tilde{U}(\tilde{r})=-U(r),\,\,\,g_{rr}^{(s)}=\frac{1}{U(r)},\,\,\, g_{xx}^{(s)}=g_{yy}^{(s)}=e^{2\tilde{V}(\tilde{r})}=\frac{e^{2V(r)}}{v(b)},\,\,\, g_{zz}^{(s)}=e^{2\tilde{W}(\tilde{r})}=\frac{e^{2W(r)}}{w(b)}.
\label{eq:DKtransf}
\end{align}
Our numerical results for the magnetic field induced anisotropic drag forces (normalized by the isotropic SYM result at zero magnetic field given in Eq. \eqref{eq:dragsym}), valid for arbitrary values of the dimensionless ratio $\mathcal{B}/T^2$, are displayed in Figs.\ \ref{fig:DKdrag1} and \ref{fig:DKdrag2}. We see that both drag forces increase with increasing magnetic field relatively to the isotropic zero magnetic field case with the drag force being generally stronger in the transverse plane to the magnetic field direction, which indicates that probes traversing the plasma in the perpendicular plane to the magnetic field lose more energy than probes moving along the magnetic field direction. Furthermore, we note that the dependence of both drag forces with the magnetic field is enhanced at higher speeds, which shows that faster/lighter probes are more affected by drag force effects in a medium with a strong magnetic field.

The stronger magnetic field dependence of the configuration with $\vec{v} \perp \vec{\mathcal{B}}$ is compatible with the expectation that, for strong magnetic fields, the system described in the gauge theory may be effectively spatially decomposed into a 2-dimensional system lying in the plane transverse to the magnetic field and an 1-dimensional system along the magnetic field direction, with the transverse system being more sensitive to the Landau levels induced by the magnetic field.

From Figs.\ \ref{fig:DKdrag1} and \ref{fig:DKdrag2}, one observes that as the system approaches the ultraviolet fixed point in the limit $\mathcal{B}/T^2\ll 1$ the anisotropy induced by the magnetic field ceases and both drag forces converge to the SYM result given in Eq. \eqref{eq:dragsym}, as expected. Moreover, in the opposite infrared limit, $F_{\textrm{drag}}^{(v\parallel B)}$ ($F_{\textrm{drag}}^{(v\perp B)}$) acquires a constant (linear) dependence on the dimensionless ratio $\mathcal{B}/T^2$ in accordance with the analytical results obtained in Ref. \cite{Li:2016bbh} in the limit $\mathcal{B}/T^2\gg 1$. Moreover, we note that $F_{\textrm{drag}}^{(v\parallel B)}$ reaches its asymptotic behavior in the infrared only for much larger values of the magnetic field than in the case of $F_{\textrm{drag}}^{(v\perp B)}$, especially for large $v$.

\subsection{Langevin diffusion coefficients}

\begin{figure}[h]
\begin{center}
\begin{tabular}{c}
\includegraphics[width=0.45\textwidth]{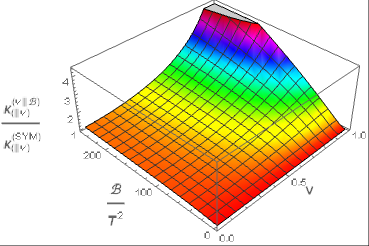} 
\end{tabular}
\begin{tabular}{c}
\includegraphics[width=0.45\textwidth]{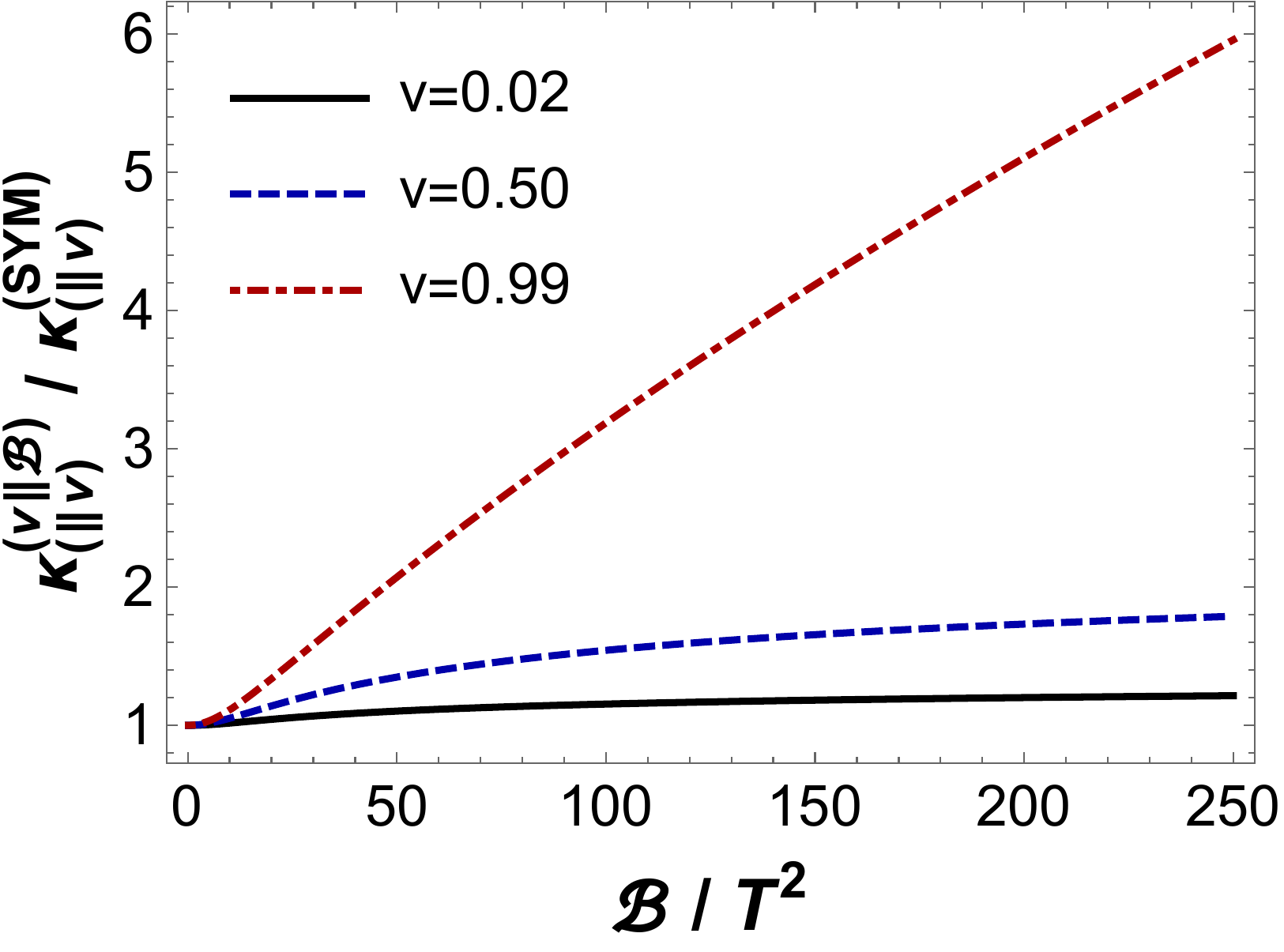} 
\end{tabular}
\end{center}
\caption{(Color online) Anisotropic Langevin diffusion coefficient $\kappa_{(\parallel v)}^{(v\parallel \mathcal{B})}$ in the magnetic brane model normalized by the SYM result at zero magnetic field. \emph{Left:} 3D plot as function of $\mathcal{B}/T^2$ and $v$. \emph{Right:} as a function of $\mathcal{B}/T^2$ for some fixed values of $v$.}
\label{fig:DK_Langevin_kappaL}
\end{figure}

\begin{figure}[h]
\begin{center}
\begin{tabular}{c}
\includegraphics[width=0.45\textwidth]{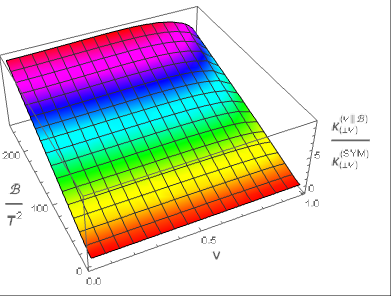} 
\end{tabular}
\begin{tabular}{c}
\includegraphics[width=0.45\textwidth]{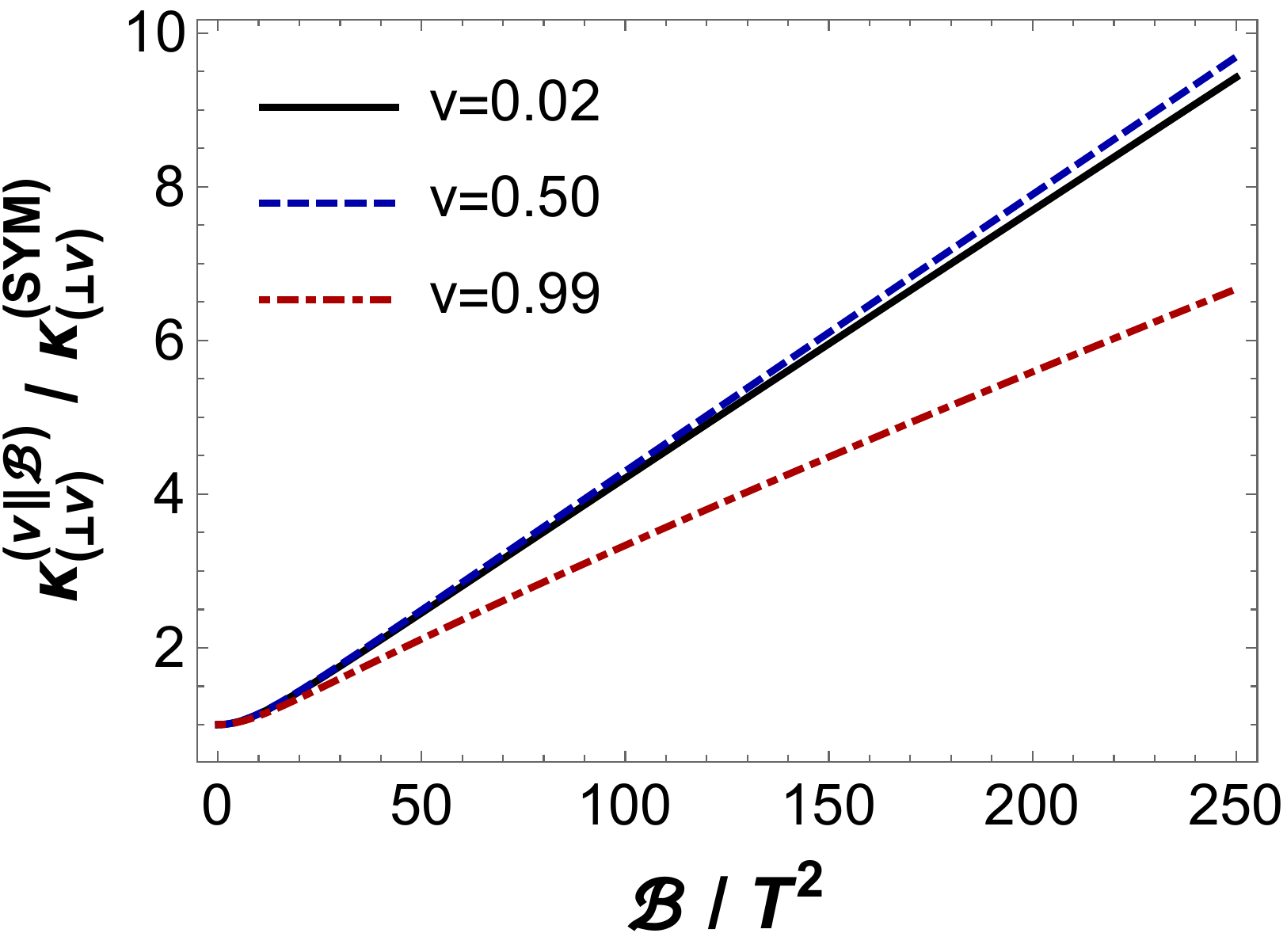} 
\end{tabular}
\end{center}
\caption{(Color online) Anisotropic Langevin diffusion coefficient $\kappa_{(\perp v)}^{(v\parallel \mathcal{B})}$ in the magnetic brane model normalized by the SYM result at zero magnetic field. \emph{Left:} 3D plot as function of $\mathcal{B}/T^2$ and $v$. \emph{Right:} as a function of $\mathcal{B}/T^2$ for some fixed values of $v$.}
\label{fig:DK_Langevin_kappaT}
\end{figure}

\begin{figure}[h]
\begin{center}
\begin{tabular}{c}
\includegraphics[width=0.45\textwidth]{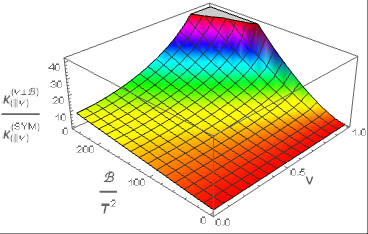} 
\end{tabular}
\begin{tabular}{c}
\includegraphics[width=0.45\textwidth]{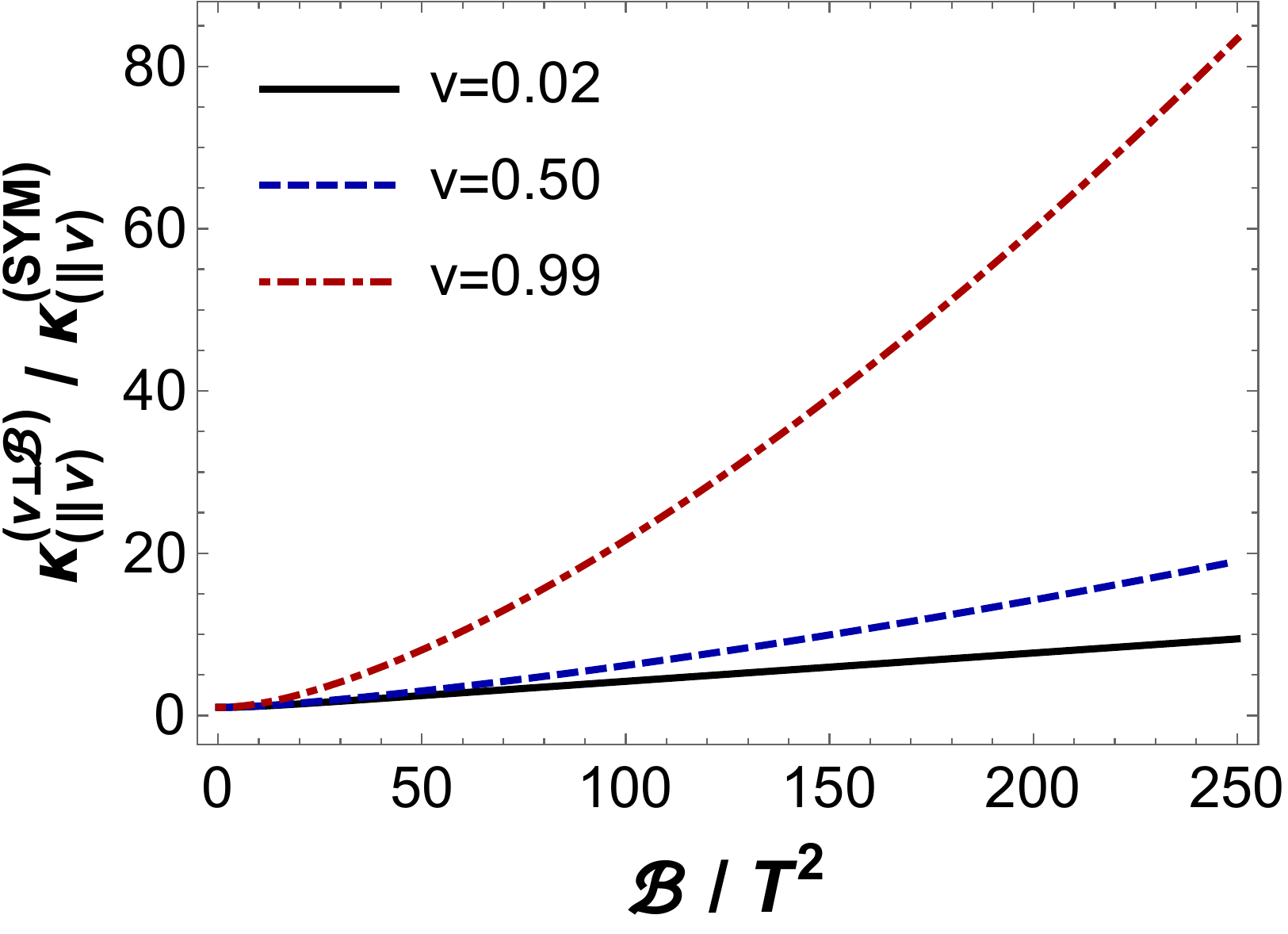} 
\end{tabular}
\end{center}
\caption{(Color online) Anisotropic Langevin diffusion coefficient $\kappa_{(\parallel v)}^{(v\perp \mathcal{B})}$ in the magnetic brane model normalized by the SYM result at zero magnetic field. \emph{Left:} 3D plot as function of $\mathcal{B}/T^2$ and $v$. \emph{Right:} as a function of $\mathcal{B}/T^2$ for some fixed values of $v$.}
\label{fig:DK_Langevin_kappax}
\end{figure}

\begin{figure}[h]
\begin{center}
\begin{tabular}{c}
\includegraphics[width=0.45\textwidth]{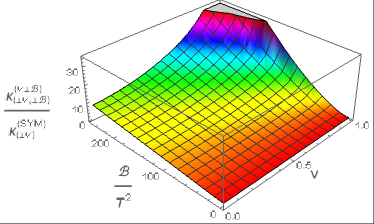} 
\end{tabular}
\begin{tabular}{c}
\includegraphics[width=0.45\textwidth]{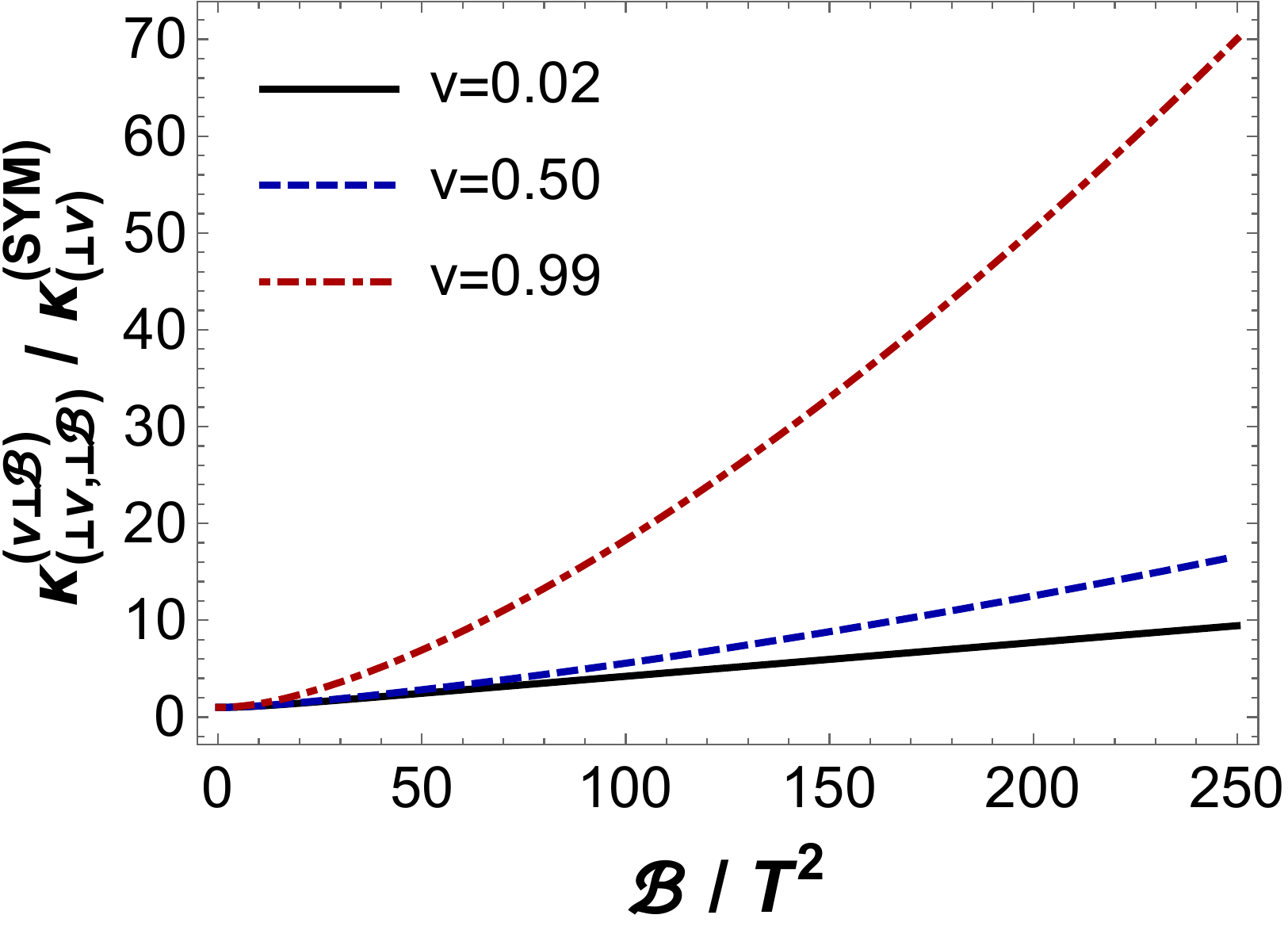} 
\end{tabular}
\end{center}
\caption{(Color online) Anisotropic Langevin diffusion coefficient $\kappa_{(\perp v, \perp \mathcal{B})}^{(v\perp \mathcal{B})}$ in the magnetic brane model normalized by the SYM result at zero magnetic field. \emph{Left:} 3D plot as function of $\mathcal{B}/T^2$ and $v$. \emph{Right:} as a function of $\mathcal{B}/T^2$ for some fixed values of $v$.}
\label{fig:DK_Langevin_kappay}
\end{figure}

\begin{figure}[h]
\begin{center}
\begin{tabular}{c}
\includegraphics[width=0.45\textwidth]{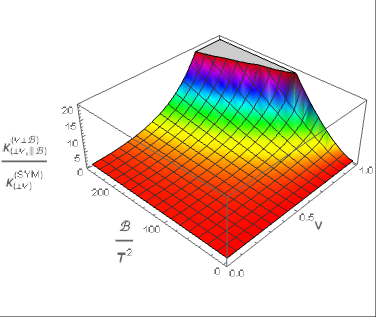} 
\end{tabular}
\begin{tabular}{c}
\includegraphics[width=0.45\textwidth]{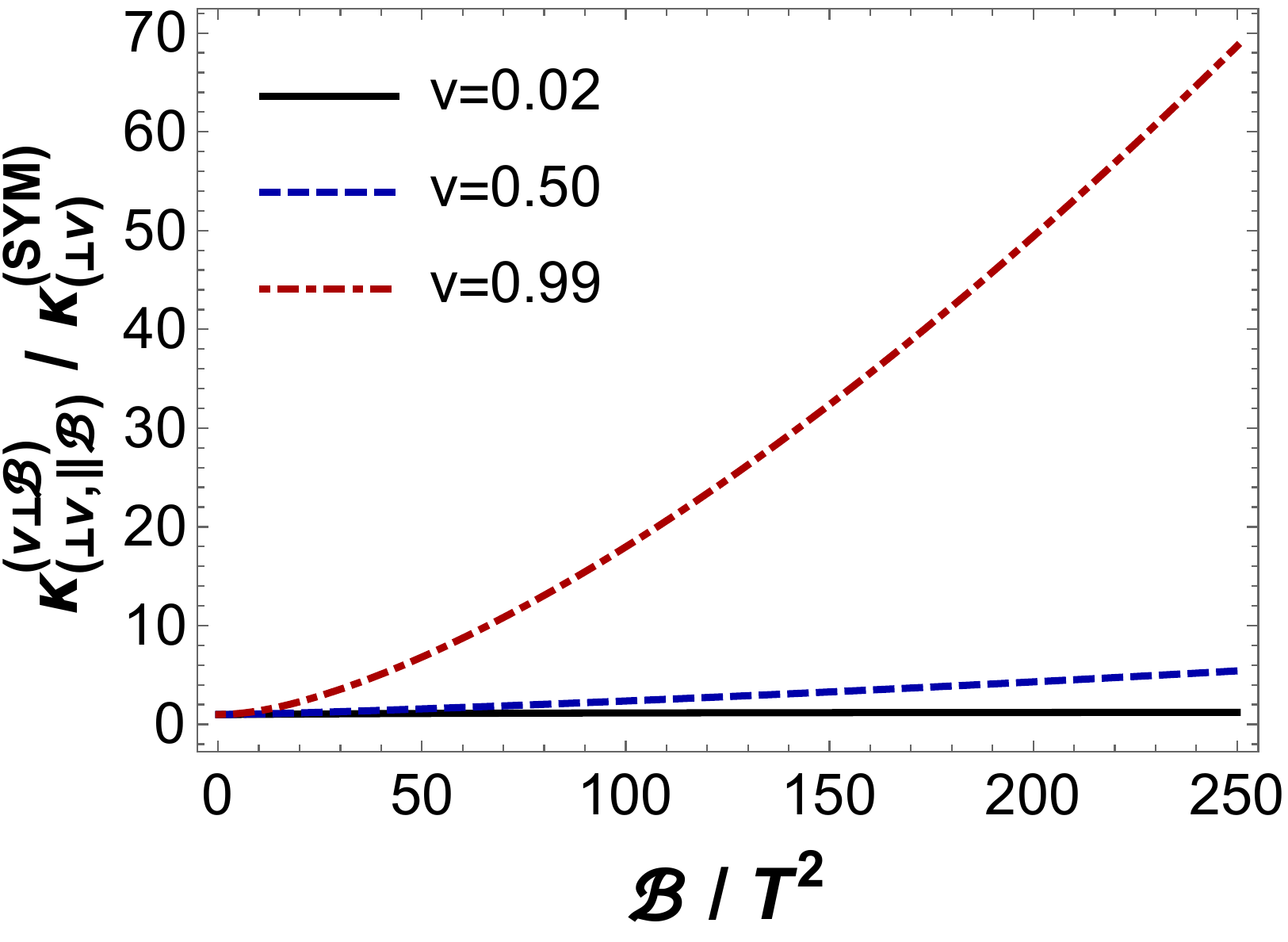} 
\end{tabular}
\end{center}
\caption{(Color online) Anisotropic Langevin diffusion coefficient $\kappa_{(\perp v, \parallel \mathcal{B})}^{(v\perp \mathcal{B})}$ in the magnetic brane model normalized by the SYM result at zero magnetic field. \emph{Left:} 3D plot as function of $\mathcal{B}/T^2$ and $v$. \emph{Right:} as a function of $\mathcal{B}/T^2$ for some fixed values of $v$.}
\label{fig:DK_Langevin_kappaz}
\end{figure}

The anisotropic Langevin diffusion coefficients described by Eqs.\ \eqref{eq:dif1} to \eqref{eq:dif2} and Eqs.\ \eqref{eq:dif3} to \eqref{eq:dif5} may be computed in the magnetic brane backgrounds by considering the relations in Eq.\ \eqref{eq:DKtransf}. Our numerical results for the magnetic field induced anisotropic Langevin coefficients, normalized by the SYM results at zero magnetic field given by the relations in Eq.\ \eqref{eq:LangSYM}, which are valid for arbitrary values of the dimensionless ratio $\mathcal{B}/T^2$, are displayed in Figs.\ \ref{fig:DK_Langevin_kappaL} to \ref{fig:DK_Langevin_kappaz}. We see that all the Langevin coefficients are enhanced in the presence of an external magnetic field relative to the zero magnetic field case, which means that diffusion through the plasma is facilitated by the presence of a magnetic field. One also notes that momentum diffusion in directions transverse to the magnetic field is generally larger than in the direction of the field. This is line with what is observed in the drag force: the probe loses more energy and diffuses more momentum when moving along directions in the plane perpendicular to the magnetic field. Furthermore, the coefficients $\kappa_{(\parallel v)}^{(v\parallel \mathcal{B})}$, $\kappa_{(\parallel v)}^{(v\perp \mathcal{B})}$, $\kappa_{(\perp v, \perp \mathcal{B})}^{(v \perp \mathcal{B})}$, and $\kappa_{(\perp v,\parallel \mathcal{B})}^{(v \perp \mathcal{B})}$ always increase with increasing velocity, contrary to what happens with the coefficient $\kappa_{(\perp v)}^{(v\parallel \mathcal{B})}$ at large $v$.

We also checked that in the infrared limit, the Langevin diffusion coefficients agree with the analytical behavior derived in Ref. \cite{Li:2016bbh} in the limit $\mathcal{B}/T^2\gg 1$.

\subsection{Shear viscosity}
\label{sec3.4}

\begin{figure}[h]
\begin{center}
\includegraphics[width=0.55\textwidth]{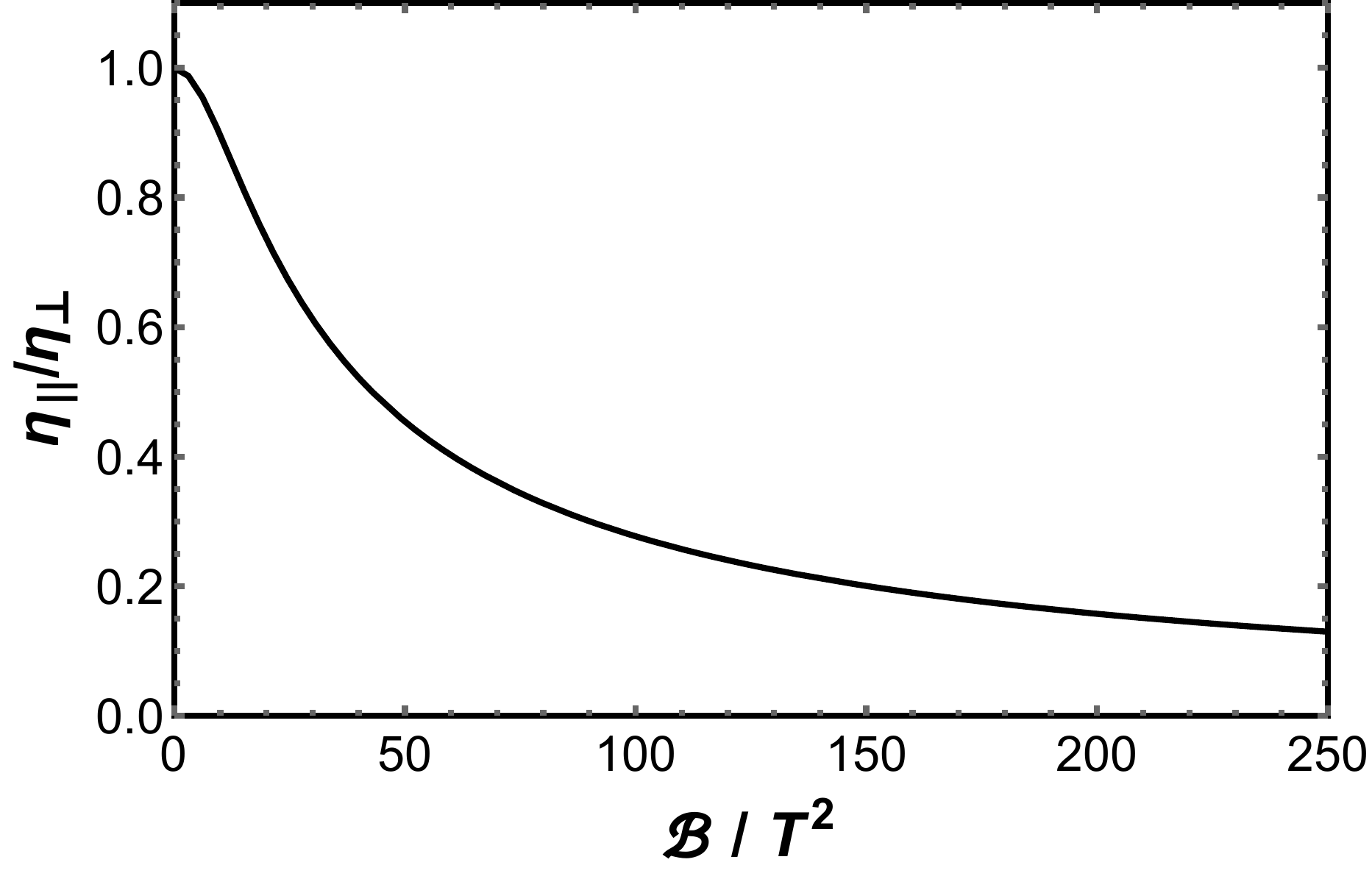} 
\end{center}
\caption{Ratio between the parallel and perpendicular shear viscosities in the magnetic brane model as a function of the dimensionless ratio $\mathcal{B}/T^2$.}
\label{fig:DK_eta}
\end{figure}

The anisotropic shear viscosities described by Eqs.\ \eqref{eq:eta_perp_form} and \eqref{eq:eta_par_form} may be computed in the magnetic brane backgrounds by considering the relations in Eq.\ \eqref{eq:DKtransf}. In Fig.\ \ref{fig:DK_eta} we plot our numerical results for the ratio between the parallel and perpendicular shear viscosities, which was originally obtained in Ref.\ \cite{DK-applications2} and it is displayed here for completeness.

One can see that the shear viscosity is reduced in the direction of the magnetic field in comparison to the value of viscosity perpendicular to the field, which indicates that the strongly coupled anisotropic plasma becomes even closer to the perfect fluid limit along the magnetic field direction.

\section{The magnetic Einstein-Maxwell-dilaton model}
\label{sec4.0}

In this section, we begin by briefly reviewing and updating the phenomenologically constructed, QCD-like, bottom-up magnetic EMD model originally proposed in Ref.\ \cite{Rougemont:2015oea} (we refer the interested reader to consult this reference for further details and discussions). We then obtain the updated results for the equation of state at finite $B$ and also calculate the anisotropic drag forces, Langevin diffusion coefficients, and shear viscosities.

The update of the magnetic EMD model we shall introduce in the present work refers to the fact that this time we will dynamically fix the free parameters of the bottom-up model by performing a global matching of the holographic equation of state (including the entropy and internal energy densities, the speed of sound squared, the pressure, and the trace anomaly) and the magnetic susceptibility at zero magnetic field with the latest $(2+1)$-flavor lattice QCD data with physical quark masses from Refs. \cite{Borsanyi:2013bia} and \cite{latticedata2}, respectively. Previously, in Ref.\ \cite{Rougemont:2015oea}, some of us used lattice data just for the speed of sound squared and the pressure at $B=0$ from the older Ref. \cite{latticedata1} to fix the dilaton potential of the EMD model. The simple improvements referred above to fix the free parameters of the model at $B=0$ will result in a much better quantitative agreement between the finite $B$ holographic and lattice QCD equations of state (as previously suggested in Ref.\ \cite{Rougemont:2015oea}). We shall also present holographic predictions for the entropy density and the crossover temperature in the presence of a magnetic field in a wider region of the $(T,B)$ phase diagram that has not yet been explored in lattice simulations.

It is also important to mention that even though the EMD setting worked out here does not explicitly introduce fundamental flavors in the gauge theory by means of the standard holographic dictionary (which would require the use of flavor D-branes), the dilaton potential used here is such that the holographic equation of state for the black brane at $B=0$ matches the corresponding lattice QCD results with $(2+1)$ flavors. In this sense, the setup used here corresponds to an \emph{emulator} which is able to \emph{mimic} some of the relevant properties of QCD with dynamical flavors, such as the crossover transition. Such an effective approach was originally proposed in Ref.\ \cite{Gubser:2008ny} (see also \cite{Yaresko:2015ysa} for more recent discussions) where it was also shown that different parametrizations for the dilaton potential can mimic not only the QCD crossover but also first and second order phase transitions, which may be relevant for different applications ranging from the QGP to condensed matter physics.

\subsection{The model and its thermodynamics}
\label{sec4.1}

\textit{Action.} The bulk action for the EMD model is given by
\begin{align}
S&=\frac{1}{16\pi G_5}\int_{\mathcal{M}_5}d^5x\sqrt{-g}\left[R-\frac{1}{2}(\partial_\mu\phi)^2-V(\phi) -\frac{f(\phi)}{4}F_{\mu\nu}^2\right],
\end{align}
which is supplemented by boundary terms, as before. However, as before these boundary terms will not be needed for the calculations carried out in the present work and, therefore, we omit them from our discussion. We shall dynamically fix the dilaton potential $V(\phi)$ and the gravitational constant $\kappa^2\equiv 8\pi G_5$ by matching the holographic equation of state at $B=0$ with the corresponding lattice data for $(2+1)$-flavor QCD from Ref.\ \cite{Borsanyi:2013bia}. The dynamical dilaton field breaks conformal symmetry in the infrared (the interior of the bulk) where it acquires a nontrivial profile, while close to the boundary it vanishes and the theory goes back to the ultraviolet fixed point corresponding to the AdS$_5$ geometry. The potential $V(\phi)$ has the near-boundary ultraviolet expansion\footnote{As before, we set to unity the radius $L$ of the asymptotically AdS$_5$ geometries.} $V(\phi\to 0) \approx -12 + m^2\phi^2/2$, where the mass $m$ of the scalar field $\phi$ defines the scaling dimension $\Delta$ of the dual operator in the gauge theory through the relation $m^2 = -\nu \Delta$, where $\nu = 4-\Delta$. For the dilaton potential we shall fix in what follows, the Breitenlohner-Freedman bound \cite{Breitenlohner:1982jf,Breitenlohner:1982bm} is satisfied. The Maxwell-dilaton gauge coupling $f(\phi)$ will be dynamically fixed by matching the holographic magnetic susceptibility at $B=0$ to the corresponding lattice data for $(2+1)$-flavor QCD from Ref.\ \cite{latticedata2}.

\vspace{8pt}

\textit{Ansatz and equations of motion.} The ansatz for the anisotropic magnetic backgrounds in the so-called standard coordinates, denoted as before by a tilde, is given by
\begin{align}
d\tilde{s}^2&=e^{2\tilde{a}(\tilde{r})}\left[-\tilde{h}(\tilde{r})d\tilde{t}^2+d\tilde{z}^2\right]+ e^{2\tilde{c}(\tilde{r})}(d\tilde{x}^2+d\tilde{y}^2)+\frac{e^{2\tilde{b}(\tilde{r})} d\tilde{r}^2}{\tilde{h}(\tilde{r})},\nonumber\\
\tilde{\phi}&=\tilde{\phi}(\tilde{r}),\,\,\, \tilde{A}=\tilde{A}_\mu d\tilde{x}^\mu=\hat{B}\tilde{x}d\tilde{y}\Rightarrow \tilde{F}=d\tilde{A}=\hat{B}d\tilde{x}\wedge d\tilde{y},
\label{2.14}
\end{align}
where the hat in $\hat{B}$ is used to denote the magnetic field in units of the inverse of the asymptotically AdS$_5$ radius squared (we shall use $B$ to denote the boundary magnetic field in physical units, as we are going to discuss in a moment). The boundary of the asymptotically $\mathrm{AdS}_5$ backgrounds lies at $\tilde{r} \to \infty$ while the horizon is at $\tilde{r} = \tilde{r}_H$. The equation of motion for the dilaton field is given by
\begin{align}
\tilde{\phi}'' + \left(2\tilde{a}'+2\tilde{c}'-\tilde{b}'+\frac{\tilde{h}'}{\tilde{h}} \right) \tilde{\phi}' - \frac{e^{2\tilde{b}}}{\tilde{h}} \left(\frac{\partial V}{\partial\tilde{\phi}} + \frac{\hat{B}^2 e^{-4\tilde{c}}}{2} \frac{\partial f}{\partial\tilde{\phi}} \right) = 0,
\end{align}
while Einstein's equations are given by
\begin{align}
\tilde{a}''+\left( \frac{14}{3} \tilde{c}' - \tilde{b}' + \frac{4}{3} \frac{\tilde{h}'}{h} \right) \tilde{a}' + \frac{8}{3} \tilde{a}'^2 + \frac{2}{3} \tilde{c}'^2 + \frac{2}{3} \frac{\tilde{h}'}{\tilde{h}} \tilde{c}' + \frac{2}{3} \frac{e^{2\tilde{b}}}{\tilde{h}} V - \frac{1}{6} \tilde{\phi}'^2 & =0, \\
\tilde{c}'' - \left(\frac{10}{3} \tilde{a}'+\tilde{b}'+\frac{1}{3} \frac{\tilde{h}'}{\tilde{h}} \right) + \frac{2}{3} \tilde{c}'^2 - \frac{4}{3} \tilde{a}'^2 - \frac{2}{3} \frac{\tilde{h}'}{\tilde{h}} \tilde{a}' - \frac{1}{3}\frac{e^{2\tilde{b}}}{\tilde{h}} V + \frac{1}{3} \tilde{\phi}'^2 & = 0, \\
\tilde{h}''+(2\tilde{a}'+2\tilde{c}'-\tilde{b}')\tilde{h}' & = 0.
\end{align}
One can derive one last useful equation from the above, which is taken as a constraint on initial data:
\begin{equation}
\tilde{a}'^2 + \tilde{c}'^2 - \frac{1}{4} \tilde{\phi}'^2 + \left( \frac{\tilde{a}'}{2} + \tilde{c}' \right)\frac{\tilde{h}'}{\tilde{h}} + 4\tilde{a}'\tilde{c}' + \frac{e^{2\tilde{b}}}{2\tilde{h}} \left( V + \frac{\hat{B}^2 e^{-4\tilde{c}}}{2} f \right) = 0.
\end{equation}
Maxwell's equations are automatically satisfied by the ansatz \eqref{2.14}; also, the function $\tilde{b}(\tilde{r})$ has no equation of motion to satisfy and, in fact, it may be gauge-fixed at will using invariance under reparametrizations of the radial coordinate. In the following, we set $\tilde{b}(\tilde{r})=0$.

\vspace{8pt}

\textit{Numerical coordinates.} As it was done before in this paper, in order to numerically solve the equations of motion we introduce numerical coordinates which are represented without the tildes. The background fields are expressed in the numerical coordinates as follows (using again the $b(r)=0$ gauge),
\begin{align}
ds^2&=e^{2a(r)}\left[-h(r)dt^2+dz^2\right]+e^{2c(r)}(dx^2+dy^2)+\frac{dr^2}{h(r)},\nonumber\\
\phi&=\phi(r),\,\,\,A=A_\mu dx^\mu=\mathcal{B}xdy\Rightarrow F=dA=\mathcal{B}dx\wedge dy.
\end{align}

Let $X(r)$ be any of the background functions $a(r)$, $c(r)$, $h(r)$, or $\phi(r)$, and take near-horizon Taylor expansions,
\begin{equation}
X(r) = X_0 + X_1(r-r_H) + X_2 (r-r_H)^2 + \ldots
\end{equation}
Working with Taylor expansions up to second order results in a total of 12 coefficients to be determined in order to start the numerical integration of the equations of motion. One of these infrared Taylor coefficients is the value of the dilaton field at the horizon, $\phi_0$, which is taken as one of the initial conditions of the system (the second initial condition corresponds to the value of the magnetic field in the numerical coordinates, which we denote in this section by $\mathcal{B}$; we are going to derive later in this paper the relation between $\hat{B}$ and $\mathcal{B}$). As discussed in Ref.\ \cite{Rougemont:2015oea}, one may use the freedom to rescale the bulk spacetime coordinates in order to fix the horizon location and three of the infrared Taylor coefficients as follows: $r_H = 0$, $a_0 = c_0 = 0$, and $h_1 = 1$, while $h_0 = 0$ follows from the defining properties of the blackening function $h(r)$. The 7 remaining infrared Taylor coefficients are fixed on-shell as functions of the pair of initial conditions $(\phi_0,\mathcal{B})$ by substituting these near-horizon expansions into the equations of motion expressed in the numerical coordinates and solving the resulting coupled system of algebraic equations. In order to perform the numerical integration of the equations of motion in the numerical coordinates we start a little above the horizon\footnote{This is done in order to avoid the singular point of the equations of motion corresponding to the radial location of the black hole horizon.}, at $r_{\textrm{start}} = 10^{-8}$, and integrate up to $r_{\textrm{max}} = 10$, which is a large value of the radial coordinate where all the geometries generated in the present work have already reached the ultraviolet fixed point corresponding to the AdS$_5$ geometry.\footnote{For the set of initial conditions considered in the present work the resulting black hole geometries generally reach the ultraviolet fixed point already before $r_{\textrm{conformal}} = 2$ but we consider integrations up to larger values of $r$ for technical reasons involving the calculation of the holographic magnetic susceptibility (see the discussion in Ref.\ \cite{Rougemont:2015oea}) and also the fixing of the leading coefficient of the near-boundary, ultraviolet expansion for the dilaton field.} Then, each value of the pair of initial conditions $(\phi_0,\mathcal{B})$ will generate numerical black hole geometries dual to a definite physical state in the gauge theory.

\begin{figure}[h]
\begin{tabular}{c}
\includegraphics[width=0.5\textwidth]{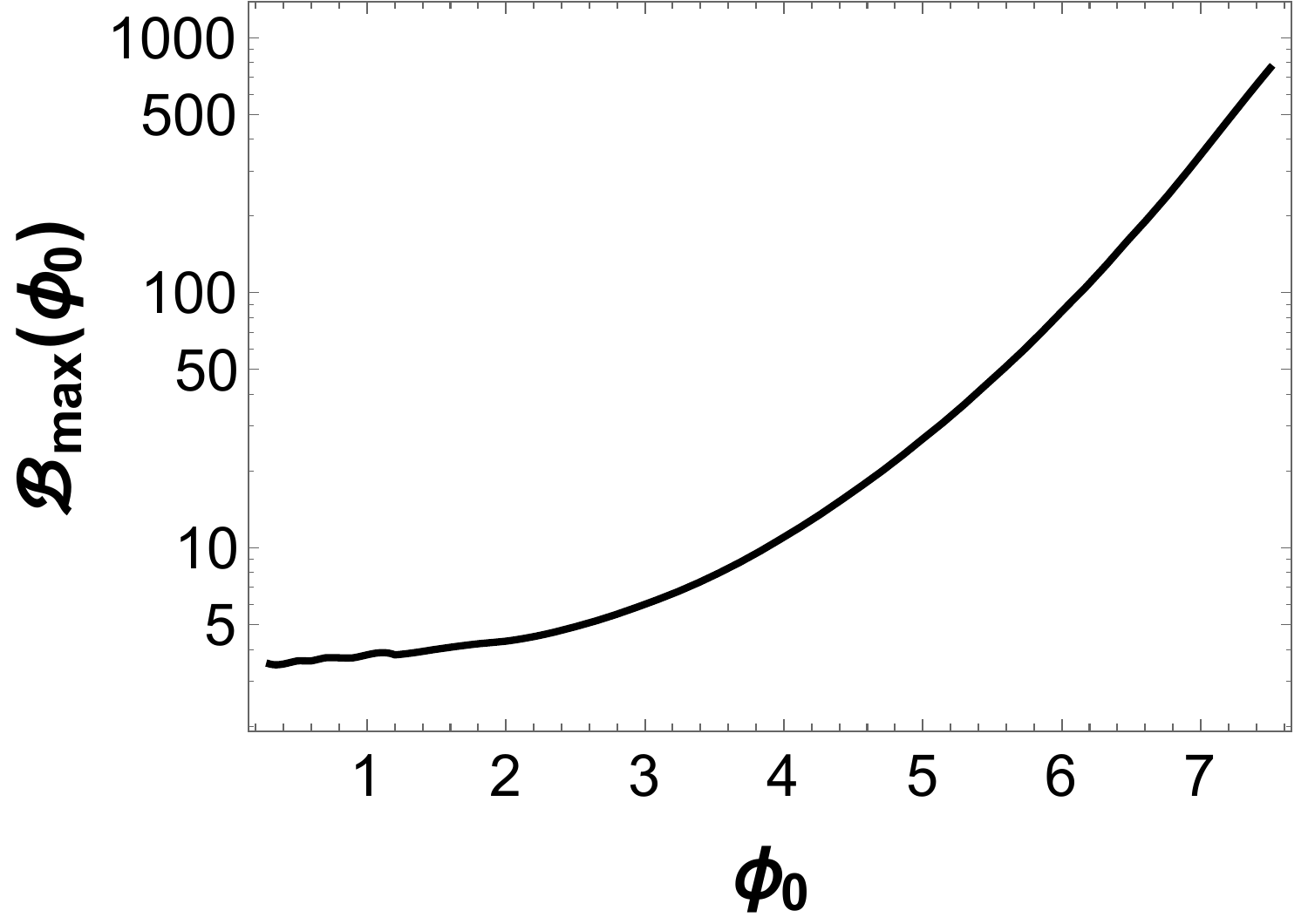} 
\end{tabular}
\caption{Numerical bound on the maximum value for the initial condition $\mathcal{B}$ given some initial condition $\phi_0$: initial conditions below the curve give asymptotically AdS$_5$ solutions.}
\label{fig:EMDbound}
\end{figure}

It is also important to remark that, as discussed in Ref. \cite{Rougemont:2015oea},\footnote{See also a similar discussion in Ref.\ \cite{gubser2} though in the different context of the EMD model at $B=0$ and finite baryon chemical potential.} there is an upper bound on the value of the initial condition $\mathcal{B}$, given some initial condition $\phi_0$, in order to generate asymptotically AdS$_5$ geometries. This bound may be derived numerically and we display it in Fig.\ \ref{fig:EMDbound}.

In order to determine the thermodynamic and transport observables to be presented in the following sections we generated 850,000 different anisotropic black hole geometries by constructing a rectangular grid in the plane of initial conditions $(\phi_0,\mathcal{B})$, with 1000 points in the $\phi_0$ direction varying in equally spaced steps within the interval $[0.3,7.5]$, and for each value of $\phi_0$, we considered 850 points in the $\mathcal{B}/\mathcal{B}_\textrm{max}(\phi_0)$ direction varying in equally spaced steps within the interval $[0,0.99]$. These settings were enough to obtain smooth results within the phenomenologically interesting region of the physical $(T,B)$ plane studied below.

\vspace{8pt}

\textit{Asymptotics and coordinate transformations.} Physical observables are naturally computed in the gravity theory using the standard coordinates while numerical solutions for the background black hole geometries are obtained in the numerical coordinates defined above. Therefore, we need to relate these different coordinate systems and here this is done by matching the near-boundary, ultraviolet expansions for the EMD fields in each case.

In the standard coordinates, the ultraviolet asymptotics attained at large $\tilde{r}$ are given by \cite{DeWolfe:2010he,Rougemont:2015oea}
\begin{align}
\tilde{a} (\tilde{r}) = \tilde{r} + \ldots, \,\,\,
\tilde{c} (\tilde{r}) = \tilde{r} + \ldots, \,\,\,
\tilde{h} (\tilde{r}) = 1 + \ldots, \,\,\,
\tilde{\phi} (\tilde{r}) = e^{- \nu \tilde{r}} + \ldots\,.
\end{align}

On the other hand, in the numerical coordinates, the near-boundary asymptotics are given by \cite{Rougemont:2015oea}
\begin{align}
a(r) & = \alpha(r) + \ldots, \\
c(r) & = \alpha(r) + c^{\mathrm{far}}_0 - a^{\mathrm{far}}_0 + \ldots, \\
h(r) & = h^{\mathrm{far}}_0 + h^{\mathrm{far}}_4 e^{-4 \alpha(r)} + \ldots, \\
\phi(r) & = \phi_A e^{-\nu \alpha(r)} + \phi_B e^{-\Delta \alpha(r)} + \ldots,
\end{align}
where $\alpha(r) = a^{\mathrm{far}}_0 + r/\sqrt{h^{\mathrm{far}}_0}$. The coefficients $a^{\mathrm{far}}_0$, $c^{\mathrm{far}}_0$, $h^{\mathrm{far}}_0$, and $\phi_A$, which are required to calculate most of the physical quantities, may be found by fitting the numerical solutions close to the boundary using the ultraviolet asymptotics above. We found that an accurate fitting procedure may be specified as follows: $h_0^{\textrm{far}}=h(r_{\textrm{conformal}})$, while $a^{\mathrm{far}}_0$ and $c^{\mathrm{far}}_0$ may be extracted by employing the fitting profiles $a(r)=a^{\mathrm{far}}_0+r/\sqrt{h_0^{\textrm{far}}}$ and $c(r)=c^{\mathrm{far}}_0+r/\sqrt{h_0^{\textrm{far}}}$ within the interval $r\in [r_{\textrm{conformal}}-1,r_{\textrm{conformal}}]$. The ultraviolet coefficient $\phi_A$ is the most difficult to extract due to the fact that the dilaton field vanishes exponentially fast in the near-boundary region. In the present work, we employed the following procedure to fix this coefficient:\footnote{For the magnetic EMD model used in Ref.\ \cite{Rougemont:2015oea} this procedure gives the same results found by the simpler procedure discussed in that reference. However, for the choice of $V(\phi)$ and $f(\phi)$ used here, the procedure discussed in Ref.\ \cite{Rougemont:2015oea} can only cover a very limited region of the plane of initial conditions while this new numerical procedure discussed in this paper can cover a much broader region. We also checked that the present procedure works fairly well with a wide variety of different choices for $V(\phi)$ and $f(\phi)$.} first, one defines the following adaptive variables, $r_{\textrm{IR}}(\phi_0,\mathcal{B})\equiv \phi^{-1}(10^{-3})$ and $r_{\textrm{UV}}(\phi_0,\mathcal{B})\equiv \phi^{-1}(10^{-5})$, using next the fitting profile $\phi(r)=\phi_A e^{-\nu a(r)}$ within the adaptive interval $r\in [r_{\textrm{IR}},r_{\textrm{UV}}]$ to reliably extract $\phi_A$ from the numerical solutions generated by different pairs of initial conditions $(\phi_0,\mathcal{B})$.

By comparing both sets of asymptotic solutions in the ultraviolet, one finds a dictionary relating the standard and the numerical coordinates \cite{Rougemont:2015oea}:
\begin{align}
\tilde{r}&=\frac{r}{\sqrt{h_0^{\textrm{far}}}}+a_0^{\textrm{far}}-\ln\left(\phi_A^{1/\nu}\right),\,\,\,
\tilde{t}=\phi_A^{1/\nu}\sqrt{h_0^{\textrm{far}}}t,\,\,\,
\tilde{x}=\phi_A^{1/\nu}e^{c_0^{\textrm{far}}-a_0^{\textrm{far}}}x,\nonumber\\
\tilde{y}&=\phi_A^{1/\nu}e^{c_0^{\textrm{far}}-a_0^{\textrm{far}}}y,\,\,\,
\tilde{z}=\phi_A^{1/\nu}z,\,\,\,
\tilde{a}(\tilde{r})=a(r)-\ln\left(\phi_A^{1/\nu}\right),\nonumber\\
\tilde{c}(\tilde{r})&=c(r)-(c_0^{\textrm{far}}-a_0^{\textrm{far}})-\ln\left(\phi_A^{1/\nu}\right),\,\,\,
\tilde{h}(\tilde{r})=\frac{h(r)}{h_0^{\textrm{far}}},\,\,\,
\tilde{\phi}(\tilde{r})=\phi(r),\nonumber\\
\hat{B}&=\frac{e^{2(a_0^{\textrm{far}}-c_0^{\textrm{far}})}}{\phi_A^{2/\nu}}\mathcal{B}.
\label{eq:RelationsEMD-Stan-Num}
\end{align}

\newpage

\vspace{8pt}

\textit{Thermodynamics at zero magnetic field.} By employing the usual formulas for the Hawking temperature of the black hole horizon and the Bekenstein-Hawking relation for the black hole entropy density in the standard coordinates, and then going over to the numerical coordinates, one obtains that
\begin{equation}
\hat{T} = \frac{1}{4\pi \phi_A^{1/\nu} \sqrt{h^{\mathrm{far}}_0}} \quad \textrm{and} \quad
\hat{s} = \frac{2\pi e^{2(a_0^{\textrm{far}}-c_0^{\textrm{far}})}}{\kappa^2\phi_A^{3/\nu}}.
\label{eq:Tands}
\end{equation}
As mentioned before, the hats in Eqs.\ \eqref{eq:RelationsEMD-Stan-Num} and \eqref{eq:Tands} denote the physical observables in units of powers of the inverse of the AdS$_5$ radius $L$, which we have already set to unity. In order to express observables in physical units, we introduce a scaling factor $\Lambda$ with units of MeV such that any observable $\hat{X}$ with mass dimension $p$ is expressed in physical units as $X=\hat{X}\Lambda^p$ [MeV$^p$] \cite{hydro,finitemu,Rougemont:2015ona,Rougemont:2015oea,Finazzo:2015xwa}. Note that, by doing so, we do not introduce any new free parameter in the theory, since this procedure just exchanges the freedom in the choice of $L$ with the freedom in the choice of the scaling factor $\Lambda$. Therefore, one has for instance, $B=\hat{B}\Lambda^2$ [MeV$^2$], $T=\hat{T}\Lambda$ [MeV], and $s=\hat{s}\Lambda^3$ [MeV$^3$].

As discussed in detail in Ref.\ \cite{Rougemont:2015oea}, the holographic formula for the magnetic susceptibility at zero magnetic field in the numerical coordinates is given by\footnote{Ideally, one should take $T_{\textrm{small}}=0$, however, due to technical difficulties in subtracting numerical geometries at finite $T$ and the vacuum geometry, in practice we numerically subtracted a zero magnetic field geometry with a very small but nonzero temperature.}
\begin{align}
\chi(T,B=0)=-\frac{1}{2\kappa^2}\left[\left(\frac{1}{\sqrt{h_0^{\textrm{far}}}} \int_{r_{\textrm{start}}}^{r^{\textrm{var}}_{\textrm{max}}} dr f(\phi(r))\right)\biggr|_{T,B=0}-\left(\textrm{same}\right)\biggr|_{T_{\textrm{small}},B=0} \right]_{\textrm{on-shell}},
\label{2.23}
\end{align}
where $r^{\textrm{var}}_{\textrm{max}}\equiv\sqrt{h_0^{\textrm{far}}}\left[\tilde{r}^{\textrm{fixed}}_{\textrm{max}}- a_0^{\textrm{far}}+\ln\left(\phi_A^{1/\nu}\right)\right]$, with $\tilde{r}^{\textrm{fixed}}_{\textrm{max}}$ being a fixed numerical ultraviolet cutoff which must be chosen in such a way that the upper limits of integration in Eq.\ \eqref{2.23} satisfy the condition $r_{\textrm{conformal}}\le r^{\textrm{var}}_{\textrm{max}}\le r_{\textrm{max}}$ for all the geometries considered.

\begin{figure}[h]
\begin{tabular}{c}
\includegraphics[width=0.45\textwidth]{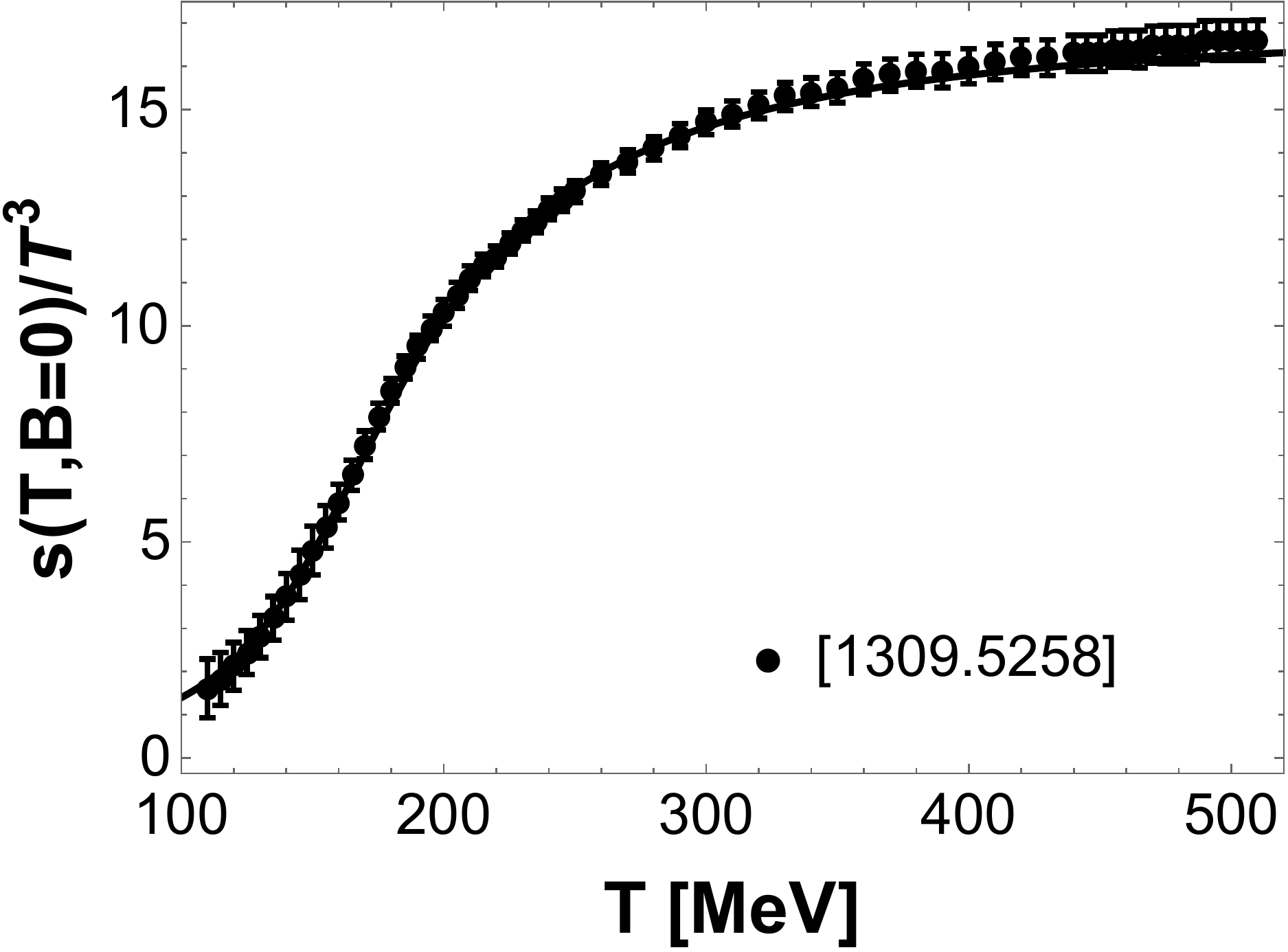} 
\end{tabular}
\begin{tabular}{c}
\includegraphics[width=0.45\textwidth]{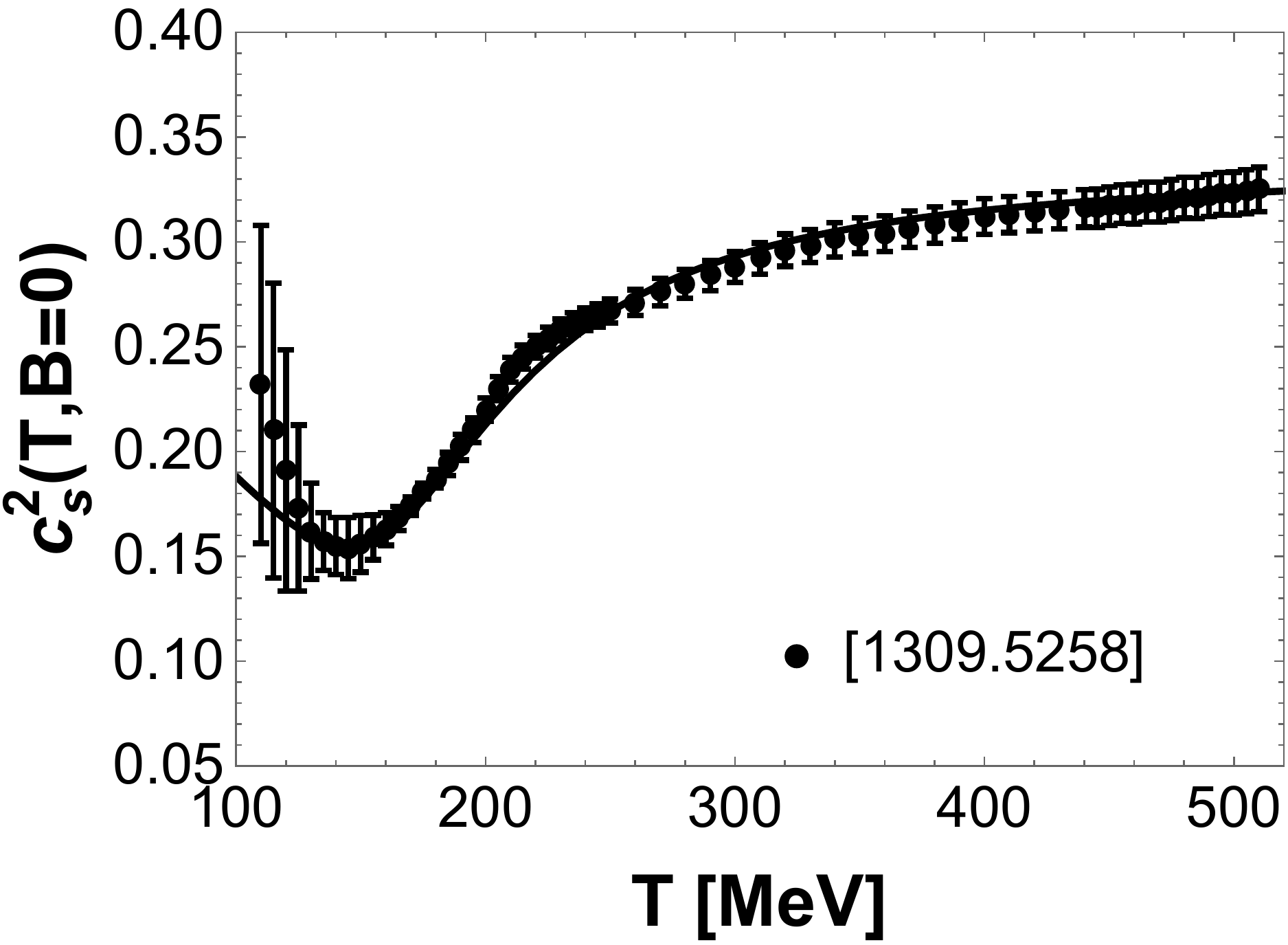} 
\end{tabular}
\begin{tabular}{c}
\includegraphics[width=0.45\textwidth]{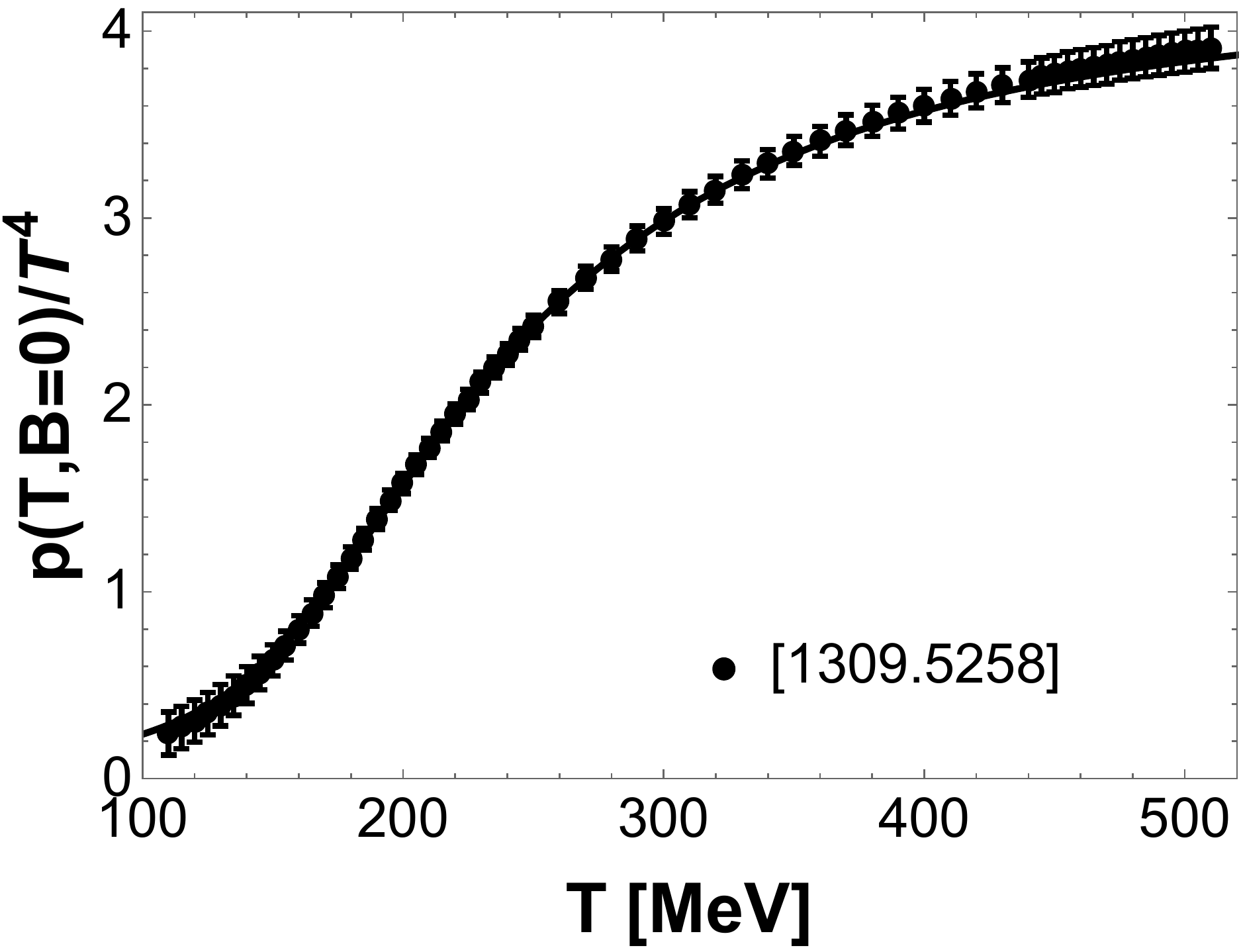} 
\end{tabular}
\begin{tabular}{c}
\includegraphics[width=0.45\textwidth]{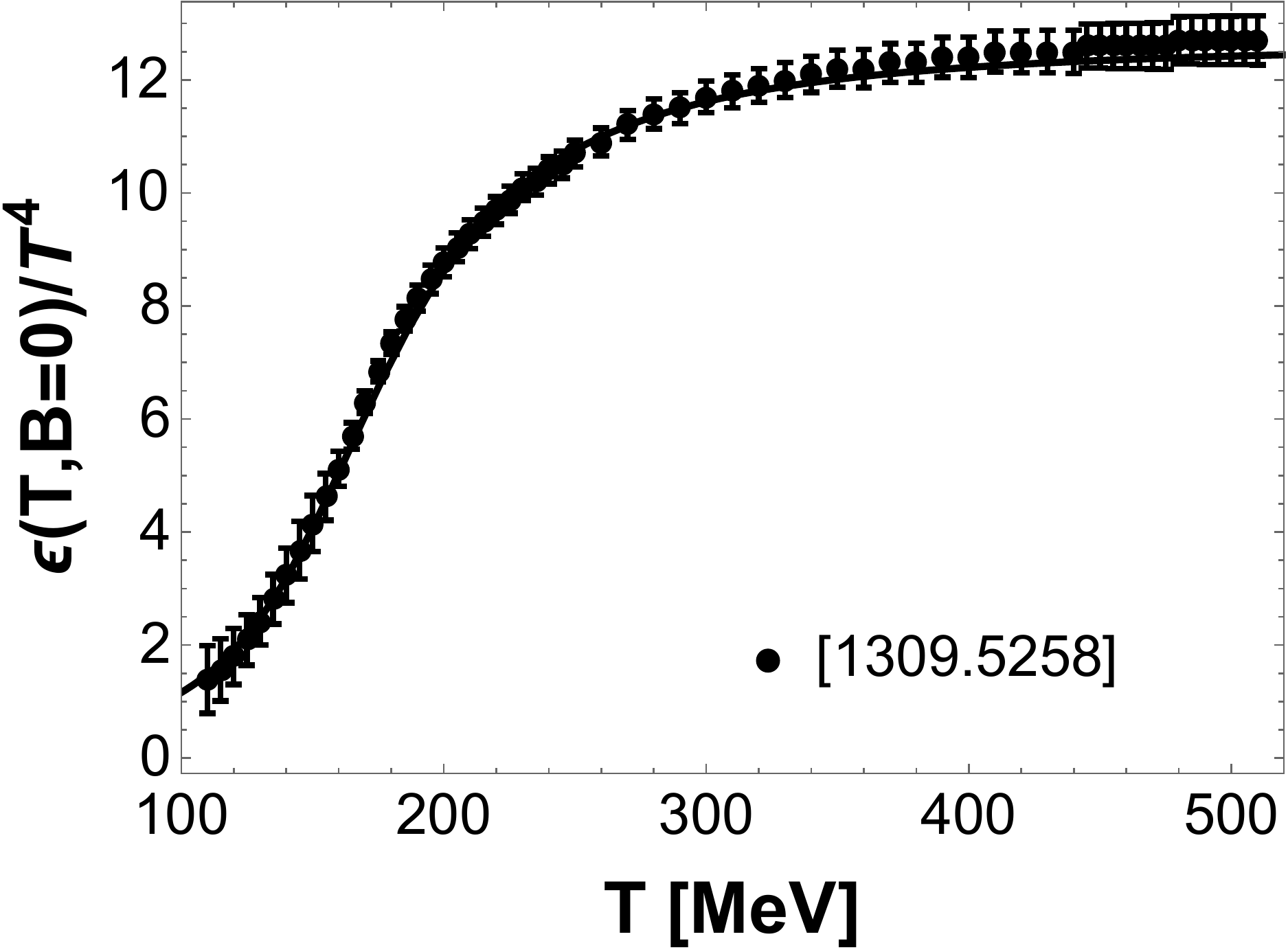} 
\end{tabular}
\begin{tabular}{c}
\includegraphics[width=0.45\textwidth]{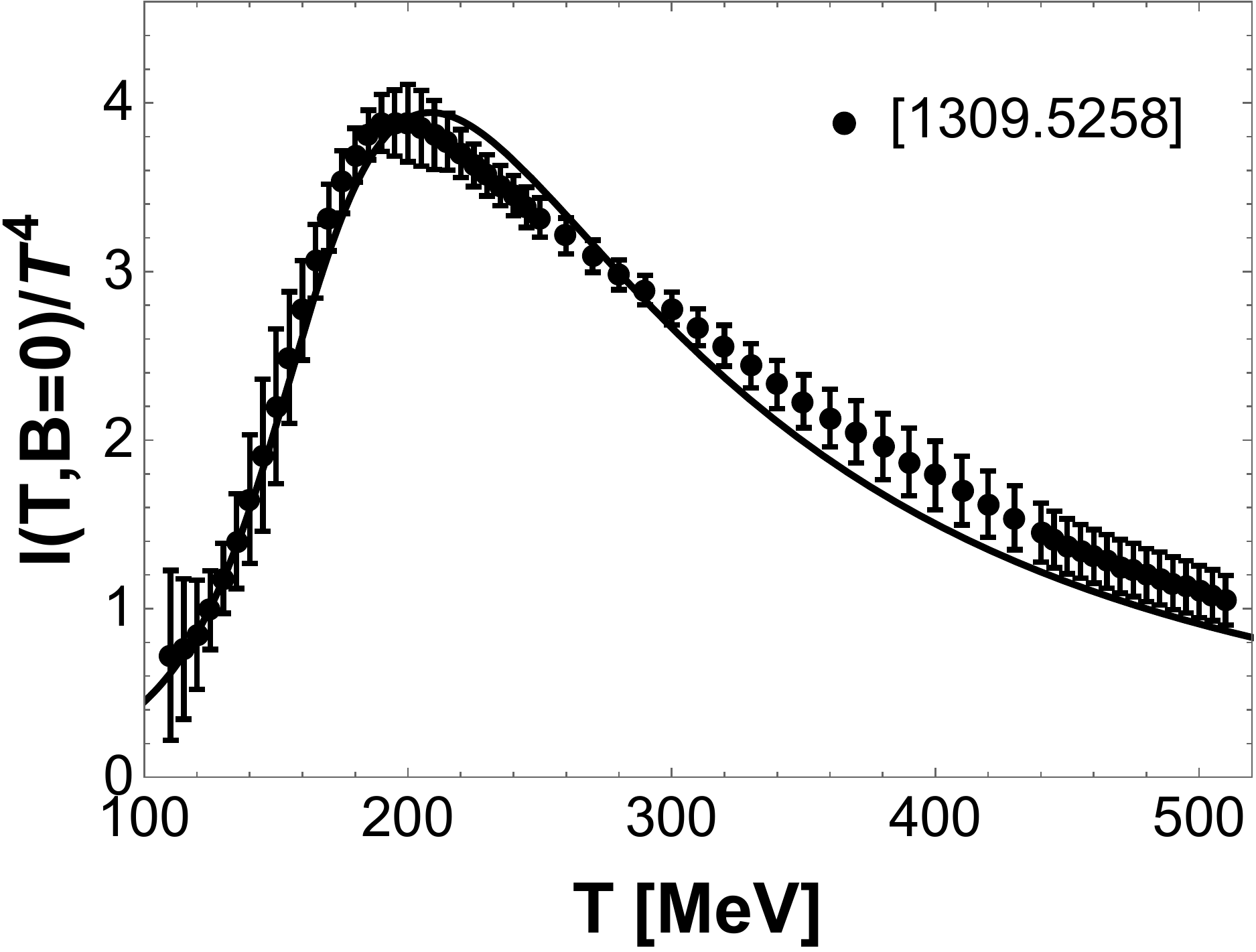} 
\end{tabular}
\begin{tabular}{c}
\includegraphics[width=0.45\textwidth]{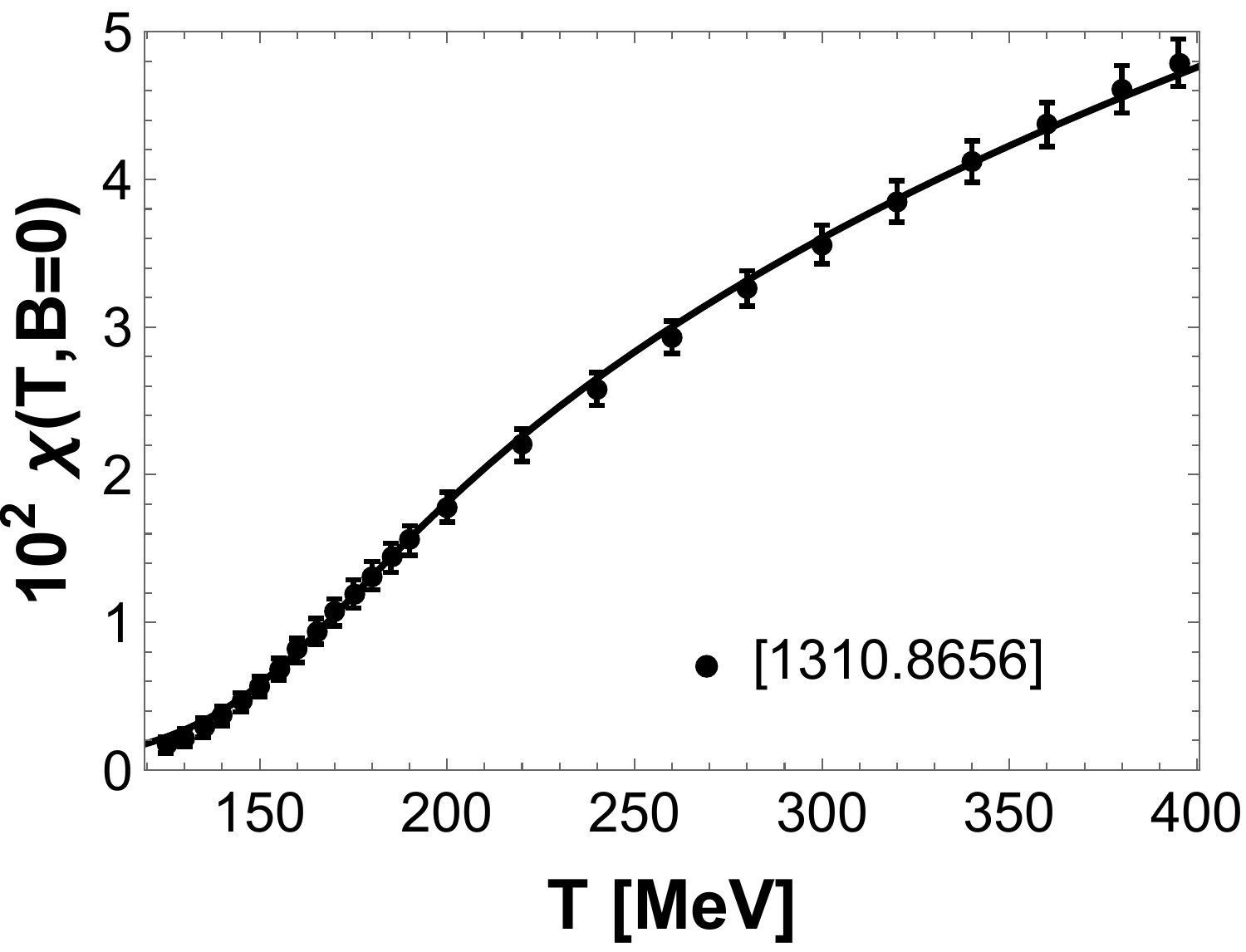} 
\end{tabular}
\caption{Thermodynamics of the magnetic EMD model at $B=0$ matched to lattice data in order to fix the free parameters of the holographic model. \emph{Top left:} entropy density. \emph{Top right:} speed of sound squared. \emph{Middle left:} pressure. \emph{Middle right:} internal energy density. \emph{Bottom left:} trace anomaly. \emph{Bottom right:} magnetic susceptibility.}
\label{fig:EMDthermoB0}
\end{figure}

Now we have all the necessary ingredients to dynamically fix the dilaton potential $V(\phi)$, the gravitational constant $\kappa^2=8\pi G_5$, the scaling factor $\Lambda$, and the Maxwell-dilaton gauge coupling function $f(\phi)$, which are the free parameters of the bottom-up EMD model. We do so by matching the holographic equation of state (represented by the temperature dependence of the entropy and internal energy densities, the speed of sound squared, the pressure, and the trace anomaly) and the holographic magnetic susceptibility at $B=0$ with the corresponding lattice results from Refs.\ \cite{Borsanyi:2013bia} and \cite{latticedata2}, respectively. The results are shown in Eq.\ \eqref{eq:fits} and Fig.\ \ref{fig:EMDthermoB0}.
\begin{align}
V(\phi)&=-12\cosh(0.63\phi)+0.65\phi^2-0.05\phi^4+0.003\phi^6,\nonumber\\
\kappa^2&=8\pi G_5=8\pi(0.46), \quad \Lambda=1058.83\,\textrm{MeV},\nonumber\\
f(\phi)&=0.95\,\textrm{sech}(0.22\phi^2-0.15\phi-0.32).
\label{eq:fits}
\end{align}
From the dilaton potential above one can obtain that the scaling dimension of the dual operator in the gauge theory is $\Delta\approx 2.73$ (a relevant operator).

\vspace{8pt}

\begin{figure}[h]
\begin{tabular}{c}
\includegraphics[width=0.45\textwidth]{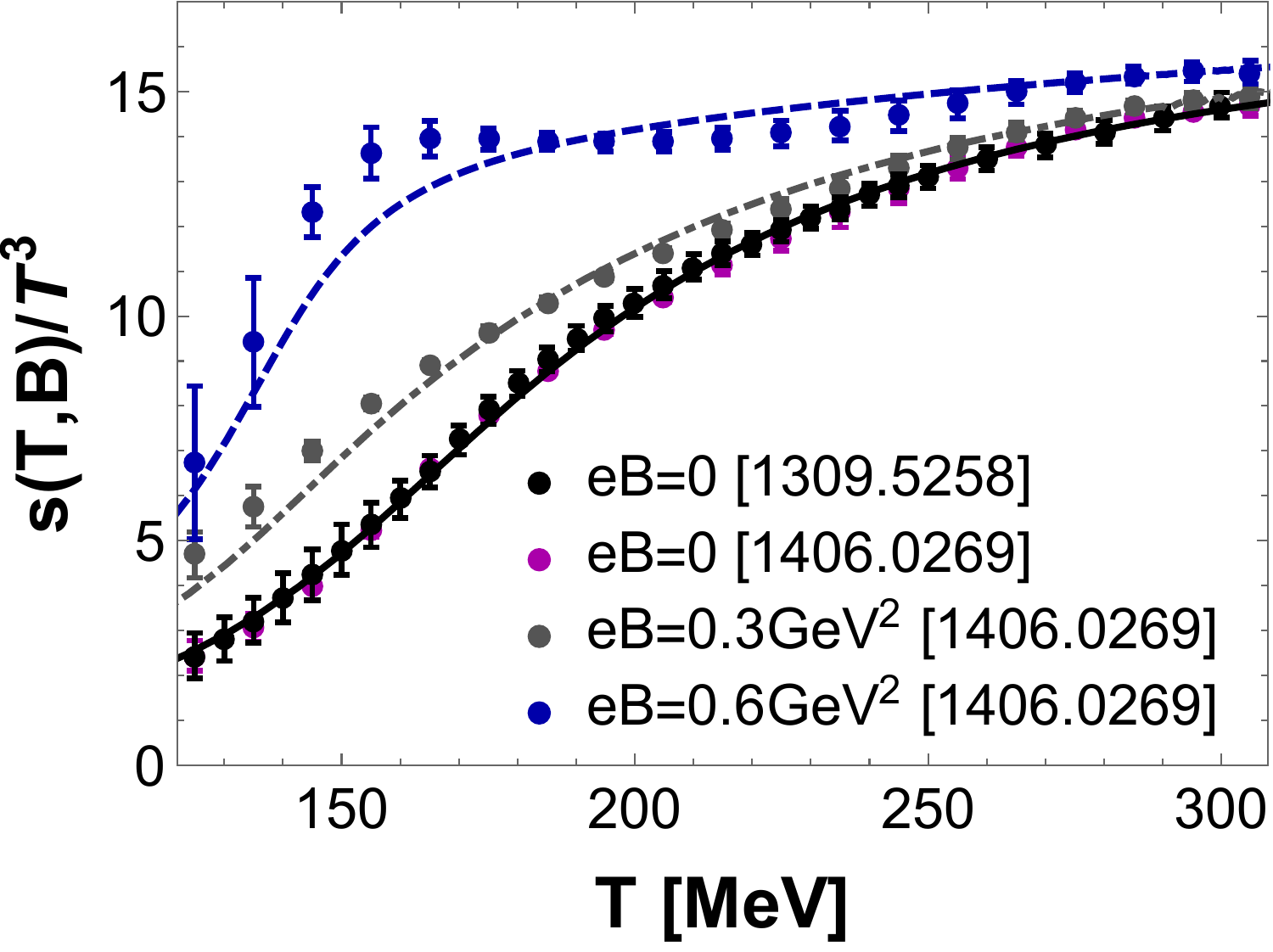} 
\end{tabular}
\begin{tabular}{c}
\includegraphics[width=0.45\textwidth]{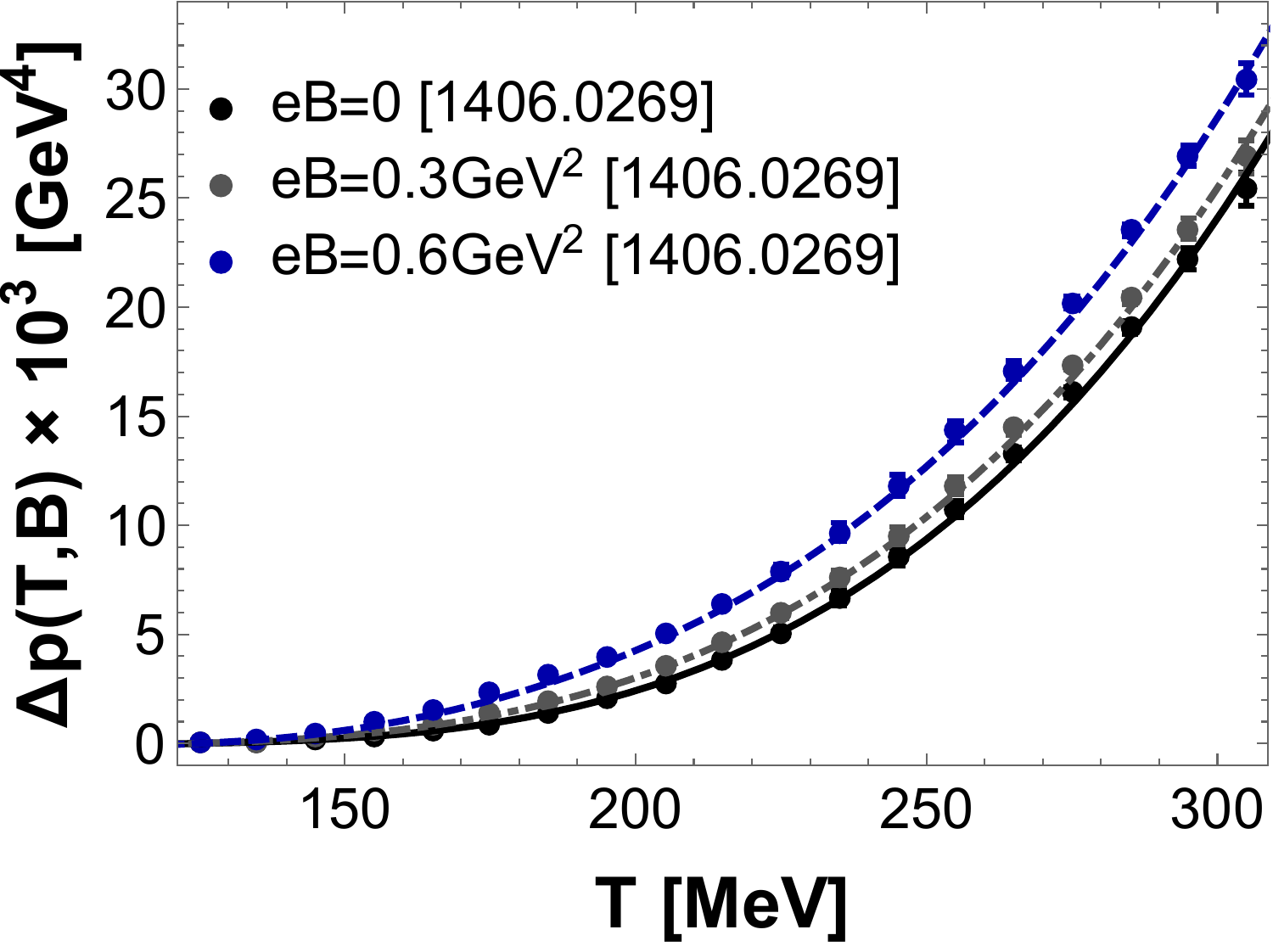} 
\end{tabular}
\caption{(Color online) Thermodynamics of the magnetic EMD model at nonzero $B$ compared to lattice data. \emph{Left:} entropy density as function of temperature for some fixed values of the magnetic field. \emph{Right:} pressure difference (in the direction of the magnetic field), $\Delta p(T,B)\equiv p(T,B) - p(T_{\textrm{ref}}=125\textrm{MeV},B)$, as a function of temperature for some fixed values of the magnetic field.}
\label{fig:EMDthermo}
\end{figure}

\textit{Thermodynamics at nonzero magnetic field.} In Fig.\ \ref{fig:EMDthermo} we show our numerical results for the entropy density, $s(T,B)/T^3$, and the pressure differences, $\Delta p(T,B)\equiv p(T,B) - p(T_{\textrm{ref}}=125\textrm{MeV},B)$, compared to the corresponding lattice data from Ref. \cite{latticedata3}. The pressure was obtained by integrating the entropy density with respect to temperature keeping the magnetic field fixed. Therefore, it corresponds to the anisotropic pressure in the direction of the magnetic field when the magnetic flux is kept fixed under compressions, which is equal to the isotropic pressure obtained by keeping instead the magnetic field fixed under compressions \cite{latticedata3}. We remark that while the results for the thermodynamics at $B=0$ displayed in Fig.\ \ref{fig:EMDthermoB0} are not predictions of the EMD model, since they were dynamically fixed in order to determine the free parameters of the model in Eq.\ \eqref{eq:fits}, the thermodynamic results at nonzero $B$ shown in Fig.\ \ref{fig:EMDthermo} constitute genuine predictions of the holographic model. A fairly good quantitative agreement is obtained with the current lattice data up to $eB=0.6$ GeV$^2$.

\begin{figure}[h]
\begin{tabular}{c}
\includegraphics[width=0.48\textwidth]{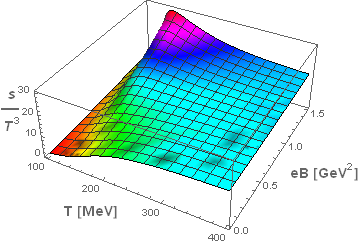} 
\end{tabular}
\begin{tabular}{c}
\includegraphics[width=0.45\textwidth]{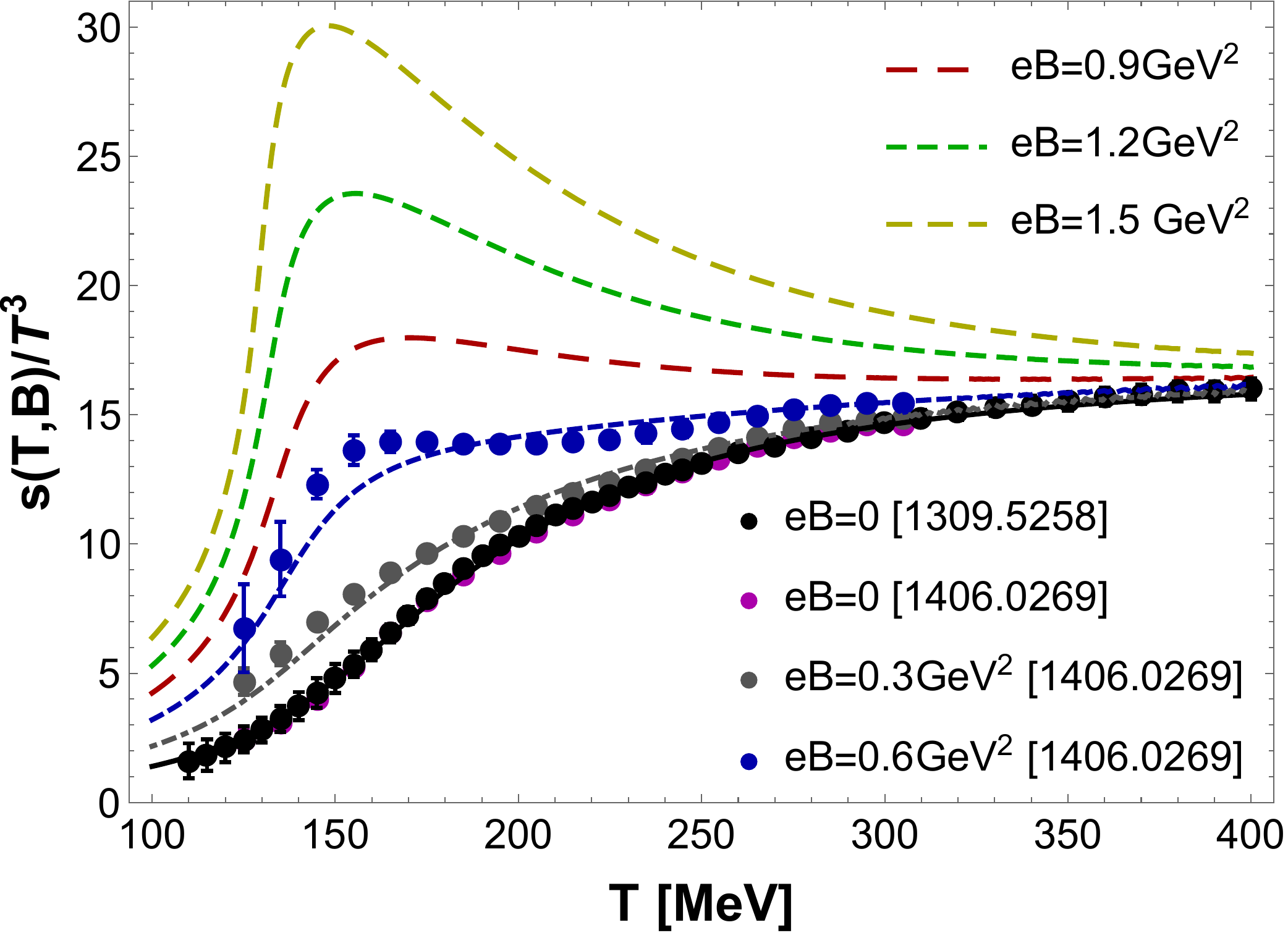} 
\end{tabular}
\begin{tabular}{c}
\includegraphics[width=0.45\textwidth]{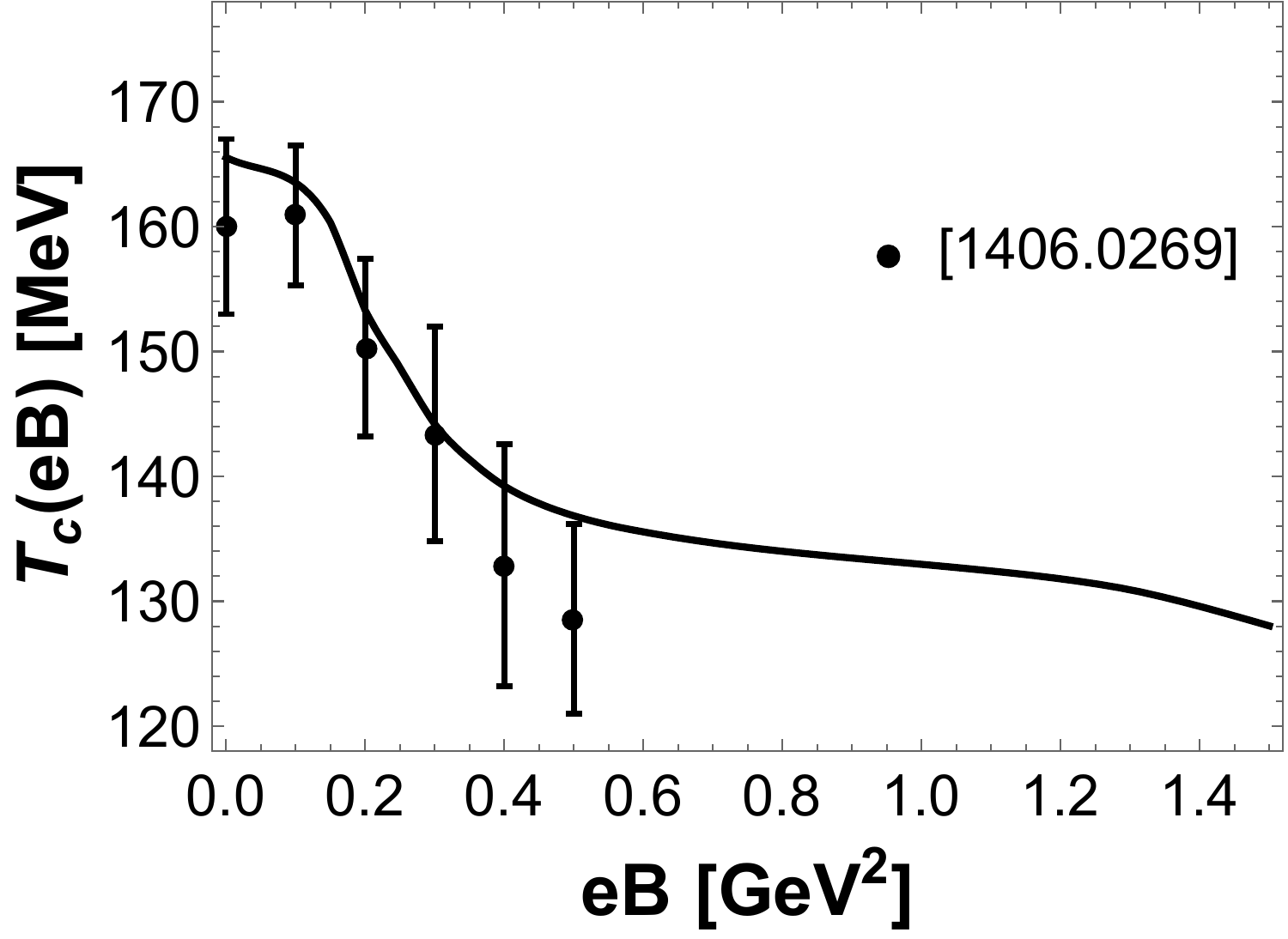} 
\end{tabular}
\caption{(Color online) New holographic predictions from the magnetic EMD model for the entropy density and the crossover temperature in an extended region of the $(T,B)$ phase diagram not yet covered on the lattice. \emph{Top:} entropy density. \emph{Bottom:} crossover temperature extracted from the inflection point of $s/T^3$ compared to the corresponding lattice data.}
\label{fig:extendedEMDthermo}
\end{figure}

In Fig.\ \ref{fig:extendedEMDthermo} we present new holographic predictions for the entropy density and the crossover temperature extracted from its inflection point in a much wider region of the $(T,B)$ phase diagram, which has not yet been covered by lattice simulations. One observes that the crossover temperature decreases with increasing magnetic field in quantitative agreement with the lattice QCD results for the values of the magnetic field already simulated on the lattice.

We plan to further explore, in an upcoming work, the phase diagram of the model in the $(T,B)$ plane by accessing much larger values of magnetic fields in order to look for possible signs of real phase transitions (instead of just the smooth crossover observed here) in the magnetic medium.

\subsection{Drag force}
\label{sec4.2}

The anisotropic drag forces described by Eqs.\ \eqref{eq:dragpal} to \eqref{eq:rperp} may be computed in the magnetic EMD backgrounds by considering
\begin{align}
g_{tt}^{(s)}&= -\tilde{h}(\tilde{r})e^{2\tilde{a}(\tilde{r})+\sqrt{2/3}\,\tilde{\phi}(\tilde{r})}= -\frac{h(r)}{h_0^{\textrm{far}}} \frac{e^{2a(r)+\sqrt{2/3}\,\phi(r)}}{\phi_A^{2/\nu}},\nonumber\\
g_{rr}^{(s)}&= \frac{e^{\sqrt{2/3}\,\tilde{\phi}(\tilde{r})}}{\tilde{h}(\tilde{r})} = \frac{h_0^{\textrm{far}}e^{\sqrt{2/3}\,\phi(r)}}{h(r)},\nonumber\\
g_{xx}^{(s)}&=g_{yy}^{(s)}= e^{2\tilde{c}(\tilde{r})+\sqrt{2/3}\,\tilde{\phi}(\tilde{r})}= \frac{e^{2(c(r)-c_0^{\textrm{far}}+a_0^{\textrm{far}})+\sqrt{2/3}\,\phi(r)}}{\phi_A^{2/\nu}},\nonumber\\
g_{zz}^{(s)}&= e^{2\tilde{a}(\tilde{r})+\sqrt{2/3}\,\tilde{\phi}(\tilde{r})}= \frac{e^{2a(r)+\sqrt{2/3}\,\phi(r)}}{\phi_A^{2/\nu}},
\label{eq:EMDtransf}
\end{align}
where one must be also careful in correctly applying the chain rule for radial derivatives when passing from the standard to the numerical coordinates, $'\equiv\partial_{\tilde{r}}=\sqrt{h_0^{\textrm{far}}}\partial_r$.

\begin{figure}[h]
\begin{center}
\begin{tabular}{c}
\includegraphics[width=0.45\textwidth]{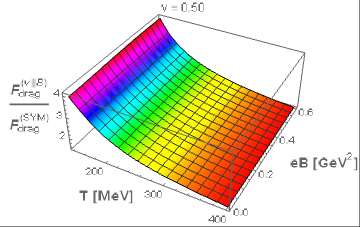} 
\end{tabular}
\begin{tabular}{c}
\includegraphics[width=0.45\textwidth]{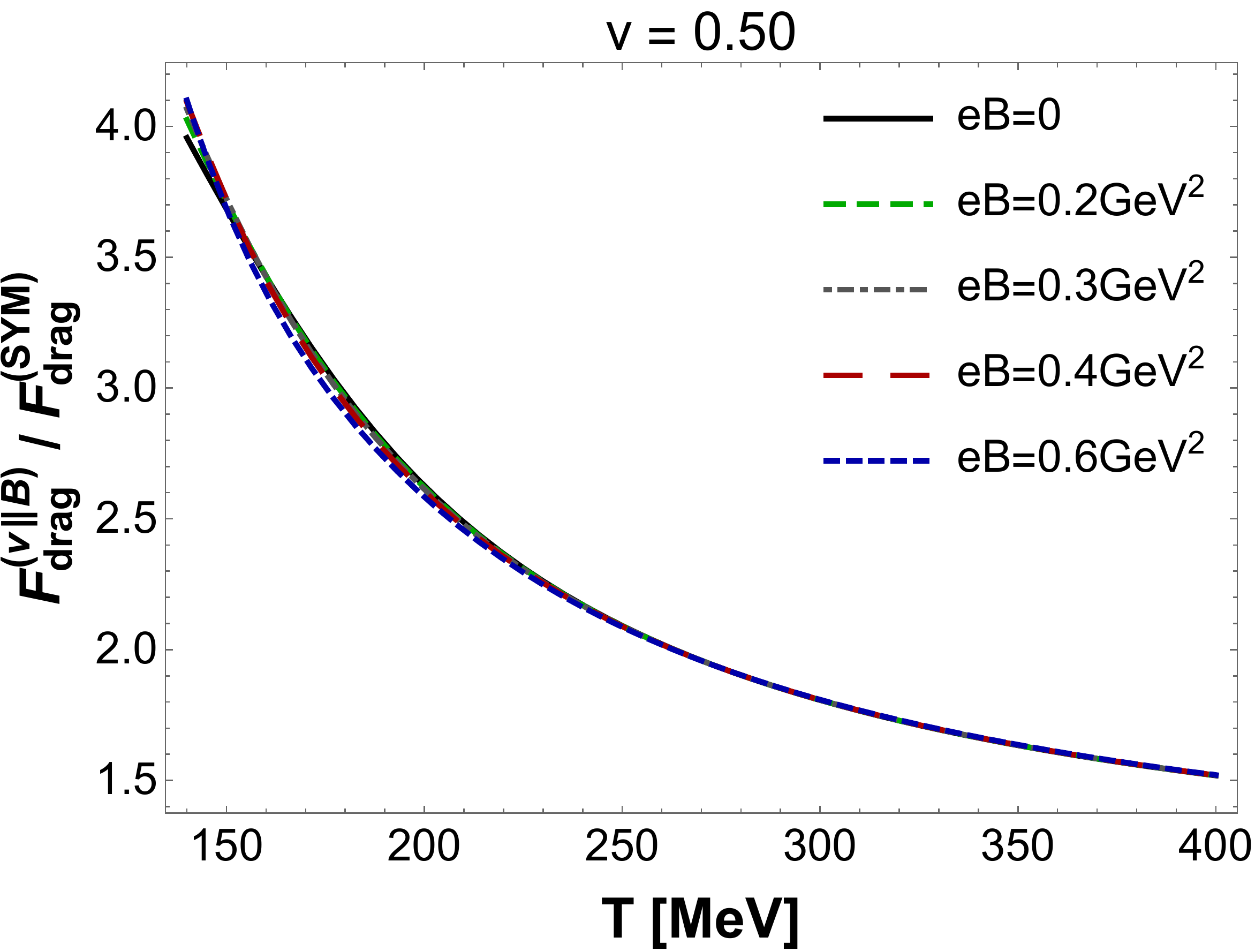} 
\end{tabular}
\end{center}
\begin{center}
\begin{tabular}{c}
\includegraphics[width=0.45\textwidth]{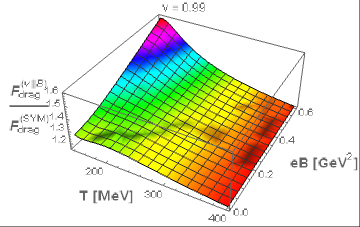} 
\end{tabular}
\begin{tabular}{c}
\includegraphics[width=0.45\textwidth]{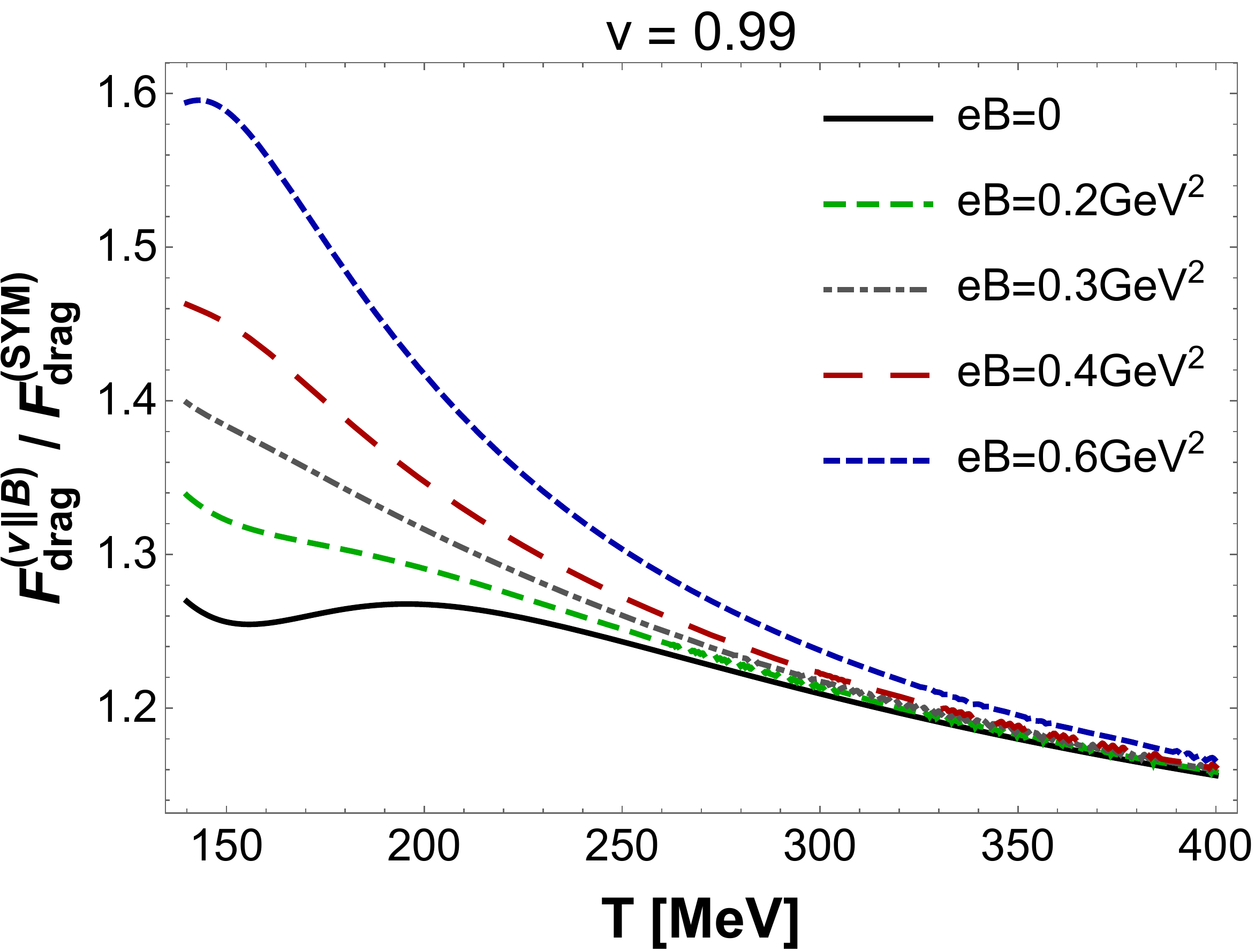} 
\end{tabular}
\end{center}
\caption{(Color online) Anisotropic drag force $F_{\textrm{drag}}^{(v\parallel B)}$ in the magnetic EMD model normalized by the isotropic SYM result at zero magnetic field. \emph{Top:} results for the heavy quark velocity $v=0.50$. \emph{Bottom:} results in the ultrarelativistic limit $v=0.99$.}
\label{fig:EMDdrag1}
\end{figure}

\begin{figure}[h]
\begin{center}
\begin{tabular}{c}
\includegraphics[width=0.45\textwidth]{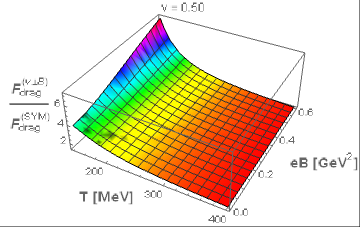} 
\end{tabular}
\begin{tabular}{c}
\includegraphics[width=0.45\textwidth]{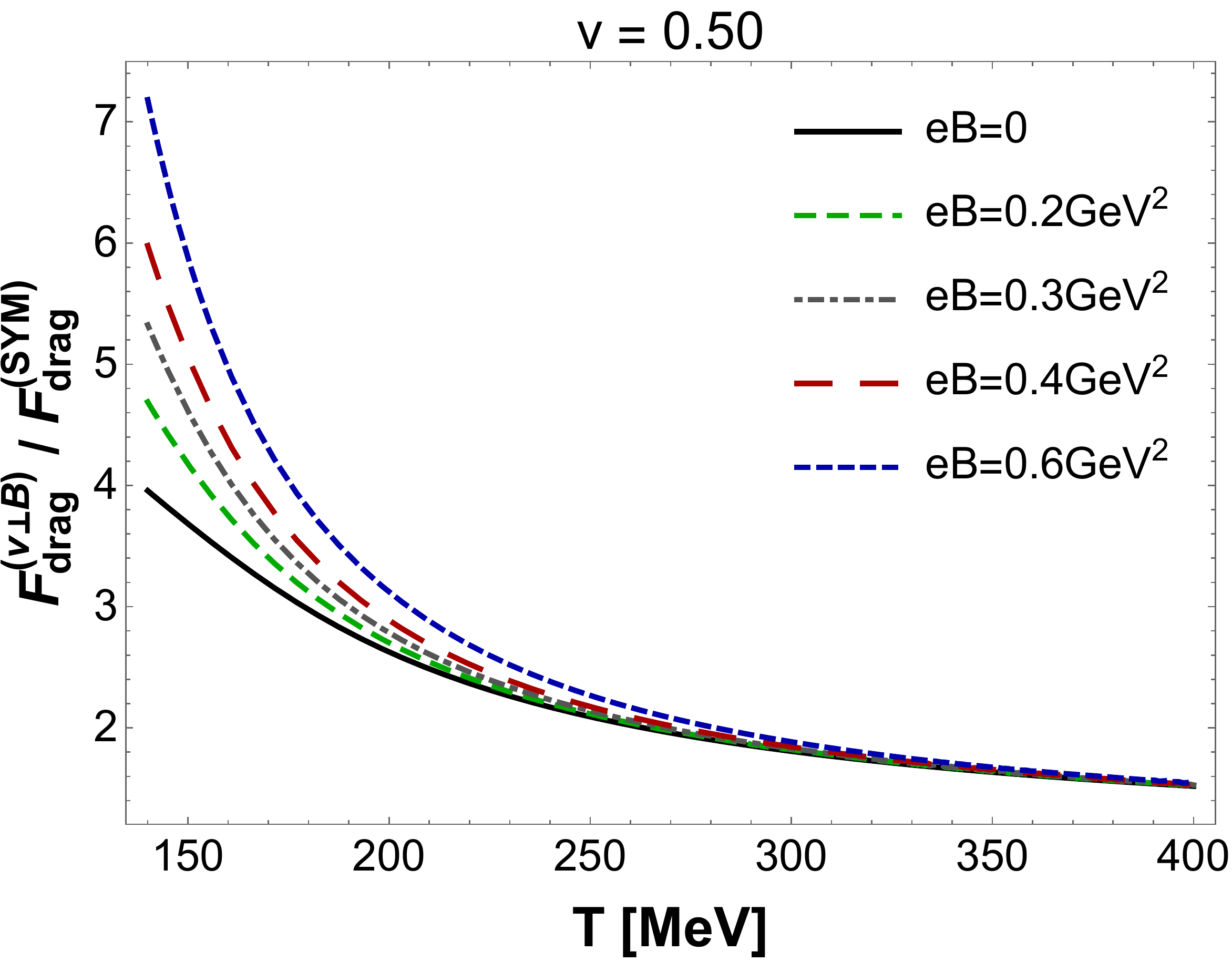} 
\end{tabular}
\end{center}
\begin{center}
\begin{tabular}{c}
\includegraphics[width=0.45\textwidth]{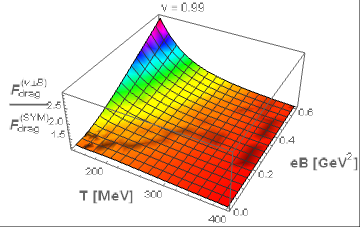} 
\end{tabular}
\begin{tabular}{c}
\includegraphics[width=0.45\textwidth]{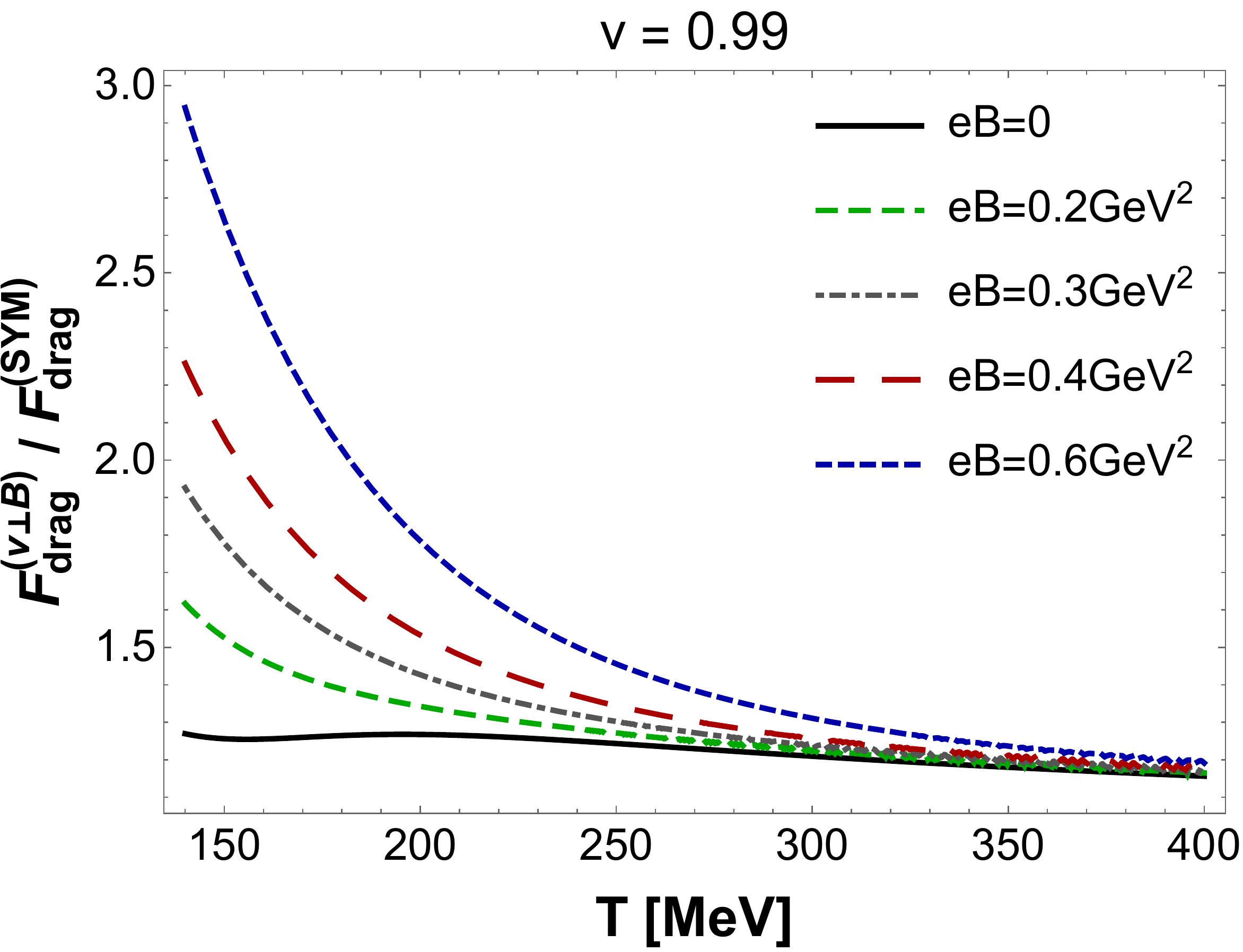} 
\end{tabular}
\end{center}
\caption{(Color online) Anisotropic drag force $F_{\textrm{drag}}^{(v\perp B)}$ in the magnetic EMD model normalized by the isotropic SYM result at zero magnetic field. \emph{Top:} results for the heavy quark velocity $v=0.50$. \emph{Bottom:} results in the ultrarelativistic limit $v=0.99$.}
\label{fig:EMDdrag2}
\end{figure}

Our numerical results for the magnetic field induced anisotropic drag forces, normalized by the isotropic SYM result at zero magnetic field given in Eq.\ \eqref{eq:dragsym}, are displayed in Figs.\ \ref{fig:EMDdrag1} and \ref{fig:EMDdrag2}. Note that in the EMD model we have three independent variables, $T$, $B$, and $v$ that may be varied independently. From Fig.\ \ref{fig:EMDdrag1} we see that $F_{\textrm{drag}}^{(v\parallel B)}$ is not affected by the magnetic field for moderate velocities while it increases with $B$ for ultrarelativistic velocities. On the other hand, its overall magnitude decreases with increasing $v$. This means that faster/lighter (charm) quarks moving in the direction of the magnetic field are more sensitive to the effects of a nonzero $B$ than slower/heavier (bottom) quarks.
Furthermore, we note that $F_{\textrm{drag}}^{(v\parallel B)}$ acquires a non-monotonic behavior as function of $T$ in the crossover region only in the case of ultrarelativistic velocities.

Additionally, from Fig.\ \ref{fig:EMDdrag2} one observes that $F_{\textrm{drag}}^{(v\perp B)}$ is affected by a nonzero magnetic field both at moderate and ultrarelativistic speeds and that it increases with $B$. Also, in the perpendicular channel the drag force is reduced with increasing $v$. In the transverse plane to the magnetic field direction, at lower temperatures, bottom quarks feel more the effects of a nonzero $B$ than charm quarks, with this tendency being inverted at higher temperatures. Moreover, we find that $F_{\textrm{drag}}^{(v\perp B)}$ develops a non-monotonic behavior as function of $T$ in the crossover region just for large $v$ and low $B$.

Figs.\ \ref{fig:EMDdrag1} and \ref{fig:EMDdrag2} also show that in general, $F_{\textrm{drag}}^{(v\perp B)} > F_{\textrm{drag}}^{(v\parallel B)}$, i.e., the energy loss is larger in the transverse plane than along the magnetic field direction as also noticed previously in the magnetic brane setup. The discussion in this section illustrates how rich (and complex) heavy quark energy loss may become once spatial isotropy is broken by a magnetic field.

\subsection{Langevin diffusion coefficients}
\label{sec4.3}

\begin{figure}[h]
\begin{center}
\begin{tabular}{c}
\includegraphics[width=0.45\textwidth]{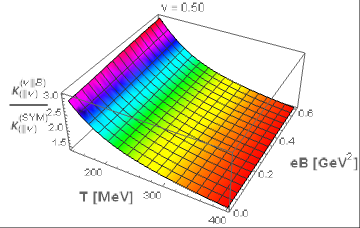} 
\end{tabular}
\begin{tabular}{c}
\includegraphics[width=0.45\textwidth]{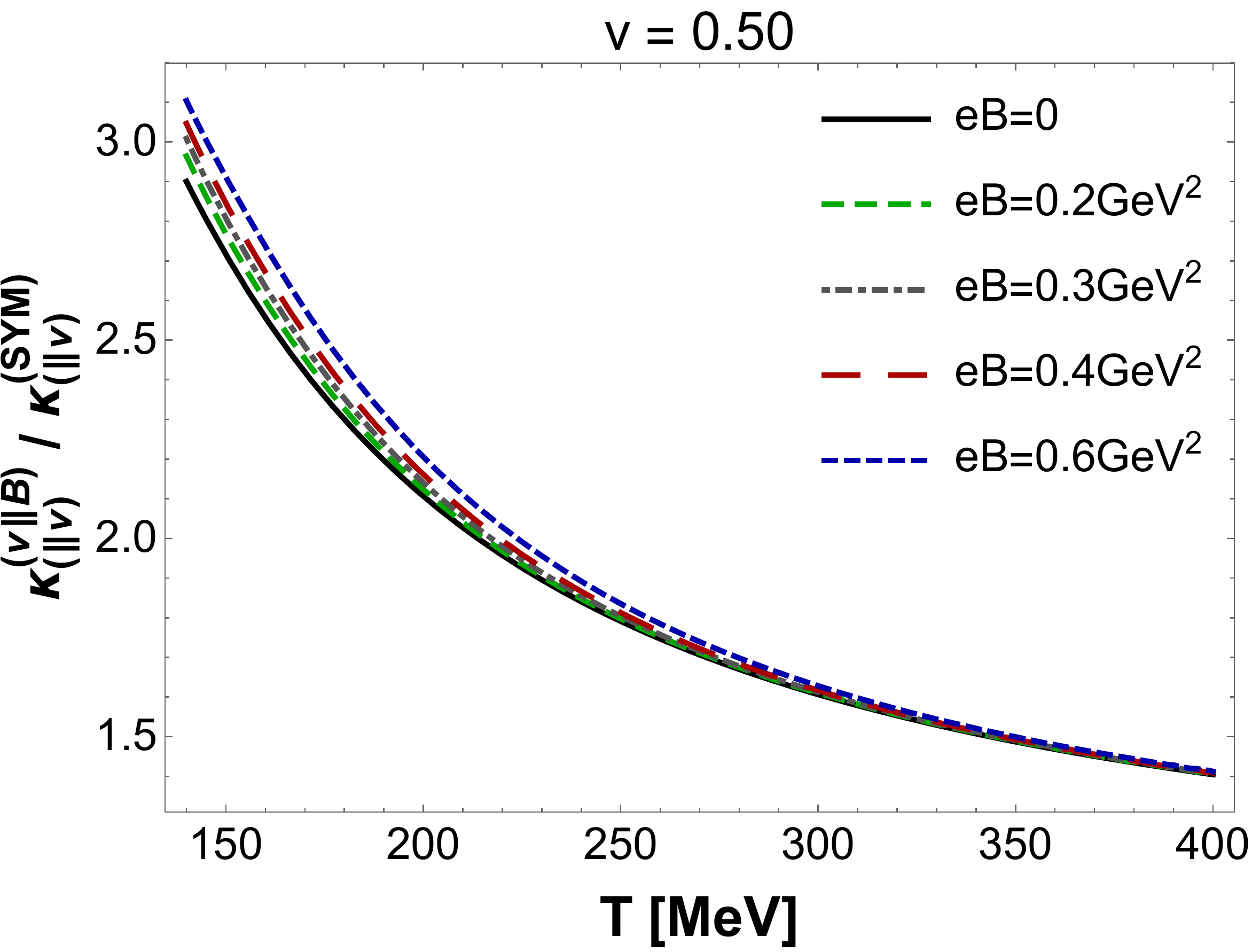} 
\end{tabular}
\end{center}
\begin{center}
\begin{tabular}{c}
\includegraphics[width=0.45\textwidth]{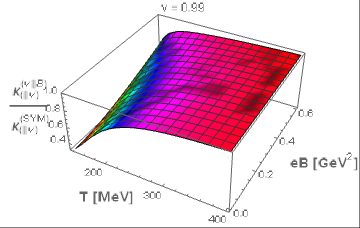} 
\end{tabular}
\begin{tabular}{c}
\includegraphics[width=0.45\textwidth]{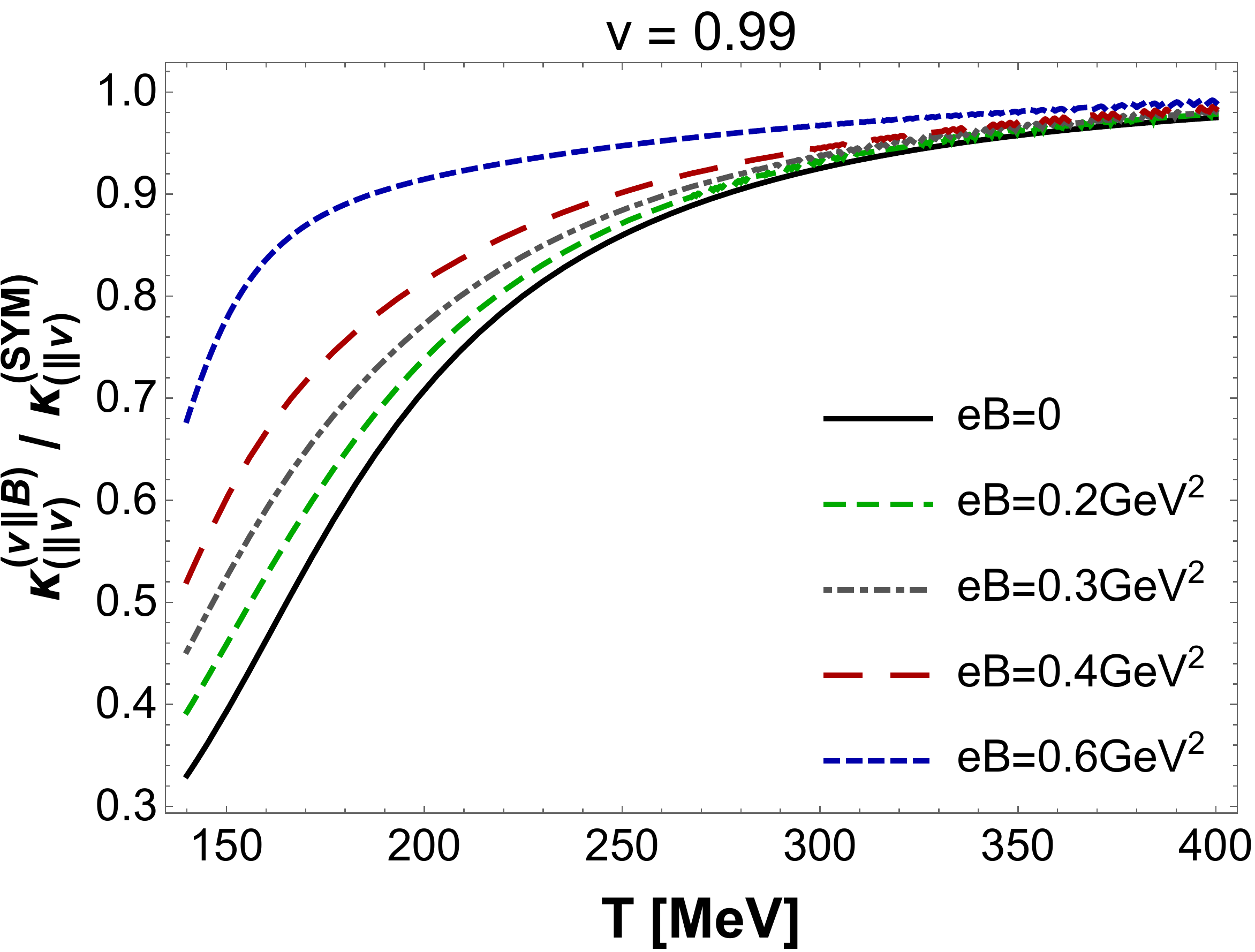} 
\end{tabular}
\end{center}
\caption{(Color online) Anisotropic Langevin diffusion coefficient $\kappa_{(\parallel v)}^{(v\parallel B)}$ in the magnetic EMD model normalized by the SYM result at zero magnetic field. \emph{Top:} results for the heavy quark velocity $v=0.50$. \emph{Bottom:} results in the ultrarelativistic limit $v=0.99$.}
\label{fig:EMD_kappaL}
\end{figure}

\begin{figure}[h]
\begin{center}
\begin{tabular}{c}
\includegraphics[width=0.45\textwidth]{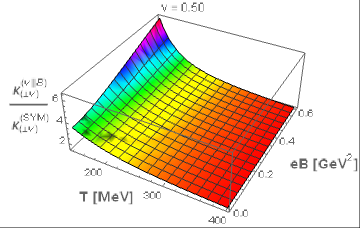} 
\end{tabular}
\begin{tabular}{c}
\includegraphics[width=0.45\textwidth]{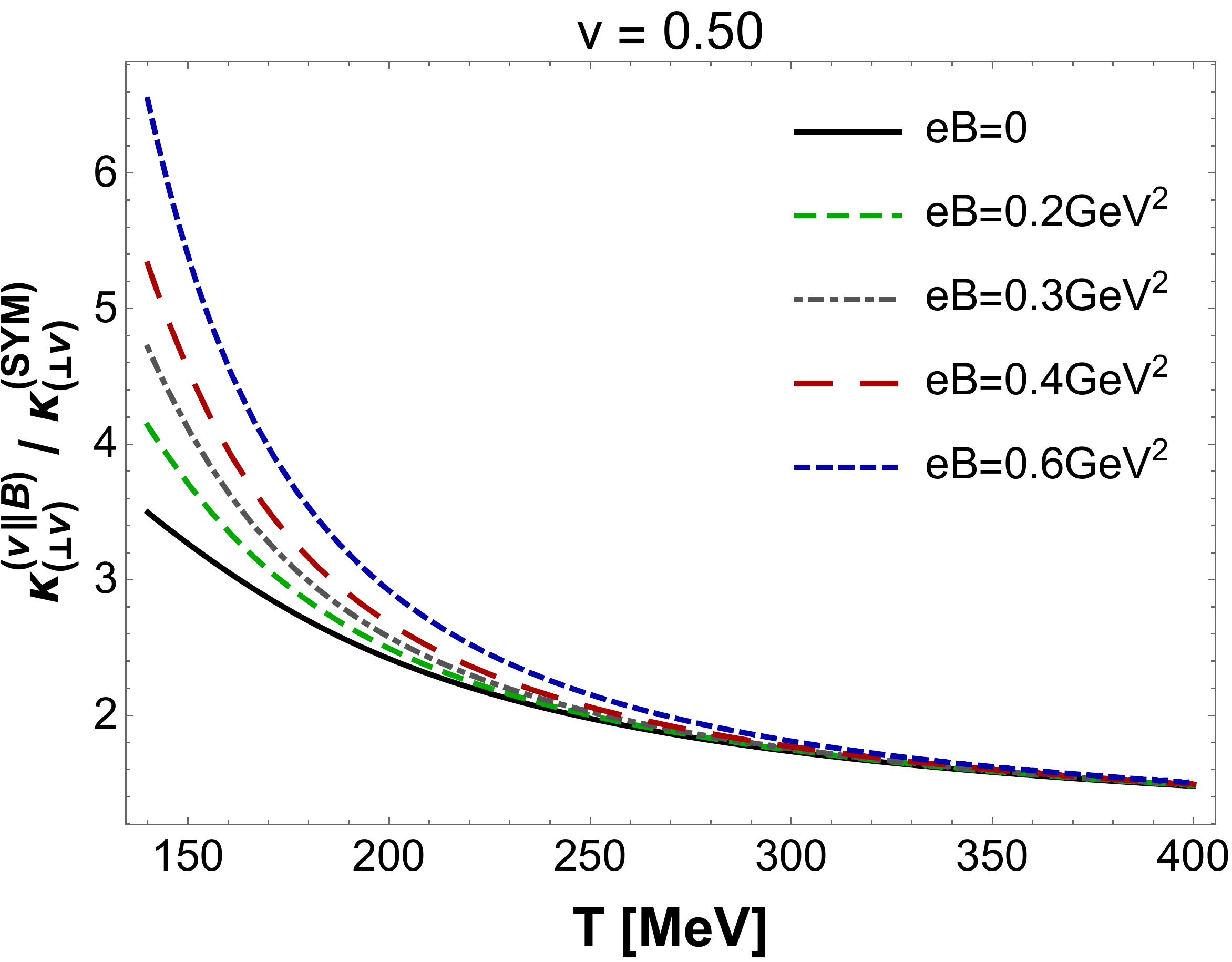} 
\end{tabular}
\end{center}
\begin{center}
\begin{tabular}{c}
\includegraphics[width=0.45\textwidth]{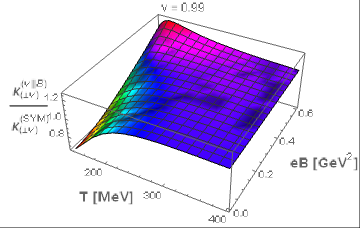} 
\end{tabular}
\begin{tabular}{c}
\includegraphics[width=0.45\textwidth]{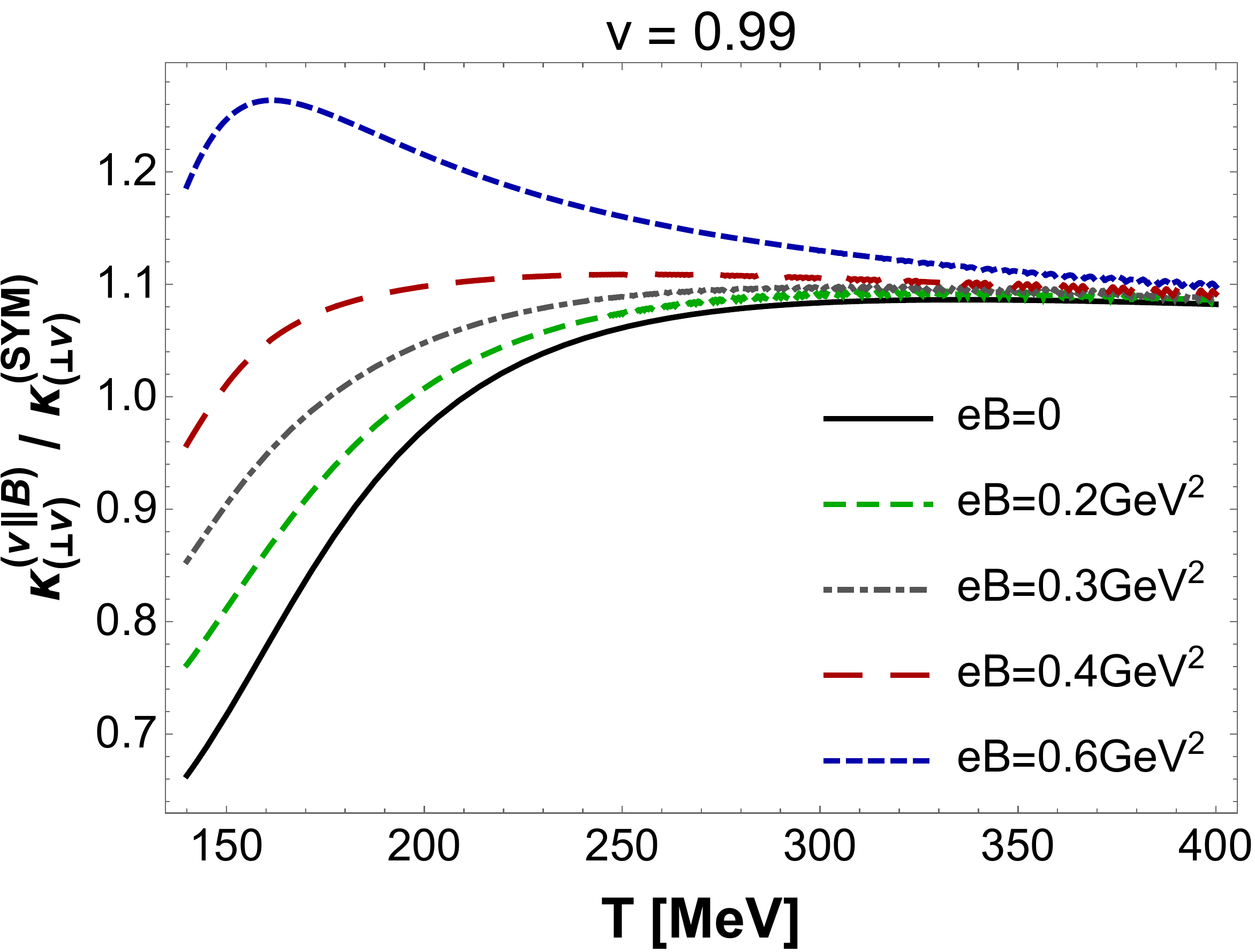} 
\end{tabular}
\end{center}
\caption{(Color online) Anisotropic Langevin diffusion coefficient $\kappa_{(\perp v)}^{(v\parallel B)}$ in the magnetic EMD model normalized by the SYM result at zero magnetic field. \emph{Top:} results for the heavy quark velocity $v=0.50$. \emph{Bottom:} results in the ultrarelativistic limit $v=0.99$.}
\label{fig:EMD_kappaT}
\end{figure}

\begin{figure}[h]
\begin{center}
\begin{tabular}{c}
\includegraphics[width=0.45\textwidth]{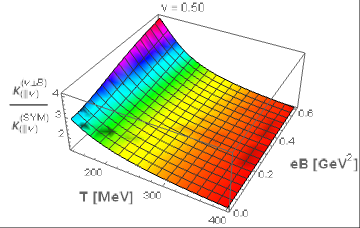} 
\end{tabular}
\begin{tabular}{c}
\includegraphics[width=0.45\textwidth]{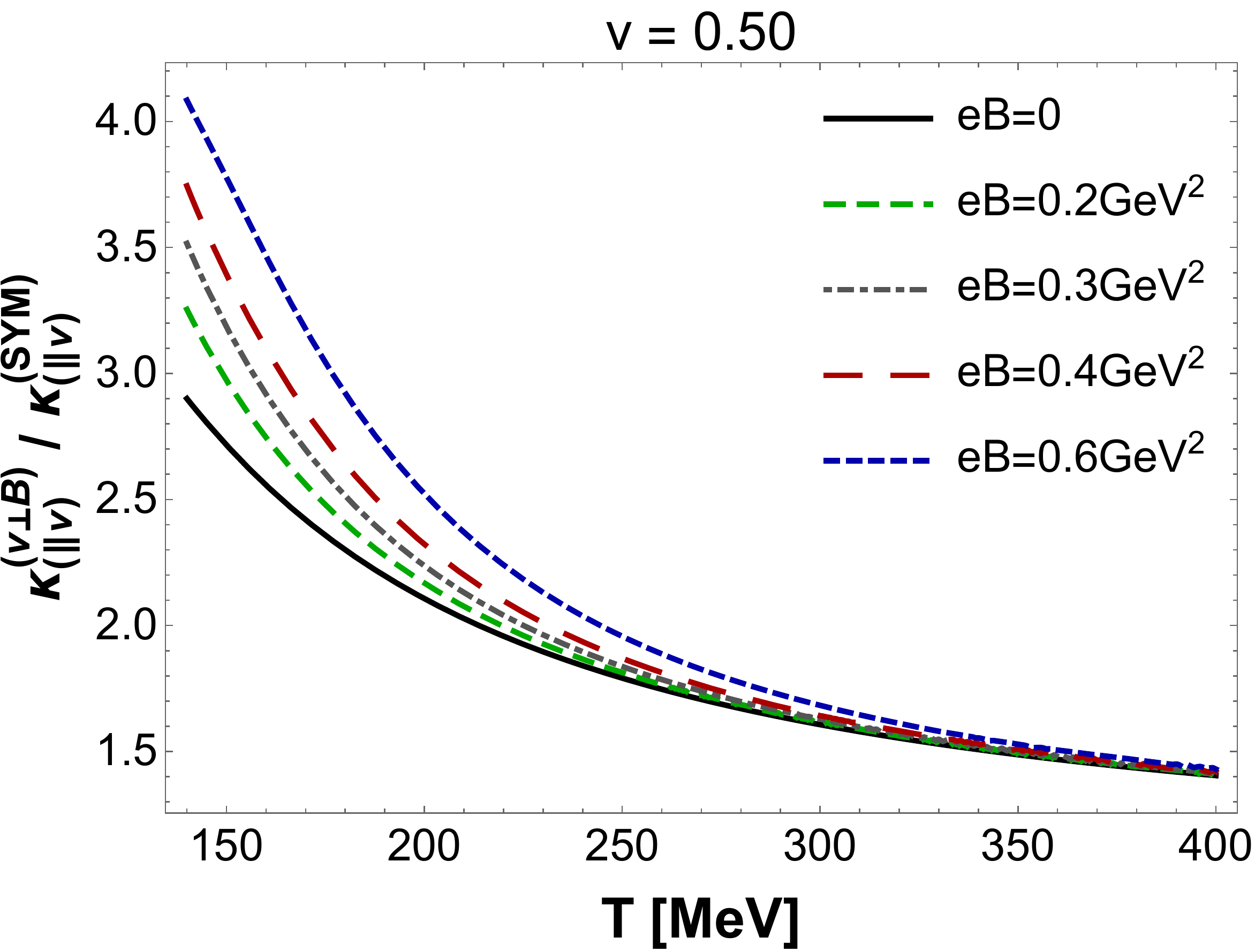} 
\end{tabular}
\end{center}
\begin{center}
\begin{tabular}{c}
\includegraphics[width=0.45\textwidth]{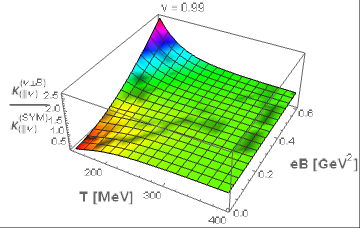} 
\end{tabular}
\begin{tabular}{c}
\includegraphics[width=0.45\textwidth]{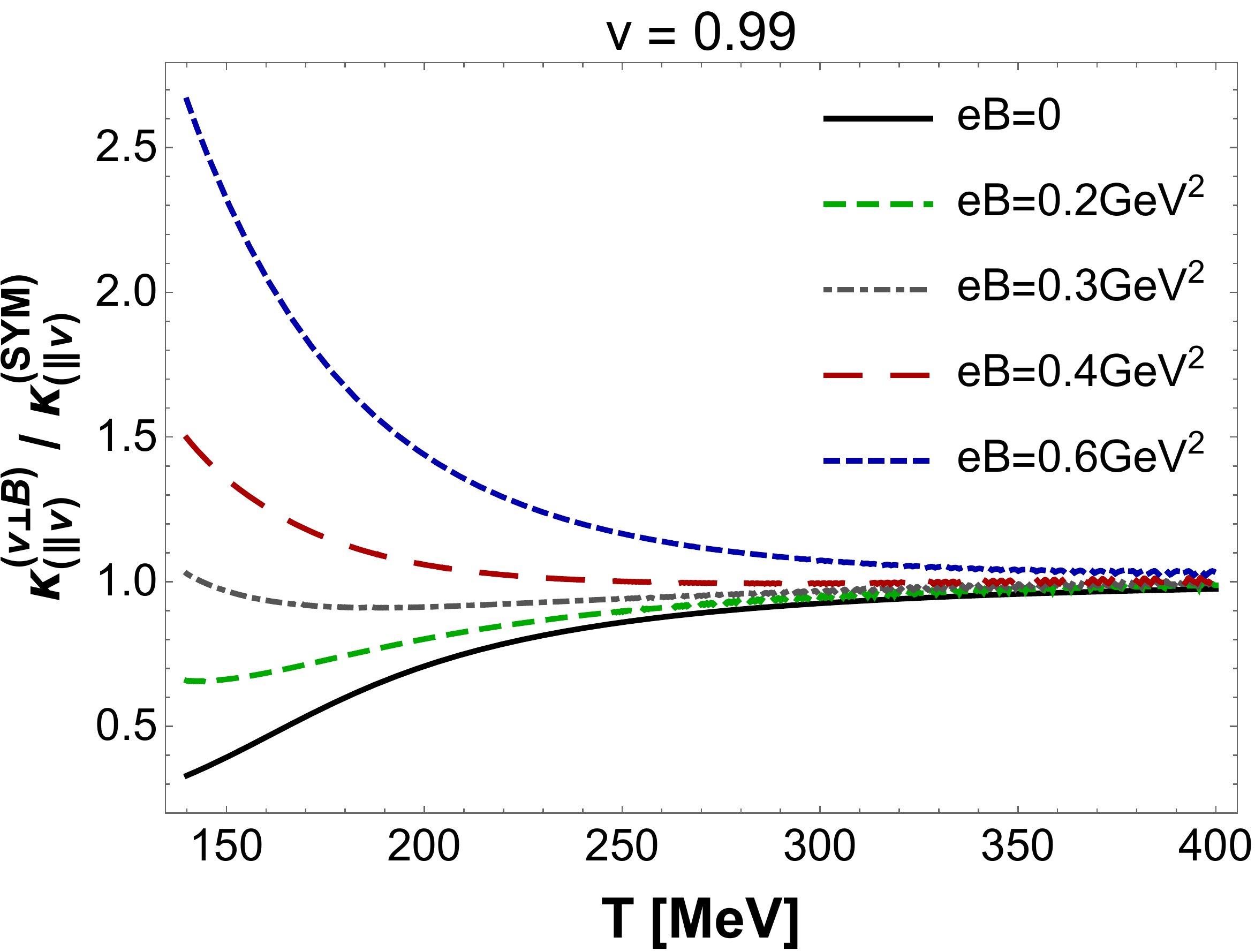} 
\end{tabular}
\end{center}
\caption{(Color online) Anisotropic Langevin diffusion coefficient $\kappa_{(\parallel v)}^{(v\perp B)}$ in the magnetic EMD model normalized by the SYM result at zero magnetic field. \emph{Top:} results for the heavy quark velocity $v=0.50$. \emph{Bottom:} results in the ultrarelativistic limit $v=0.99$.}
\label{fig:EMD_kappax}
\end{figure}

\begin{figure}[h]
\begin{center}
\begin{tabular}{c}
\includegraphics[width=0.45\textwidth]{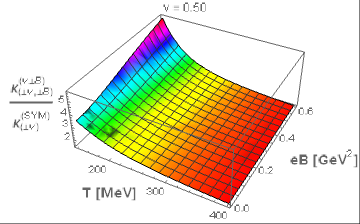} 
\end{tabular}
\begin{tabular}{c}
\includegraphics[width=0.45\textwidth]{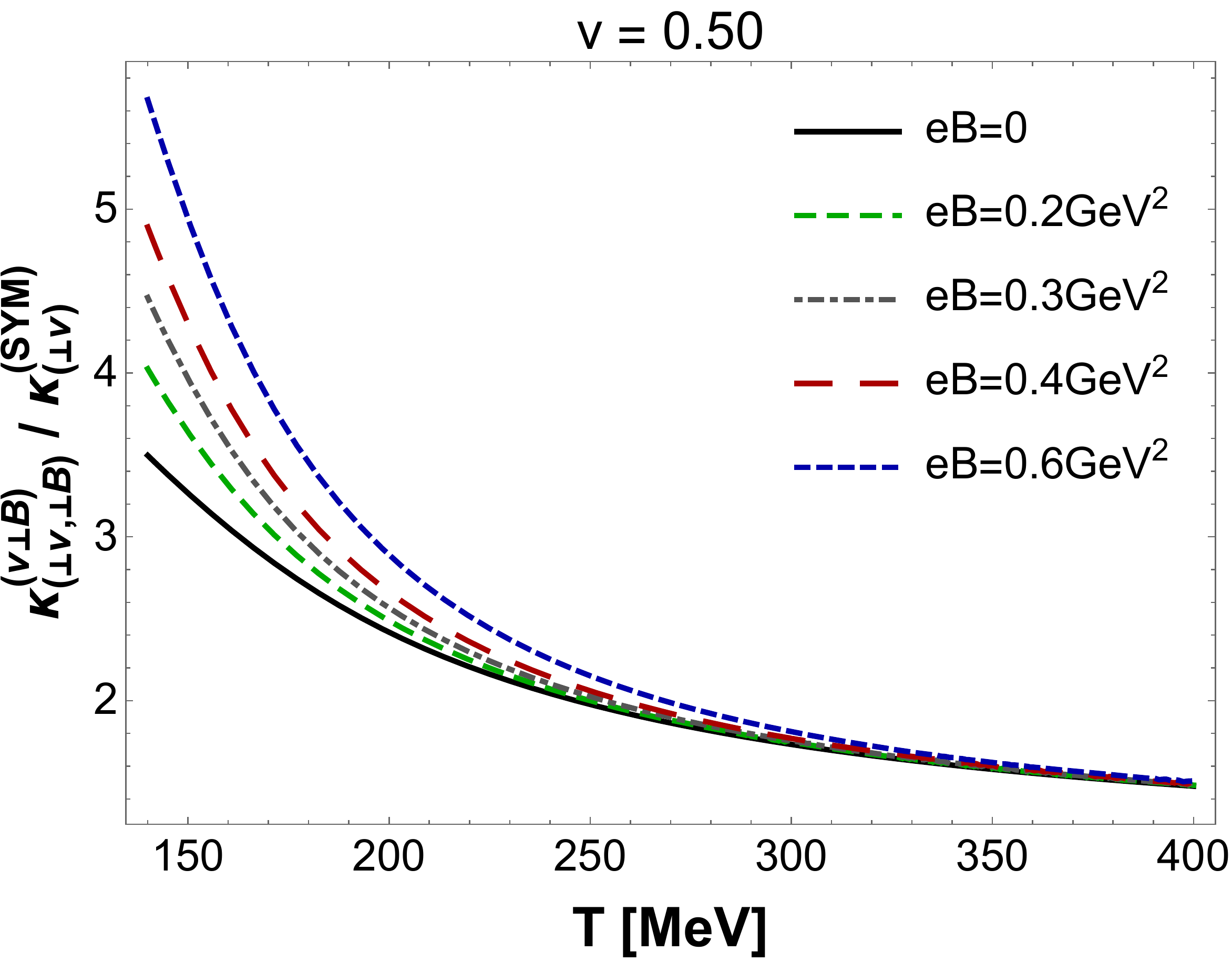} 
\end{tabular}
\end{center}
\begin{center}
\begin{tabular}{c}
\includegraphics[width=0.45\textwidth]{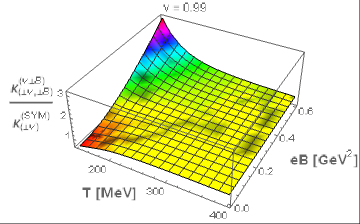} 
\end{tabular}
\begin{tabular}{c}
\includegraphics[width=0.45\textwidth]{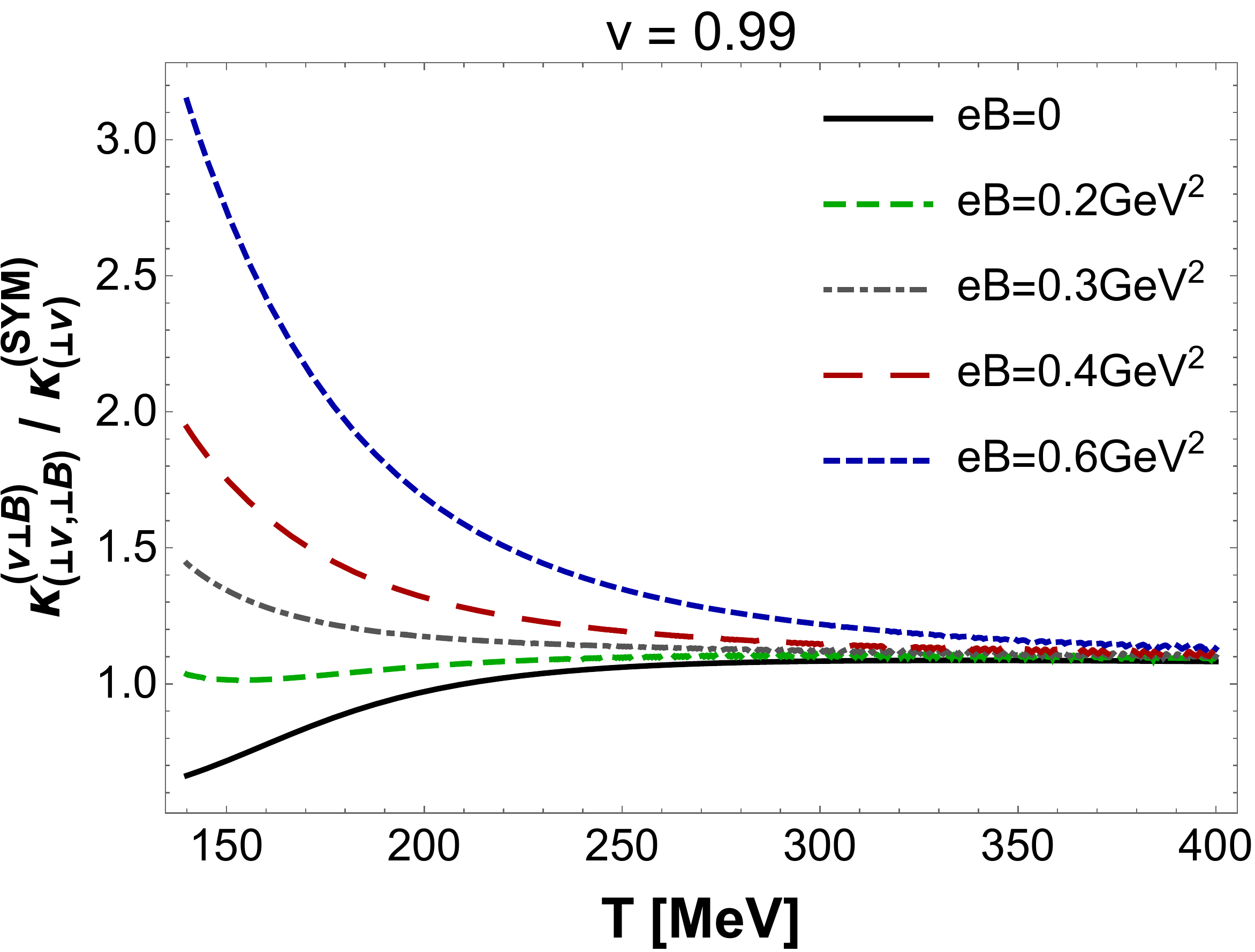} 
\end{tabular}
\end{center}
\caption{(Color online) Anisotropic Langevin diffusion coefficient $\kappa_{(\perp v,\perp B)}^{(v\perp B)}$ in the magnetic EMD model normalized by the SYM result at zero magnetic field. \emph{Top:} results for the heavy quark velocity $v=0.50$. \emph{Bottom:} results in the ultrarelativistic limit $v=0.99$.}
\label{fig:EMD_kappay}
\end{figure}

\begin{figure}[h]
\begin{center}
\begin{tabular}{c}
\includegraphics[width=0.45\textwidth]{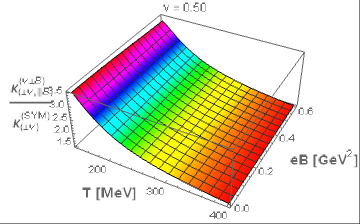} 
\end{tabular}
\begin{tabular}{c}
\includegraphics[width=0.45\textwidth]{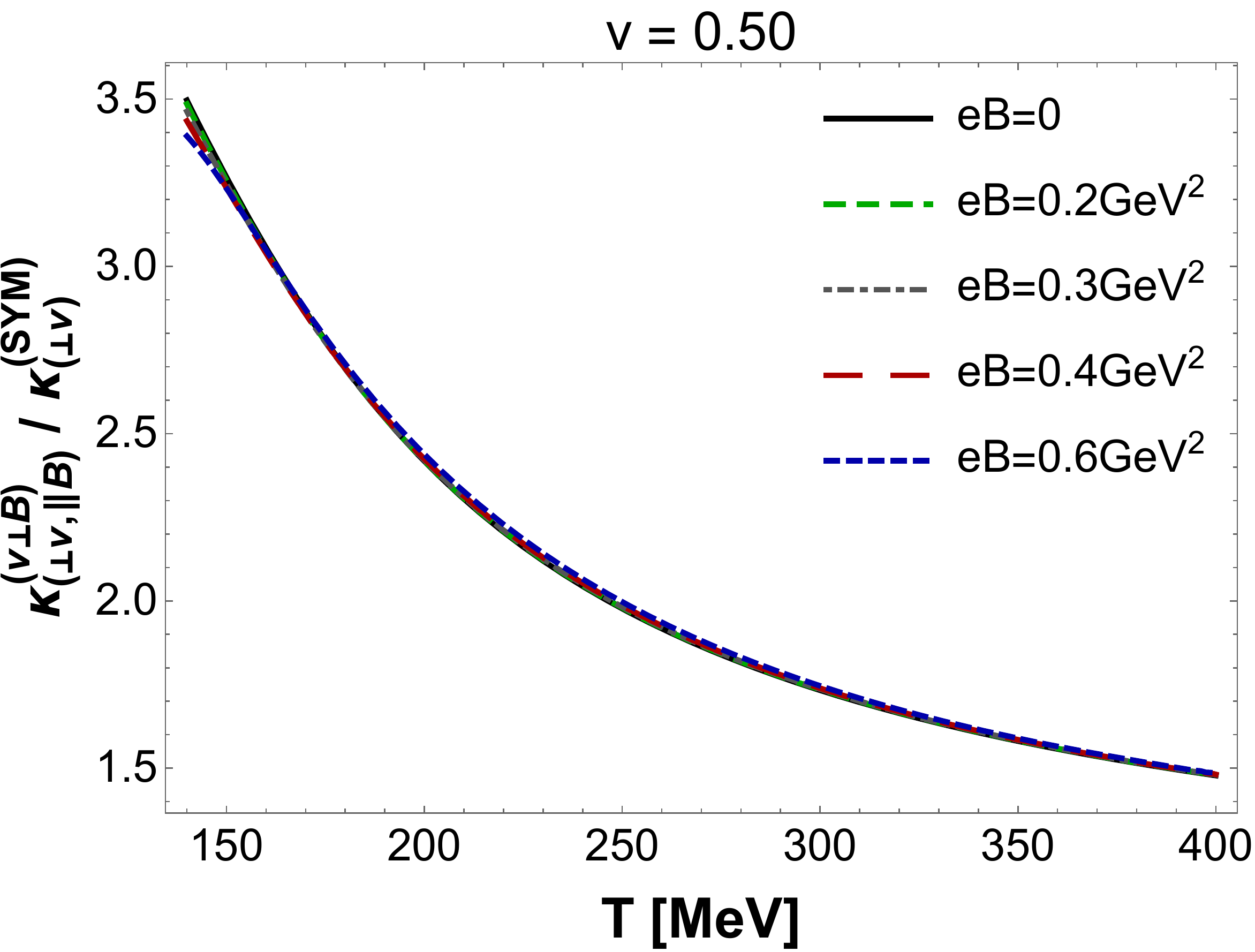} 
\end{tabular}
\end{center}
\begin{center}
\begin{tabular}{c}
\includegraphics[width=0.45\textwidth]{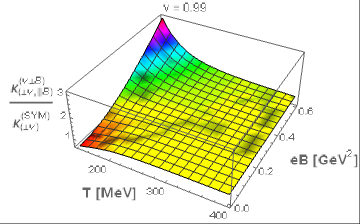} 
\end{tabular}
\begin{tabular}{c}
\includegraphics[width=0.45\textwidth]{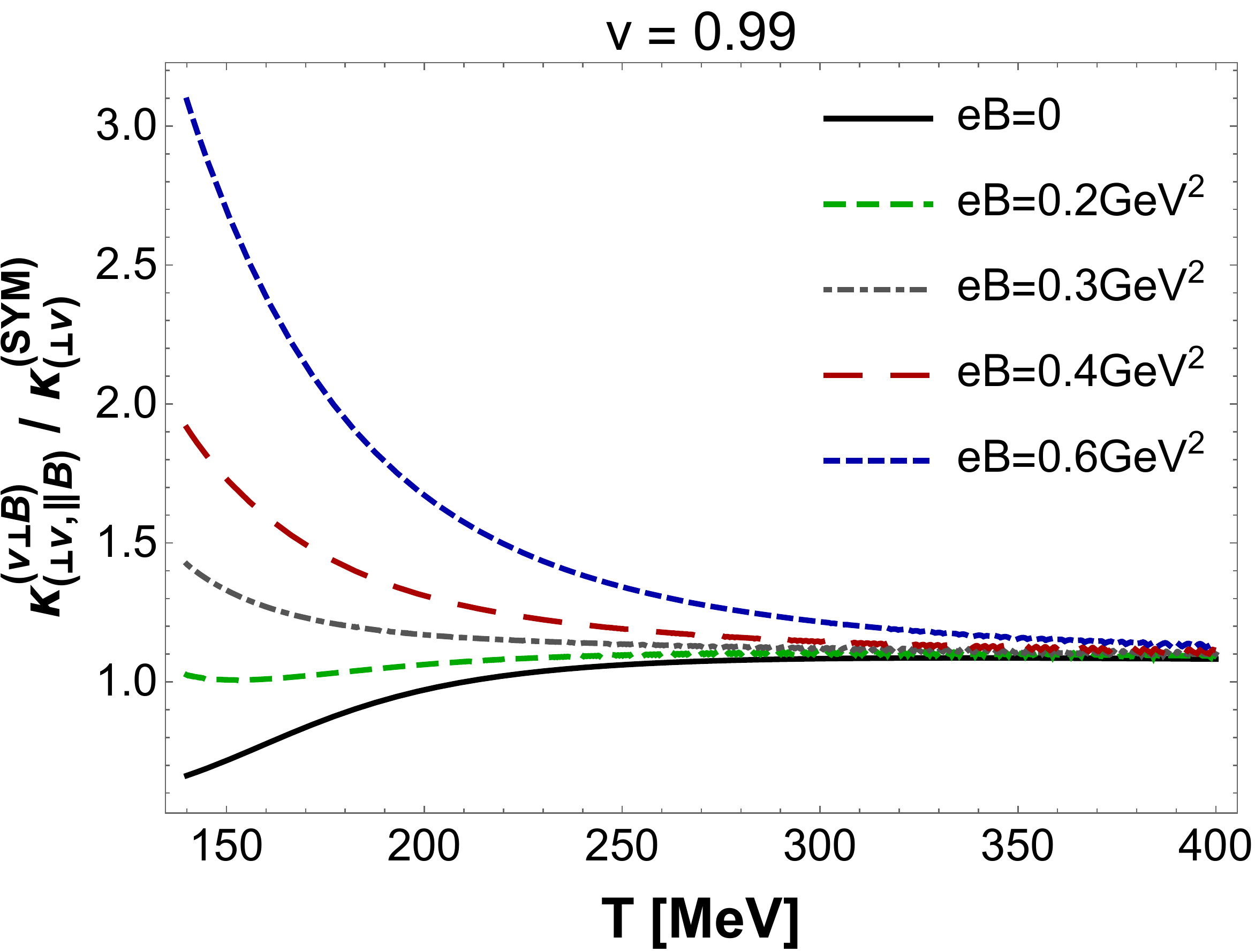} 
\end{tabular}
\end{center}
\caption{(Color online) Anisotropic Langevin diffusion coefficient $\kappa_{(\perp v,\parallel B)}^{(v\perp B)}$ in the magnetic EMD model normalized by the SYM result at zero magnetic field. \emph{Top:} results for the heavy quark velocity $v=0.50$. \emph{Bottom:} results in the ultrarelativistic limit $v=0.99$.}
\label{fig:EMD_kappaz}
\end{figure}

The anisotropic Langevin diffusion coefficients described by Eqs.\ \eqref{eq:dif1} to \eqref{eq:dif2} and Eqs. \eqref{eq:dif3} to \eqref{eq:dif5} may be computed in the magnetic EMD background by considering the relations in Eq.\ \eqref{eq:EMDtransf}. Our numerical results for the magnetic field induced anisotropic Langevin coefficients, normalized by the SYM results at zero magnetic field given by the relations in Eq.\ \eqref{eq:LangSYM}, are displayed in Figs.\ \ref{fig:EMD_kappaL} to \ref{fig:EMD_kappaz}.

Our results show that momentum diffusion is always reduced with increasing velocity\footnote{For fixed energy, bottom quarks are slower than charm quarks and, thus, the latter should experience less momentum diffusion than the former.}. In the $\vec{v}\parallel\vec{B}$ channel, $\kappa_{(\parallel v)}^{(v\parallel B)}$ always increases with $B$ while it decreases with $T$ at lower speeds (bottom quarks), whilst increasing with $T$ for larger velocities (charm quarks). The same observations hold also for $\kappa_{(\perp v)}^{(v\parallel B)}$, but in general, $\kappa_{(\perp v)}^{(v\parallel B)} > \kappa_{(\parallel v)}^{(v\parallel B)}$. On the other hand, in the $\vec{v}\perp\vec{B}$ channel, $\kappa_{(\parallel v)}^{(v\perp B)}$ and $\kappa_{(\perp v,\perp B)}^{(v\perp B)}$ always increase with $B$ while decreasing with $T$ at lower velocities, whilst increasing [decreasing] with $T$ at higher velocities for lower [higher] values of the magnetic field.

Regarding $\kappa_{(\perp v,\parallel B)}^{(v\perp B)}$, we see that it is not affected by the magnetic field at lower velocities (bottom quarks) while it increases with $B$ for larger velocities (charm quarks). This quantity decreases with $T$ at lower velocities while it increases [decreases] with $T$ for larger velocities and lower [larger] values of the magnetic field. Thus, associating slower moving probes with bottom quarks one finds that in this case $\kappa_{(\perp v,\perp B)}^{(v\perp B)} > \kappa_{(\parallel v)}^{(v\perp B)} > \kappa_{(\perp v,\parallel B)}^{(v\perp B)}$, while for more rapidly moving heavy probes (charm quarks), $\kappa_{(\perp v,\perp B)}^{(v\perp B)} > \kappa_{(\parallel v)}^{(v\perp B)} \sim \kappa_{(\perp v,\parallel B)}^{(v\perp B)}$. Consequently, one concludes that heavy quark momentum diffusion in directions transverse to the magnetic field is generally larger than in the direction of the field, as also noticed before in the magnetic brane setup.

\subsection{Shear viscosity}
\label{sec4.4}

\begin{figure}
\begin{center}
\begin{tabular}{c}
\includegraphics[width=0.5\textwidth]{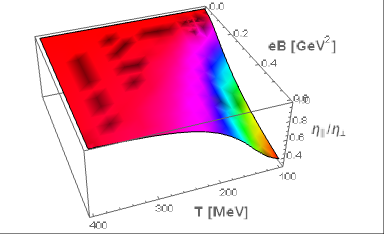} 
\end{tabular}
\begin{tabular}{c}
\includegraphics[width=0.47\textwidth]{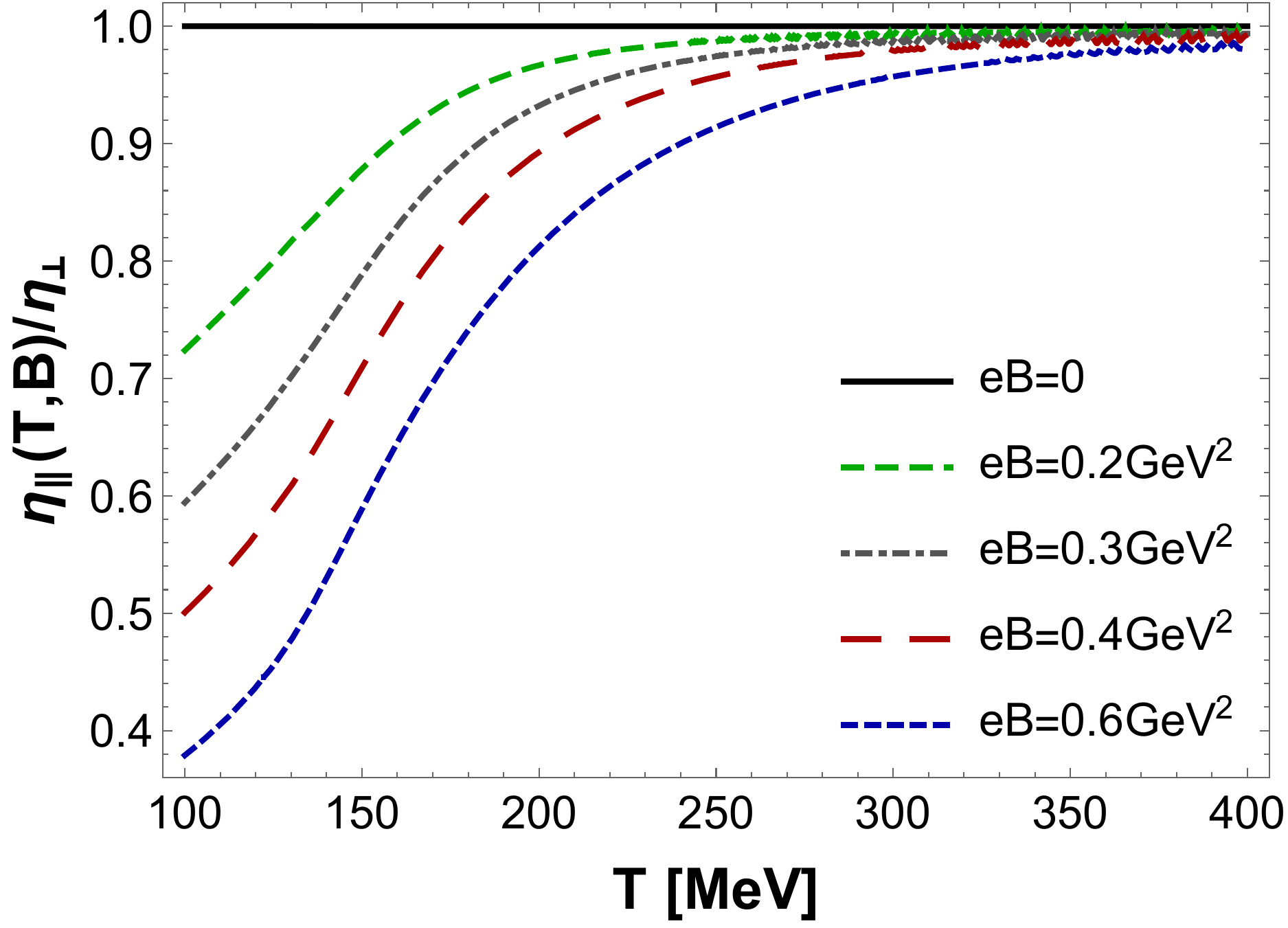} 
\end{tabular}
\end{center}
\caption{(Color online) Ratio between the parallel and the perpendicular shear viscosities in the magnetic EMD model as a function of $T$ and $B$.}
\label{fig:shearnonconfor}
\end{figure}

The anisotropic shear viscosities described by Eqs.\ \eqref{eq:eta_perp_form} and \eqref{eq:eta_par_form} may be computed in the magnetic EMD background by considering the relations in Eq.\ \eqref{eq:EMDtransf}. In Fig.\ \ref{fig:shearnonconfor} we plot our numerical results for the ratio between the parallel and the perpendicular shear viscosities.

As previously found for the magnetic brane model, also in the EMD model one observes that the shear viscosity is reduced in the direction of the magnetic field relative to the viscosity transverse to the field. This shows that the anisotropic plasma becomes a more perfect fluid along the magnetic field direction. Note also the strong $T$ dependence of this ratio near the crossover region, which could not be studied before in the context of the magnetic brane model discussed in the previous sections. Moreover, the behavior of $\eta_{\parallel}$ is similar to the one found within a Boltzmann equation calculation \cite{Tuchin:2011jw} in which the anisotropic viscosities decrease with increasing magnetic field.  

With the values of the anisotropic shear viscosities for the QCD-like model at hand, one may use them in realistic numerical viscous magnetohydrodynamics calculations to investigate the effects of a magnetic field on the hydrodynamic evolution of the QGP (assuming that the magnetic field remains strong enough after $\sim 1$ fm/c to affect its evolution). Even though the current state-of-the-art of relativistic hydrodynamics \cite{reviewQGP1} has not yet incorporated viscous magnetohydrodynamic effects in a stable and causal manner\footnote{See, for instance, Refs.\ \cite{Martinez:2010sc,Martinez:2012tu,Florkowski:2013lza,Tinti:2013vba,Strickland:2014pga,Bazow:2015cha,Molnar:2016vvu} for efforts towards deriving a causal an stable dissipative theory of anisotropic fluid dynamics.}, it must be possible to express the corresponding anisotropic viscosities in any hydrodynamic theory via Kubo formulas - and this is exactly what was done here. Furthermore, it may be possible to investigate qualitatively new effects from the (anisotropic) viscosity by extending previous works that exploit simple flow patterns, such as the Bjorken flow studied already in the context of ideal magnetohydrodynamics \cite{Roy:2015kma,Pu:2016ayh,Pu:2016bxy}. Additionally, it would be interesting to check how the magnetic field affects the bulk viscosity of the medium (see Appendix \ref{apb} for a discussion of the corresponding Kubo formulas), which was shown to be relevant in heavy ion collision simulations in \cite{Noronha-Hostler:2013gga,Noronha-Hostler:2014dqa,Gardim:2014tya} and, more recently, in \cite{Ryu:2015vwa}.

\section{Conclusions}
\label{conclusion}

In the present work, we conducted a systematic investigation of momentum transport in strongly coupled anisotropic plasmas in the presence of strong magnetic fields. We studied two different holographic settings, one corresponding to a top-down deformation of SYM theory triggered by an external magnetic field, and the other one corresponding to a phenomenologically motivated bottom-up EMD model, which is able to quantitatively reproduce $(2+1)$ flavor lattice QCD thermodynamics with physical quark masses at both zero and nonzero magnetic fields. The main conclusions of the present work regarding transport properties in strongly coupled magnetized media hold for both models. Namely, energy loss and momentum diffusion are generally enhanced in the presence of a magnetic field being larger in transverse directions than in the direction parallel to the magnetic field. Moreover, the anisotropic shear viscosity was found to be lower in the direction of the magnetic field than in the plane perpendicular to the field, which indicates that strongly coupled magnetized plasmas can become even closer to the idealized perfect fluid limit along the magnetic field direction.

Compared to other anisotropic models available in the literature, such as the Einstein-axion-dilaton model from Refs. \cite{Mateos:2011ix,Mateos:2011tv}, our results for the Einstein-Maxwell and EMD models have some interesting common features and some qualitative differences as well. For instance, the anisotropic shear viscosity $\eta_\parallel/s\le\eta_\perp/s=1/4\pi$ always decreases as the magnitude of the source of anisotropy (a magnetic field or a nontrivial bulk axion profile) increases; indeed, this is the main topic of Ref. \cite{Jain:2015txa}. On the other hand, the anisotropic drag force in the transverse plane is larger than in the direction of the magnetic field, which is the opposite behavior found in the Einstein-axion-dilaton model for an axion driven anisotropy \cite{Giataganas:2012zy}.

The phenomenological consequences of our results to the current effort in the heavy ion community toward finding direct effects of strong magnetic fields on the QGP will be discussed elsewhere. However, an immediate consequence of the suppression of the shear viscosity along the magnetic field direction found here seems to be that it provides a potential source of suppression for the elliptic flow coefficient in strongly magnetized media. Naively, such a suppression would occur since the flow along the magnetic field direction (which should be, on average, perpendicular to the 2nd harmonic event plane \cite{noncentralB6}) is nearly dissipationless while the viscosity in the direction perpendicular to the field can be twice as large. However, a  realistic estimate of such an effect requires numerical magnetohydrodynamic calculations\footnote{For a discussion of other effects of strong magnetic fields on elliptic flow in ideal hydrodynamics see \cite{Pang:2016yuh}.}. In this regard, our results for the phenomenologically realistic magnetic EMD model, which constitute the main outcome of the present work, could be readily used as inputs in numerical codes for magnetohydrodynamics.

We also presented, in the context of the QCD-like magnetic EMD model, new holographic predictions for the behavior of the entropy density and the crossover temperature extracted from its inflection point in a broader region of the $(T,B)$ phase diagram which has not yet been covered by lattice simulations. In particular, the crossover temperature, which is also in quantitative agreement with the lattice results, was found to be reduced all the way up to $eB=1.5$ GeV$^2$.

Very recently, we used the magnetic EMD model presented here to compute the holographic Polyakov loop and heavy quark entropy in the presence of strong magnetic fields, finding quantitative agreement with lattice results in the deconfined plasma phase \cite{Critelli:2016cvq}. We also intend to study in an upcoming work the QCD phase diagram in the $(T,B)$ plane for much larger values of the magnetic field looking for possible signs of real phase transitions \cite{Endrodi:2015oba} instead of the smooth crossover observed here up to $eB=1.5$ GeV$^2$. Other projects we plan to pursue in the near future include the calculation of quasinormal modes and the anisotropic bulk viscosities in the context of the phenomenological magnetic EMD setup.

\acknowledgments

J.N. thanks G.~Endrodi for discussions about the QCD phase diagram in the presence of strong magnetic fields and also for making available the corresponding lattice data. S.I.F. was supported by the S\~{a}o Paulo Research Foundation (FAPESP) under FAPESP grant number 2015/00240-7 and Coordena\c{c}\~ao de Aperfei\c{c}oamen\-to de Pessoal de N\'{i}vel Superior (CAPES). R.C.  acknowledges financial support by Conselho Nacional de Desenvolvimento Cient\'ifico e Tecnol\'ogico (CNPq). R.R. acknowledges financial support by FAPESP under FAPESP grant number 2013/04036-0. J.N. acknowledges financial support by FAPESP and CNPq.

We also thank Pavel Kovtun and Juan Hernandez for pointing out inconsistencies in some of the Kubo relations derived in Appendix B of the previous version of this work, which we corrected in the present version.

\appendix

\section{Revision of heavy quark momentum transport}
\label{transport}

In this Appendix we review the general holographic formalism used to calculate the heavy quark drag force, the Langevin diffusion coefficients, and the shear viscosity. The discussion presented here describes both isotropic and anisotropic plasmas. In the cases we are interested in, the anisotropy will be induced by an external magnetic field acting on the plasma, which explicitly breaks $SO(3)$ spatial rotation symmetry down to $SO(2)$ rotations in the plane transverse to the direction of the magnetic field, which we shall consider to be constant and uniform in the $z$ direction.

The derivations reviewed in this section have been treated extensively in the literature and are included here for completeness and also to set a clear notation for our purposes. In the isotropic cases, the formalism for calculating the heavy quark drag force was originally discussed in Refs. \cite{Gubser:2006bz,Herzog:2006gh,Herzog:2006se,CasalderreySolana:2006rq,Gubser:2006qh,Gursoy:2009kk}, while the formalism for the Langevin diffusion coefficients was presented in Refs. \cite{Gubser:2006nz,CasalderreySolana:2007qw,Gursoy:2010aa}. The method for calculating the holographic shear viscosity in isotropic gravity duals with actions containing at most two derivatives was treated in Refs. \cite{Policastro:2001yc,Buchel:2003tz,Kovtun:2004de}. In anisotropic environments, the approach used to obtain the drag forces was discussed in Refs. \cite{Chernicoff:2012iq,Giataganas:2012zy,Misobuchi:2015ioa}, while the Langevin coefficients were studied in Refs. \cite{Giataganas:2013zaa,Giataganas:2013hwa,Chakrabortty:2013kra}. The calculation of anisotropic shear viscosities in holography was done, for instance, in Refs. \cite{Rebhan:2011vd,DK-applications2,Jain:2014vka,Jain:2015txa}. See also Ref. \cite{finitemu} for the calculation of the isotropic drag force and the Langevin diffusion coefficients in a phenomenologically realistic EMD model at finite baryon chemical potential and zero magnetic field, and Ref. \cite{Cheng:2014fza} for the calculation of the drag forces in an anisotropic holographic plasma at finite chemical potential and zero magnetic field.

Before proceeding, let us fix the notation used in this paper. We deal with five dimensional black hole backgrounds which may be written, in the Einstein frame, as follows\footnote{Since we use a mostly plus Lorentzian metric signature, $g_{tt}<0$.}
\begin{equation}
\label{eq:genmetric}
ds^2 = g_{tt}(r)dt^2+g_{rr}(r)dr^2+g_{xx}(r)(dx^2+dy^2)+g_{zz}(z)dz^2,
\end{equation}
where $r$ is the holographic radial coordinate. In the different coordinate systems we use in this work, the boundary of the asymptotically $\mathrm{AdS}_5$ geometries is always at $r\to\infty$. The geometries we consider have a black hole horizon at $r = r_H$, where $g_{tt}(r)$ and $g_{rr}^{-1}(r)$ have a simple zero. For nonzero external magnetic fields, these backgrounds display an anisotropy that differentiates the $(x,y)$ plane transverse to the magnetic field from the $z$ direction, that is, $g_{xx} \neq g_{zz}$ for $B\neq 0$, while $g_{xx} = g_{zz}$ in the isotropic $B=0$ case.

The starting point to determine the holographic drag forces and Langevin diffusion coefficients is to consider the Nambu-Goto action for a classical relativistic bosonic string which, in the string frame denoted here by a superscript $(s)$, is written as follows
\begin{equation}
\label{eq:NGaction}
S_{\textrm{NG}} = -\frac{1}{2\pi \alpha'} \int d\tau d\sigma \sqrt{-\gamma^{(s)}},
\end{equation}
where $\alpha'=l_s^2$ is the square of the string length, which may be related to the effective t' Hooft coupling in the dual gauge theory as (we set in the present work the radius of the asymptotically $\mathrm{AdS}_5$ spaces to unity) $\alpha'=\lambda_t^{-1/2}$, $\gamma^{(s)}\equiv\det\gamma_{ab}^{(s)}$, with $\gamma^{(s)}_{ab} = g_{\mu\nu}^{(s)}\partial_{a} \mathcal{X}^\mu \partial_{b} \mathcal{X}^\nu$ being the induced metric on the string worldsheet (i.e., the pullback). The string worldsheet is parametrized by the internal coordinates $a, b \in \{\tau, \sigma\}$, $\mathcal{X}^\mu(\sigma,\tau)$ is the worldsheet embedding on the target background spacetime, and $g_{\mu\nu}^{(s)}$ is the string frame metric of the background. The relation between the string frame metric, $g_{\mu\nu}^{(s)}$, and the Einstein frame metric, $g_{\mu\nu}$, is given by
\begin{equation}
g_{\mu\nu}^{(s)}=e^{\sqrt{\frac{2}{3}}\phi}g_{\mu\nu},
\end{equation}
where $\phi$ is the background dilaton field (following the general philosophy of Improved Holographic QCD \cite{Gursoy:2007cb,Gursoy:2007er,Gursoy:2010fj}).

The general idea that comes from the holographic dictionary is that, following the original proposals to calculate holographic Wilson loops \cite{Maldacena:1998im,Rey:1998ik}, any string endpoint attached to an ultraviolet brane near the boundary of the background spacetime is to be regarded as a probe quark living in the gauge theory. The motion of this string endpoint at the boundary is influenced by the bulk dynamics of the string and, therefore, by studying  its dynamics one can extract information about the behavior of the probe heavy quark in the gauge theory.

\subsection{Drag force: the trailing string}
\label{sec2.0}

\subsubsection{Isotropic case}
\label{sec2.1}

We start by reviewing the holographic formalism to calculate the drag force in isotropic media. A heavy probe moving with a constant velocity $\vec{v}$ is inserted at the boundary and one computes the force that must be applied on the probe in order to equilibrate the effect of the drag force exerted by the medium.

We will consider a heavy probe moving along the $x$ axis, i.e, $\vec{v}=v\hat{x}$. We work in the so-called static gauge for the string worldsheet where one fixes $\sigma = r$ and $\tau = t$. The \textit{trailing string} ansatz for the embedding function describing this situation is given by
\begin{align}
\label{eq:embedding}
\mathcal{X}^\mu=(\mathcal{X}^t,\mathcal{X}^r,\mathcal{X}^x,\mathcal{X}^y,\mathcal{X}^z)= (t,r,x(t,r)=vt+\xi(r),0,0),
\end{align}
where $\xi(r)$ describes the bulk string profile in the $(r,x)$ plane, which is dynamically fixed by extremizing the Nambu-Goto action \eqref{eq:NGaction} on the black brane backgrounds \eqref{eq:genmetric}.

The Nambu-Goto action with the trailing string ansatz becomes\footnote{For heavy quarks with finite masses one should consider the minimal coupling of the background Maxwell field describing the external magnetic field at the boundary with the string worldsheet, as done in Ref. \cite{Kiritsis:2011ha}. However, in the present work we consider only infinitely heavy probes and in this case the minimal coupling is $\alpha'$ suppressed relatively to the Nambu-Goto action and may be neglected at leading order in the t' Hooft coupling.}
\begin{equation}
S_{\textrm{NG}} = -\frac{1}{2\pi \alpha'} \int dr dt \sqrt{-g_{tt}^{(s)} g_{rr}^{(s)} -g_{tt}^{(s)} g_{xx}^{(s)} \xi'(r)^2 - g_{rr}^{(s)} g_{xx}^{(s)} v^2},
\end{equation}
where $\xi'(r)\equiv\partial_r\xi(r)$. This action does not depend explicitly on $\xi(r)$ and, therefore, the corresponding radial conjugate momentum,
\begin{equation}
\label{eq:pi}
\pi_{\xi} = \frac{\delta S_{\textrm{NG}}}{\delta \xi'} = - \frac{1}{2 \pi \alpha'} \frac{g_{tt}^{(s)} g_{xx}^{(s)} \xi'(r)}{\sqrt{-g^{(s)}_{tt} g^{(s)}_{rr} -g_{tt}^{(s)} g_{xx}^{(s)} \xi'(r)^2 - g_{rr}^{(s)} g_{xx}^{(s)} v^2}},
\end{equation}
is conserved in the radial direction and we may compute it at any value of $r$. We can do it by first solving the above relation for $\xi'(r)$,
\begin{equation}
\xi'(r) = \sqrt{\frac{-g_{tt}^{(s)} g_{rr}^{(s)} - g_{xx}^{(s)} g_{rr}^{(s)} v^2}{g_{tt}^{(s)} g_{xx}^{(s)} \left(1 + \frac{g_{tt}^{(s)} g_{xx}^{(s)}}{(2\pi \alpha' \pi_{\xi})^2} \right)}},
\end{equation}
which must be real\footnote{Otherwise the probe's trajectory at the boundary, $x(t,r\to\infty)=vt+\xi(r\to\infty)$, would not be real-valued.}. Therefore, in order to render the expression inside the square root nonnegative for every $r$, when the numerator changes sign, the denominator must also change sign which happens at a common value of the radial coordinate $r=r_\star$ obtained by solving the equation
\begin{equation}
\label{eq:rstar}
g_{tt}^{(s)} (r_\star) + g_{xx}^{(s)} (r_\star) v^2 = 0.
\end{equation}
As discussed in detail in Refs. \cite{Gubser:2006nz,CasalderreySolana:2007qw,Gursoy:2010aa}, the induced worldsheet metric has the form of a two dimensional black hole with a horizon precisely at $r=r_\star$. Evaluating \eqref{eq:pi} at the worldsheet horizon, one obtains
\begin{equation}
\pi_{\xi} = - \frac{\sqrt{g_{tt}^{(s)}(r_\star) g_{xx}^{(s)} (r_\star)}}{2\pi \alpha'}. 
\end{equation}

The flux of momentum from the string endpoint at the boundary to the worldsheet horizon in the interior of the bulk is equal to the drag force exerted by the plasma on the infinitely heavy probe at the boundary, which is given by $\pi_{\xi}$. Therefore, the drag force exerted by the medium on the probe reads \cite{Gursoy:2009kk}
\begin{equation}
\label{eq:Fiso}
F^{\textrm{(iso)}}_{\textrm{drag}} = \frac{dp_x}{dt} = \pi_{\xi} = -\frac{1}{2\pi \alpha'} \sqrt{-g_{tt}^{(s)} (r_\star) g_{xx}^{(s)} (r_\star)} = -\frac{\sqrt{\lambda_t}}{2\pi} g_{xx}^{(s)}(r_\star) v.
\end{equation}

Specializing the isotropic background to the AdS$_5$-Schwarzschild case, which is the gravity dual of a strongly coupled SYM plasma at $B=0$, we obtain the well-known result \cite{Gubser:2006bz,Herzog:2006gh}
\begin{equation}
\label{eq:dragsym}
\frac{F_{\textrm{drag}}^{\textrm{(SYM)}}}{\sqrt{\lambda_t} T^2} = -\frac{\pi\gamma v}{2},
\end{equation}
where $\gamma=1/\sqrt{1-v^2}$ is the Lorentz factor.

\subsubsection{Anisotropic case}

In the anisotropic case, one needs to consider the angle $\varphi$ formed between the velocity $\vec{v}$ of the heavy probe and the anisotropic direction $\hat{z}$, which in our case is the direction of the magnetic field. A general discussion which yields the drag force in terms of an arbitrary angle $\varphi$ was presented in Appendix A of \cite{Misobuchi:2015ioa}, generalizing the original derivation for transverse and longitudinal angles first presented in Appendix B of \cite{Giataganas:2012zy}. We consider below an alternative derivation based in Appendix A of Ref. \cite{Misobuchi:2015ioa}, but here we make a rotation of the $xz$ axes before computing the drag force. We then particularize our results to the transverse and longitudinal orientations, obtaining the same results as in Refs. \cite{Giataganas:2012zy,Misobuchi:2015ioa}.

Let us reorient our old $xz$ axes into new $XZ$ axes such that now it is the new $X$ direction which coincides with the probe's velocity direction. That is,
\begin{align}
dx = \cos (\theta) dX - \sin(\theta) dZ \quad \mathrm{and} \quad dz = \sin (\theta) dX + \cos(\theta) dZ,
\end{align}
where $\theta \equiv \pi/2 - \varphi$ is the angle between the $\hat{x}$ direction and the velocity of the probe $\vec{v}\parallel\hat{X}$.

The (Einstein frame) rotated metric becomes
\begin{equation}
ds^2 = g_{tt} dt^2 + g_{rr} dr^2 + g_{XX} dX^2 + g_{yy} dy^2 + g_{ZZ} dZ^2 + 2 g_{XZ} dX dZ,
\end{equation}
where
\begin{align}
g_{XX} = g_{xx} \cos^2 \theta + g_{zz} \sin^2 \theta,\,\,\, g_{ZZ} = g_{xx} \sin^2 \theta + g_{zz} \cos^2 \theta,\,\,\,
g_{XZ} = \cos\theta\sin\theta (g_{zz} - g_{xx}).
\end{align}

The derivation of the drag force carried out in the isotropic case can be repeated here with almost no modifications, except for the substitution $x \to X$, in which case one obtains that
\begin{equation}
\label{eq:anisotropicdrag}
F^{\textrm{(aniso)}}_{\textrm{drag}}(\theta) = -\frac{\sqrt{\lambda_t}}{2\pi} g_{XX}^{(s)}(\theta, r_\star(\theta)) v,
\end{equation}
where $r_\star(\theta)$ is the root of
\begin{equation}
g_{tt}(r_\star(\theta))+g_{XX}(r_\star(\theta),\theta)v^2=0.
\end{equation}

If one takes $\theta = \pi/2$, such that the quark's velocity is parallel to the magnetic field, one obtains that \cite{Giataganas:2012zy},
\begin{equation}
\label{eq:dragpal}
F_{\textrm{drag}}^{(v\parallel B)} \equiv F^{\textrm{(aniso)}}_{\textrm{drag}}(\theta =\pi/2) = - \frac{\sqrt{\lambda_t}}{2\pi}  g_{zz}^{(s)} (r_\star^{\parallel}) v,
\end{equation}
where $r_\star^{\parallel}$ is the root of
\begin{equation}
\label{eq:rpar}
g_{tt}^{(s)} (r_\star^{\parallel}) + g_{zz}^{(s)} (r_\star^{\parallel}) v^2 = 0.
\end{equation}
The other limiting case we shall consider here corresponds to a heavy quark moving perpendiculary to the magnetic field direction, which is obtained by taking $\theta = 0$ \cite{Giataganas:2012zy},
\begin{equation}
\label{eq:dragperp}
F_{\textrm{drag}}^{(v\perp B)} \equiv F^{\textrm{(aniso)}}_{\textrm{drag}}(\theta = 0) = - \frac{\sqrt{\lambda_t}}{2\pi} g_{xx}^{(s)} (r_\star^{\perp}) v,
\end{equation}
where $r_\star^{\perp}$ is the root of
\begin{equation}
\label{eq:rperp}
g_{tt}^{(s)} (r_\star^{\perp}) + g_{xx}^{(s)} (r_\star^{\perp}) v^2 = 0.
\end{equation}

\subsection{Langevin coefficients}
\label{sec2.2}

\subsubsection{Diffusion from worldsheet membrane paradigm}

The drag force derived previously in the trailing string scenario defines a matrix of friction coefficients, $\eta^D_{ij}$, $i,j\in\{x,y,z\}$, according to the following equation,
\begin{align}
\label{eq:drag}
F_i^{\textrm{drag}}=\frac{dp_i}{dt}=-\eta^D_{ij}p^j, \quad \eta^D_{ij}\equiv\frac{\sqrt{\lambda_t}}{2\pi} \frac{g_{ij}^{(s)}(r_\star)}{m_q\gamma},
\end{align}
where $m_q$ is the quark mass and $p_i=m_q\gamma v_i$ is its 4-momentum. Note that for a diagonal background metric, the friction matrix $\eta^D_{ij}$ in Eq. \eqref{eq:drag} is also diagonal.

In order to calculate the Langevin diffusion coefficients associated with the Brownian motion of the heavy quark under the influence of thermal fluctuations one needs to add a small perturbation $\delta\mathcal{X}^\mu$ for the worldsheet embedding function on top of the trailing string ansatz in the static gauge given by Eq.\ \eqref{eq:embedding},
\begin{align}
\label{eq:embedding2}
\bar{\mathcal{X}}^\mu=\mathcal{X}^\mu+\delta\mathcal{X}^\mu= (t,r,x^\ell(t,r),x^i(t,r),x^j(t,r))=(t,r, vt+\xi(r)+\delta\mathcal{X}^\ell(t,r),\delta\mathcal{X}^i(t,r),\delta\mathcal{X}^j(t,r)),
\end{align}
where we employ the spatial index $\ell$ to denote the component in the direction of the quark's velocity, $\vec{v}=v\hat{x}_\ell$.

The inclusion of thermal fluctuations modifies the equation of motion \eqref{eq:drag} for the probe quark in the gauge theory. In fact, the boundary value of the worldsheet embedding fluctuation, $\delta\mathcal{X}_i(t,r\to\infty)$, serves as a source for an operator $\mathcal{F}_i(t)$ playing the role of a random force acting on the heavy quark in the gauge theory. Assuming that for long times the temporal correlation functions of the operator $\mathcal{F}_i(t)$ are proportional to Dirac delta distributions, with the proportionality factors defining the Langevin diffusion coefficients $\kappa_{ij}$, one may derive in this limit an effective equation of motion for the heavy quark taking into account thermal fluctuations, which has the form of a local Langevin equation \cite{Gursoy:2010aa,Giataganas:2013zaa},
\begin{align}
\frac{dp_i}{dt}=-\eta^D_{ij}p^j+\mathcal{F}_i(t),
\end{align}
and one can also show that the Langevin diffusion coefficients satisfy the following modified Einstein's relation\footnote{Eq. \eqref{eq:DefKappa} is a modification of the usual Einstein's relation since it relates the friction and the diffusion coefficients through the worldsheet temperature $T_\star$ instead of the heat bath temperature $T$. According to Ref. \cite{Gursoy:2010aa}, $T_\star$ is the temperature measured by a probe quark moving at speed $v$ inside a strongly coupled plasma.} \cite{Gursoy:2010aa,Giataganas:2013zaa},
\begin{equation}
\label{eq:DefKappa}
\kappa_{ij} = 2 T_\star \eta^D_{ij} = -2 T_\star \lim_{\omega\rightarrow 0} \frac{\text{Im}\,G^{R}_{ij}(\omega)}{\omega},
\end{equation}
where $G^{R}_{ij}(\omega)$ is the retarded propagator of the random force operator $\mathcal{F}_i(t)$ and $T_\star$ is the worldsheet Hawking's temperature, to be discussed in the sequel.

The friction coefficients $\eta^D_{ij}=-\lim_{\omega\rightarrow 0}\text{Im}\,G^{R}_{ij}(\omega)/\omega$ may be directly extracted from the quadratic action for the worldsheet embedding fluctuations to be derived next using the membrane paradigm \cite{Iqbal:2008by} applied to the worldsheet horizon, as we are going to discuss below.

\subsubsection{Thermal fluctuations on the trailing string}

In this subsection we review the results of Refs. \cite{Giataganas:2013zaa, Giataganas:2013hwa} concerning the derivation of the quadratic fluctuated action for the string worldsheet. The disturbed metric induced on the string worldsheet in the string frame, associated with the perturbed ansatz \eqref{eq:embedding2}, reads
\begin{equation}
\bar{\gamma}^{(s)}_{ab}= \gamma^{(s)}_{ab}+\delta\gamma_{ab}^{(s)\,(1)}+\delta\gamma_{ab}^{(s)\,(2)},
\end{equation}
where $\gamma_{ab}^{(s)}$ is the unperturbed pullback and we also have linear and quadratic terms in the perturbations,
\begin{align}
\delta\gamma_{ab}^{(s)\,(1)} &=g_{\ell\ell}^{(s)}\left( \partial_{a}\mathcal{X}^{\ell}\partial_{b}\delta \mathcal{X}^{\ell}+\partial_{b}\mathcal{X}^{\ell}\partial_{a}\delta \mathcal{X}^{\ell} \right), \\
\delta\gamma_{ab}^{(s)\,(2)}& = \sum_{i}g_{ii}^{(s)} \partial_{a}\delta \mathcal{X}^{i}\partial_{b}\delta \mathcal{X}^{i}.
\end{align}

The task now is to obtain an expression for $\sqrt{-\bar{\gamma}}$ where $\bar{\gamma}$ is the determinant of the perturbed pullback. Since the fluctuations of the worldsheet embedding are small, we retain only terms up to second order in the perturbations,
\begin{equation}
\label{eq:corrections}
\sqrt{-\bar{\gamma}^{(s)}}=\sqrt{-\gamma^{(s)}}\left[1 +\frac{1}{2}\delta\gamma^{(s)(1)a}_{a}-\frac{1}{4}\delta\gamma^{(s)(1)ab}\delta\gamma^{(s)(1)}_{ab}+\frac{1}{8}(\delta\gamma^{(s)(1)a}_{a})^2+\frac{1}{2}\delta\gamma^{(s)(2)a}_{a} \right].
\end{equation}
In the above expression, we are not interested in the first two terms since they are not quadratic in the perturbations and, therefore, they do not contribute to the 2-point Green's function of the stochastic force operator $\mathcal{F}_i(t)$ sourced by the boundary value of the worldsheet embedding fluctuations. The explicit corrections contributing to this propagator are given by the last three terms in Eq. \eqref{eq:corrections}, which can be written as
\begin{align}
\frac{1}{4}\delta\gamma^{(s)(1)ab}\delta\gamma_{ab}^{(s)(1)} &= \frac{1}{2}g_{\ell\ell}^{(s)2}\left[(\gamma^{(s)cd}\partial_{c}\mathcal{X}^{\ell}\partial_{d}\mathcal{X}^{\ell})\gamma^{(s)ab}\partial_{a}\delta \mathcal{X}^{\ell}\partial_{b}\delta \mathcal{X}^{\ell} +(\gamma^{(s)cd}\partial_{c}\mathcal{X}^{\ell}\partial_{d}\delta \mathcal{X}^{\ell})^2 \right], \\
\frac{1}{8}(\delta\gamma^{(s)(1)a}_{a})^2 &= \frac{1}{2}g_{\ell\ell}^{(s)2}(\gamma^{(s)cd}\partial_{c}\mathcal{X}^{\ell}\partial_{d}\delta \mathcal{X}^{\ell})^2, \\
\frac{1}{2}\delta\gamma^{(s)(2)a}_{a} &= \gamma^{(s)ab}g_{ij}^{(s)}\partial_{a}\delta \mathcal{X}^{i}\partial_{b}\delta \mathcal{X}^{j}.
\end{align}

Plugging these corrections into the Nambu-Goto action, we obtain the following quadratic action for the worldsheet embedding fluctuations
\begin{equation}
S_2=-\frac{1}{2\pi\alpha'}\int dt dr \sqrt{-\gamma^{(s)}}\frac{\gamma^{(s)ab}}{2}\left[ (1-g_{\ell\ell}^{(s)}\gamma^{(s)cd}\partial_{c}\mathcal{X}^{\ell}\partial_{d}\mathcal{X}^{\ell})g_{\ell\ell}^{(s)}\partial_a \delta \mathcal{X}^{\ell}\partial_b \delta X^{\ell}+\sum_{i\neq l}g_{ii}^{(s)}\partial_a \delta \mathcal{X}^{i}\partial_b \delta \mathcal{X}^{i} \right].
\end{equation}
We can still make one further simplification in the kinetic term of the $\ell$-fluctuation. First, we note that
\begin{equation}
\gamma^{(s)ab}\underbrace{g_{\mu\nu}^{(s)}\partial_{a}\mathcal{X}^{\mu}\partial_{b}\mathcal{X}^{\nu}}_{=\gamma^{(s)}_{ab}} = \delta_a^a = 2 \Rightarrow 1- g_{\ell\ell}^{(s)} \gamma^{(s)ab}\partial_{a}\mathcal{X}^{\ell}\partial_{b}\mathcal{X}^{\ell} = -1 + \gamma^{(s)tt}g_{tt}^{(s)}+\gamma^{(s)rr}g_{rr}^{(s)},
\end{equation}
and since, from the holographic drag force calculation,
\begin{equation}
\gamma^{(s)tt}=\frac{\left(g_{tt}^{(s)}\right)^2-v^2(2\pi\alpha'\pi_\xi)^2}{\left(g_{tt}^{(s)}\right)^2(g_{tt}^{(s)}+v^2g_{\ell\ell})}, \ \ \ \gamma^{(s)rr} = \frac{g_{tt}^{(s)}g_{\ell\ell}^{(s)}+(2\pi\alpha'\pi_\xi)^2}{g_{tt}^{(s)}g_{rr}^{(s)}g_{\ell\ell}^{(s)}},
\end{equation}
we have that
\begin{equation}
\label{eq:DefN}
(1- g_{\ell\ell}^{(s)} \gamma^{(s)ab}\partial_{a}\mathcal{X}^{\ell}\partial_{b}\mathcal{X}^{\ell})g_{\ell\ell}^{(s)} = \frac{g_{tt}^{(s)}g_{\ell\ell}^{(s)}+(2\pi\alpha'\pi_\xi)^2}{g_{tt}^{(s)}+v^2g_{\ell\ell}^{(s)}}\equiv N(r).
\end{equation}  

Therefore, a simple expression for the quadratic action in the worldsheet embedding fluctuations can be found \cite{Giataganas:2013zaa,Giataganas:2013hwa}
\begin{equation}
\label{eq:FlucAct_nondiag}
S_2 = -\frac{1}{2\pi\alpha'}\int dt dr \sqrt{-\gamma^{(s)}}\frac{\gamma^{(s)ab}}{2}\left[ N(r)\partial_a \delta \mathcal{X}^{\ell}\partial_b \delta \mathcal{X}^{\ell}+\sum_{i\neq \ell}g_{ii}^{(s)}\partial_a \delta \mathcal{X}^{i}\partial_b \delta \mathcal{X}^{i} \right],
\end{equation}
which reduces to Eq. (3.33) of Ref. \cite{Gursoy:2010aa} in the isotropic limit.

Now we diagonalize the pullback by making a worldsheet coordinate redefinition: $dt\rightarrow dt-dr\,\gamma_{tr}/\gamma_{tt}$. The quadratic action for the fluctuations now takes the form \cite{Giataganas:2013zaa,Giataganas:2013hwa}
\begin{equation}\label{eq:FlucAct_Diag}
S_2 = -\frac{1}{2\pi\alpha'}\int dt dr \sqrt{-h^{(s)}}\frac{h^{(s)ab}}{2}\left[ N(r)\partial_a \delta \mathcal{X}^{\ell}\partial_b \delta \mathcal{X}^{\ell}+\sum_{i\neq \ell}g_{ii}^{(s)}\partial_a \delta \mathcal{X}^{i}\partial_b \delta \mathcal{X}^{i} \right],
\end{equation}
where $h_{ab}^{(s)}$ is the diagonalized pullback, whose explicit form in terms of the background metric is given by
\begin{equation}
h_{ab}^{(s)} = \text{diag}\left\lbrace g_{tt}^{(s)}+v^2g_{\ell\ell}^{(s)}, \frac{g_{tt}^{(s)}g_{rr}^{(s)}g_{\ell\ell}^{(s)}}{g_{tt}^{(s)}g_{\ell\ell}^{(s)}+v^2(g_{\ell\ell}^{(s)}(r_\star))^2} \right\rbrace.
\end{equation}
This metric has a worldsheet black hole horizon and the associated worldsheet Hawking's temperature reads \cite{Giataganas:2013zaa,Giataganas:2013hwa}
\begin{align}\label{eq:WSTemperature}
T_\star &=\frac{1}{4\pi}\sqrt{\left\vert h'^{(s)}_{tt}\left(h^{(s)rr}\right)'\right\vert_{r_\star}}, \notag \\
    &= \frac{1}{4\pi}\sqrt{\left\vert(g_{tt}^{(s)}+ v^2g_{\ell\ell}^{(s)})'\left(\frac{g_{tt}^{(s)}g_{\ell\ell}^{(s)}+ v^2(g_{\ell\ell}^{(s)}(r_\star))^2}{g_{tt}^{(s)}g_{\ell\ell}^{(s)}g_{rr}^{(s)}} \right)'\right\vert_{r_\star}},  \notag \\
    &= \frac{1}{4\pi}\sqrt{\left\vert \frac{(g'^{(s)}_{tt})^2-v^4(g'^{(s)}_{\ell\ell})^2}{g_{tt}^{(s)}g_{rr}^{(s)}} \right\vert_{r_\star}},  \notag \\
 &= \frac{1}{4\pi\sqrt{-g_{tt}^{(s)}(r_\star)g_{rr}^{(s)}(r_\star)}}\sqrt{\left\vert (g_{tt}^{(s)}g_{\ell\ell}^{(s)})'\left(\frac{g_{tt}^{(s)}}{g_{\ell\ell}^{(s)}}\right)'\right\vert_{r_\star}}.
\end{align}
Note, from the expression above, that in the anisotropic case the worldsheet temperature will depend on which direction the string propagates.

\subsubsection{Isotropic case}

In the case of zero magnetic field $g_{xx}=g_{zz}$. If we choose, without loss of generality, $\ell=x$, the action \eqref{eq:FlucAct_Diag} takes the form
\begin{equation}
\label{eq:FluAct-Lan-Sym}
S_2 =- \frac{1}{2\pi\alpha'}\int dt dr \sqrt{-h^{(s)}}\frac{h^{(s)ab}}{2}\left[ N(r)\partial_a \delta \mathcal{X}^{x}\partial_b \delta \mathcal{X}^{x}+\sum_{i=y,z}g_{ii}^{(s)}\partial_a \delta \mathcal{X}^{i}\partial_b \delta \mathcal{X}^{i} \right].
\end{equation}
The diffusion coefficients parallel and transverse to the probe's velocity are then related to the 2-point functions obtained from the fluctuations $\delta\mathcal{X}^x$ and $\delta\mathcal{X}^y$ (or $\delta\mathcal{X}^z$), respectively. The easiest way to extract them is via the membrane paradigm \cite{Iqbal:2008by}, which implies that the transport coefficient associated to the propagator of a massless fluctuation via linear response theory, in the limit of zero spatial momentum and zero frequency, can be directly extracted from the coefficient in front of the effective kinetic term for the fluctuation evaluated at the black hole horizon. More specifically, considering a quadratic action for some generic bulk massless perturbation $\phi$ in arbitrary $(d+1)$-dimensions,
\begin{equation}
\label{eq:Example-Fluc}
S_2^\phi = -\frac{1}{2}\int drd^{d}x \sqrt{-g}\frac{g^{\mu\nu}}{q(r)}\partial_\mu\phi\partial_\nu\phi,
\end{equation}
the transport coefficient $\chi$ extracted from \eqref{eq:Example-Fluc} is given by \cite{Iqbal:2008by}
\begin{equation}
\label{eq:MemPara}
\chi =- \lim_{\omega\rightarrow 0} \frac{\text{Im}\,G^{R}(\omega,\vec{k}=\vec{0})}{\omega} =\lim_{r\to r_H}\frac{1}{q(r)}\sqrt{\frac{-g(r)}{g_{tt}(r)g_{rr}(r)}}.
\end{equation}

In the case of \eqref{eq:FluAct-Lan-Sym}, one first applies the membrane paradigm to the two dimensional worldsheet black hole horizon (i.e., take $g_{\mu\nu}\to h_{ab}^{(s)}$ and $r_H\to r_\star$ in Eq. \eqref{eq:MemPara}) to obtain the friction coefficients $\eta^D_{ij}=-\lim_{\omega\rightarrow 0}\text{Im}\,G^{R}_{ij}(\omega)/\omega$ associated to the worldsheet embedding fluctuations, and then uses the modified Einstein's relation \eqref{eq:DefKappa} to obtain the corresponding Langevin coefficients \cite{Giataganas:2013zaa, Giataganas:2013hwa},\footnote{Note from Eqs. \eqref{eq:rstar} and \eqref{eq:DefN} that $N(r_\star)\to 0/0$. Therefore, one needs to use the L'Hopital's rule to calculate the limit $\lim_{r\to r_\star}N(r)$. In order to do this, one first rewrites \eqref{eq:DefN} as $N(r)=\frac{1}{g_{\ell\ell}}\times \frac{g^{(s)}_{tt}g^{(s)}_{\ell\ell}+(2\pi\alpha'\pi_\xi)^2}{(g^{(s)}_{tt}/g^{(s)}_{\ell\ell})+v^2}$, which gives $\lim_{r\to r_\star}N(r)=\frac{1}{g^{(s)}_{\ell\ell}}\times \frac{(g^{(s)}_{tt}g^{(s)}_{\ell\ell})'}{(g^{(s)}_{tt}/g^{(s)}_{\ell\ell})'}\biggr|_{r_\star}$.}
\begin{align}
\kappa_{(\parallel v)}^{\textrm{(iso)}}&=\frac{T_\star}{\pi\alpha'}\lim_{r\to r_\star}N(r) = \frac{T_\star}{\pi\alpha'}\frac{1}{g_{xx}^{(s)}(r_\star)} \frac{(g_{tt}^{(s)}g_{xx}^{(s)})'}{(g^{(s)}_{tt}/g_{xx}^{(s)})'}\biggr|_{r_\star},\\
\kappa_{(\perp v)}^{\textrm{(iso)}}&=\frac{T_\star}{\pi\alpha'} g_{xx}^{(s)}(r_\star),
\end{align}
where the subscript $(\parallel v)\,[(\perp v)]$ denotes the momentum diffusion parallel [perpendicular] to the initial probe's velocity.

Then, for a SYM plasma, one obtains the well-known results \cite{Gubser:2006nz,CasalderreySolana:2007qw}
\begin{align}
\frac{\kappa_{(\parallel v)}^{\textrm{(SYM)}}}{\sqrt{\lambda_t}T^3}=\pi\gamma^{5/2} \quad \textrm{and} \quad
\frac{\kappa_{(\perp v)}^{\textrm{(SYM)}}}{\sqrt{\lambda_t}T^3}&=\pi\gamma^{1/2}.
\label{eq:LangSYM}
\end{align}

\subsubsection{Anisotropic case}

\begin{figure}[h]
\begin{center}
\begin{tabular}{c}
\includegraphics[width=0.45\textwidth]{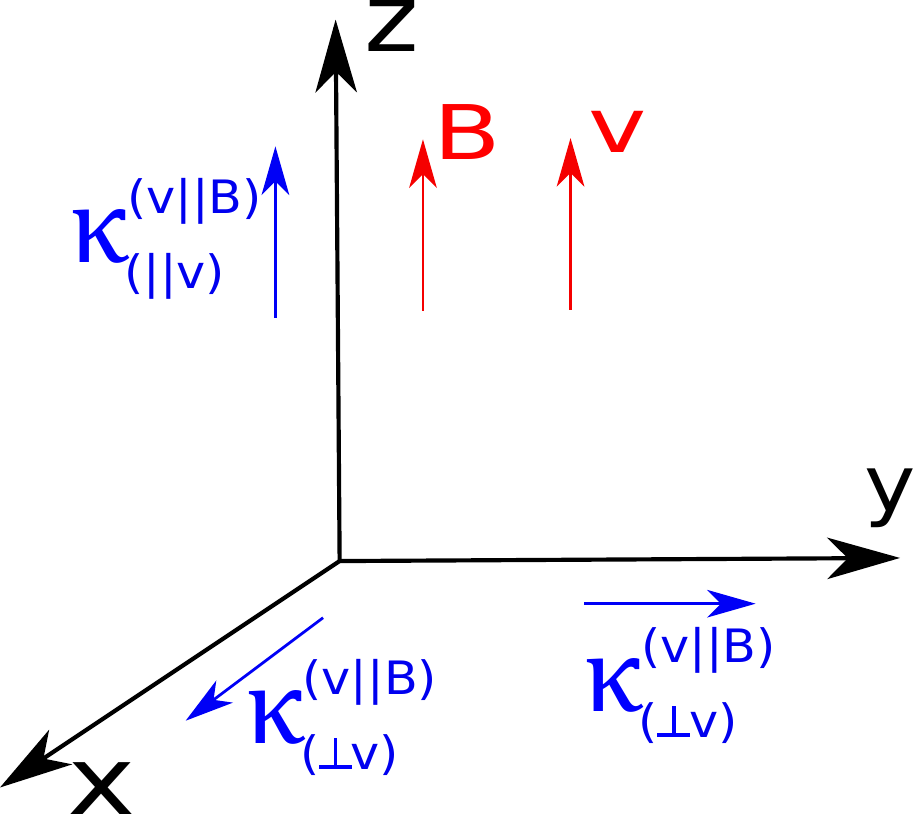} 
\end{tabular}
\begin{tabular}{c}
\includegraphics[width=0.45\textwidth]{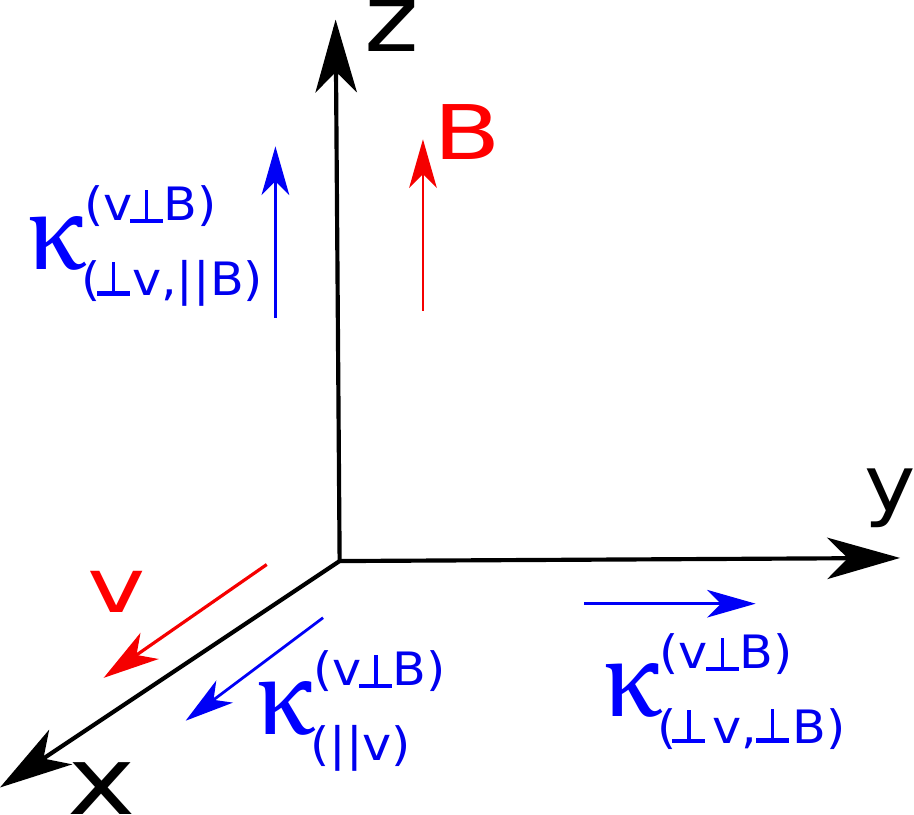} 
\end{tabular}
\end{center}
\caption{(Color online). Schematic representation of the anisotropic Langevin diffusion coefficients computed in this work. }
\label{fig:kxfig}
\end{figure}

To deal with the general anisotropic case for the diffusion coefficients using the generic diagonal background \eqref{eq:genmetric}, one would need to consider three angles: $\textrm{ang}(\vec{v},\vec{\kappa})$, $\textrm{ang}(\vec{B},\vec{\kappa})$, and $\textrm{ang}(\vec{v},\vec{B})$, where $\vec{\kappa}$ is the direction of momentum diffusion. For this work, though, we shall investigate the following cases: \textbf{(i)} $\vec{v}$ parallel to $\vec{B}$, which is characterized by two diffusion coefficients; and \textbf{(ii)} $\vec{v}$ perpendicular to $\vec{B}$, which is characterized by three diffusion coefficients. Thus, our goal here is to calculate the momentum diffusion along the spatial directions $\hat{x}$, $\hat{y}$, and $\hat{z}$ for each case, as illustrated in Fig. \ref{fig:kxfig} 

Starting with the case \textbf{(i)}, $\vec{v}\parallel\vec{B}$, with $\vec{B}=B\hat{z}$, one may write the disturbed action \eqref{eq:FlucAct_Diag} as follows \cite{Giataganas:2013zaa,Giataganas:2013hwa},
\begin{equation}\label{eq:FluAct-vparaB}
S_2^{(v\parallel B)} = -\frac{1}{2\pi\alpha'}\int dt dr \sqrt{-h^{(s)}}\frac{h^{(s)ab}}{2}\left[ N(r)\partial_a \delta \mathcal{X}^{z}\partial_b \delta \mathcal{X}^{z}+\sum_{i=x,y}g_{ii}^{(s)}\partial_a \delta \mathcal{X}^{i}\partial_b \delta \mathcal{X}^{i} \right].
\end{equation}
Proceeding with the membrane paradigm calculation, just as done in the previous subsection, one obtains the following diffusion coefficients \cite{Giataganas:2013zaa, Giataganas:2013hwa},
\begin{align}
\label{eq:dif1}
\frac{\kappa_{(\parallel v)}^{(v\parallel B)}}{\sqrt{\lambda_t}} &= \frac{T_\star^{\parallel}}{\pi g_{zz}^{(s)}(r_{\star}^{\parallel})}\left.\frac{(g_{tt}^{(s)}g_{zz}^{(s)})'}{(g_{tt}^{(s)}/g_{zz}^{(s)})'}\right\vert_{r_{\star}^{\parallel}}, \\
\frac{\kappa_{(\perp v)}^{(v\parallel B)}}{\sqrt{\lambda_t}} &= \frac{T_\star^{\parallel}}{\pi}g_{xx}^{(s)}(r_{\star}^{\parallel}),
\end{align}
where \cite{Giataganas:2013zaa, Giataganas:2013hwa},
\begin{align}
T_\star^{\parallel} &= \frac{1}{4\pi \sqrt{-g_{tt}^{(s)}(r_{\star}^{\parallel})g_{rr}^{(s)}(r_{\star}^{\parallel})}}\sqrt{\left\vert (g_{tt}^{(s)}g_{zz}^{(s)})'\left(\frac{g_{tt}^{(s)}}{g_{zz}^{(s)}}\right)'\right\vert_{r_{\star}^{\parallel}}},\\
0 &= g_{tt}^{(s)}(r_{\star}^{\parallel})+v^2g_{zz}^{(s)}(r_\star^{\parallel}).
\label{eq:dif2}
\end{align}

Considering now the case \textbf{(ii)}, $\vec{v}\perp \vec{B}$, with $\vec{v}=v\hat{x}$, the fluctuated action \eqref{eq:FlucAct_Diag} becomes \cite{Giataganas:2013zaa,Giataganas:2013hwa},
\begin{equation}
S_2^{(v\perp B)} = -\frac{1}{2\pi\alpha'}\int dt dr \sqrt{-h^{(s)}}\frac{h^{(s)ab}}{2}\left[ N(r)\partial_a \delta \mathcal{X}^{x}\partial_b \delta \mathcal{X}^{x}+g_{xx}^{(s)}\partial_a \delta \mathcal{X}^{y}\partial_b \delta \mathcal{X}^{y} +g_{zz}^{(s)}\partial_a \delta \mathcal{X}^{z}\partial_b \delta \mathcal{X}^{z}\right].
\end{equation}
From the membrane paradigm, one obtains the following diffusion coefficients \cite{Giataganas:2013zaa, Giataganas:2013hwa},
\begin{align}
\label{eq:dif3}
\frac{\kappa_{(\parallel v)}^{(v\perp B)}}{\sqrt{\lambda_t}} &= \frac{T_\star^{\perp}}{\pi g_{xx}^{(s)}(r_{\star}^{\perp})}\left.\frac{(g_{tt}^{(s)}g_{xx}^{(s)})'}{(g_{tt}^{(s)}/g_{xx}^{(s)})'}\right\vert_{r_{\star}^{\perp}}, \\
\frac{\kappa_{(\perp v, \perp B)}^{(v\perp B)}}{\sqrt{\lambda_t}} &= \frac{T_\star^{\perp}}{\pi}g_{xx}^{(s)}(r_{\star}^{\perp}),\\
\frac{\kappa_{(\perp v, \parallel B)}^{(v\perp B)}}{\sqrt{\lambda_t}} &= \frac{T_\star^{\perp}}{\pi} g_{zz}^{(s)}(r_{\star}^{\perp}),
\end{align}
where \cite{Giataganas:2013zaa, Giataganas:2013hwa},
\begin{align}
T_\star^{\perp} &= \frac{1}{4\pi \sqrt{-g_{tt}^{(s)}(r_\star^{\perp})g_{rr}^{(s)}(r_\star^{\perp})}}\sqrt{\left\vert (g_{tt}^{(s)}g_{xx}^{(s)})'\left(\frac{g_{tt}^{(s)}}{g_{xx}^{(s)}}\right)'\right\vert_{r_\star^{\perp}}},\label{eq:dif4}\\
0 &= g_{tt}^{(s)}(r_{\star}^{\perp})+v^2g_{xx}^{(s)}(r_\star^{\perp}).\label{eq:dif5}
\end{align}

\subsection{Shear viscosity}
\label{sec2.3}

Along with the drag force and the Langevin diffusion coefficients, we will also compute how the shear viscosity coefficient $\eta$ changes with the inclusion of a magnetic field.

Any quantum field theory that possess an isotropic gravity dual whose action contains terms up to two derivatives has the following value for the shear viscosity to entropy density ratio,
\begin{equation}
\label{eq:taxa}
\frac{\eta}{s}=\frac{1}{4\pi},
\end{equation}
which was previously conjectured \cite{Kovtun:2004de} to be a lower bound for the value of the ratio $\eta/s$ in Nature, which is known in the literature as the Kovtun-Son-Starinets (KSS) bound. In order to obtain departures and possible violations of the KSS bound \eqref{eq:taxa} in holographic settings, one may include higher order curvature terms in the gravity action \cite{Kats:2007mq,Brigante:2007nu,Brigante:2008gz}. Furthermore, these higher order derivative corrections may be also employed in conjunction to a nontrivial dilaton potential breaking conformal symmetry in the infrared to provide a non-constant temperature profile for the ratio $\eta/s$ \cite{Cremonini:2012ny}.

Another way to violate the bound \eqref{eq:taxa} in holographic setups is to break rotational or translational symmetries. The first calculation of anisotropic shear viscosities was done in Ref.\ \cite{Rebhan:2011vd} for the case of an anisotropic plasma created by a spatially dependent bulk axion profile originally proposed in Refs. \cite{Mateos:2011ix,Mateos:2011tv}. The result is similar to the one obtained in Ref. \cite{DK-applications2} in the context of the magnetic brane model originally proposed in Refs. \cite{DK1,DK2,DK3} - see also Ref. \cite{Liu:2016njg} for results concerning $p$-form magnetically charged branes. One may also find in the literature models with a dilaton driven anisotropy \cite{Jain:2014vka,Jain:2015txa}, anisotropic $SU(2)$ Einstein-Yang-Mills models used as gravity duals of holographic superfluids \cite{Natsuume:2010ky,Erdmenger:2010xm, Erdmenger:2012zu}, and a black brane model whose temperature is modulated in the spatial directions \cite{Moskalets:2014hoa, Ovdat:2014ipa}. Recently, violations of the KSS bound were also found in isotropic theories dual to massive gravity models which break translational invariance \cite{Alberte:2016xja,Hartnoll:2016tri}, and in the context of Horndeski gravity duals \cite{Feng:2015oea}.

The shear viscosity is obtained from the imaginary part of the retarded 2-point function associated with the stress-energy tensor,
\begin{equation}
\label{eq:KuboEtaijkl}
\eta_{ijkl} = - \lim_{\omega \to 0} \frac{1}{\omega} \text{Im} \ G_{T_{ij}T_{kl}}^{R} (\omega, \vec{k} = \vec{0}) \ \text{with} \ i, j,k, l = x, y, z,
\end{equation}
where
\begin{equation}
\label{eq:Green-Stress}
G^{R}_{T_{ij}T_{kl}}(\omega,\vec{k}) \equiv - i\int d^4x e^{i(\omega t-\vec{k}\cdot \vec{x})}\theta(t) \langle \left[T_{ij}(t,\vec{x}), T_{kl}(0,\vec{0}) \right] \rangle.
\end{equation}
In isotropic and homogeneous theories there is only one shear viscosity coefficient which is obtained from the non-diagonal part of the Green's function \eqref{eq:Green-Stress}, i.e. $\eta = \eta_{xyxy}$. However, since the backgrounds to be considered here will be anisotropic one has that $\eta_{xyxy}\neq \eta_{xzxz}$, for instance. When the anisotropy in the fluid is induced by a magnetic field one has now to consider up to seven different viscosity coefficients, with five shear viscosities and two bulk viscosities \cite{Bragi,LandauKine,Huang:2009ue,Huang:2011dc,Chandra:2015iza}. In order to clarify as much as possible how one may obtain these different viscosity coefficients in an anisotropic theory induced by a magnetic field, we provide in Appendix \ref{apb} a detailed analysis of the Kubo's formulas for all the viscosities that may appear in such case.

Although the general viscous magnetohydrodynamics constructed with the viscous tensor described in Eq. \eqref{eq:ViscTensMag} of Appendix \ref{apb} may contain seven different nontrivial viscosity coefficients, for the magnetized gravity backgrounds considered here, where $SO(3)$ rotational invariance is broken down to $SO(2)$, but do not take into account, for instance, the contribution of a nontrivial angular momentum, the number of nontrivial independent viscosity coefficients will be less since we have only four independent fluctuations of the metric field which are important to compute the viscosities\footnote{Note that $h_{\mu\nu}$ in this section denotes the metric fluctuations and has nothing to do with the diagonal disturbed pullback $h^{(s)}_{ab}$ discussed in the previous section.} (cf. Eq. \eqref{eq:KuboEtaijkl}): $h_{xy}$, $h_{xz}$, $h_{xx}+h_{zz}$, and $h_{zz}$. Moreover, the diagonal fluctuations are related to the bulk viscosities (they also couple to the dilaton fluctuation $\delta\phi$ as discussed in Ref. \cite{GN2}), whose calculations are much more involved and deferred for a future work. Consequently, in the present we focus on the two different nontrivial shear viscosities appearing in our anisotropic magnetized backgrounds. Note that we will also neglect the Abelian field fluctuations $\delta A_i$ since they only couple to vector fluctuations, e.g. $h_{ti}$, which are important, for example, in the calculation of the electric conductivity.

The two nontrivial shear viscosities we calculate in the present work are given by the Kubo's formulas
\begin{align}
\eta_{\perp} \equiv \eta_{xyxy} =  - \lim_{\omega \to 0} \frac{1}{\omega} \text{Im} \ G_{T_{xy}T_{xy}}^{R} (\omega, \vec{k} = \vec{0}), \label{eq:Defeta_perp} \\
\eta_{\parallel} \equiv \eta_{xzxz} =  - \lim_{\omega \to 0} \frac{1}{\omega} \text{Im} \ G_{T_{xz}T_{xz}}^{R} (\omega, \vec{k} = \vec{0}). \label{eq:Defeta_pera}
\end{align}

We compute $\eta_{\perp}$ and $\eta_{\parallel}$ via holography using the membrane paradigm approach, as it was done with the Langevin diffusion coefficients in the previous subsection. For such a task, one needs to obtain the second order disturbed action for the metric fluctuations $h_{xy}\equiv\delta g_{xy}$ and $h_{xz}\equiv\delta g_{xz}$ coupling to $T^{xy}$ and $T^{xz}$, respectively. It was shown in Ref. \cite{DK-applications2} that the quadratic part of the disturbed actions with respect to $h_{x}^{y}=g^{xx}h_{xy}$ and $h_{x}^{z}=g^{zz}h_{zx}$ are given by the following expressions
\begin{align}
S_{2}^{(\perp)}&=-\frac{1}{16\pi G_5}\int d^5x \sqrt{-g}\frac{1}{2}(\partial h_{x}^{y})^2,\label{eq:Shxy} \\
S_{2}^{(\parallel)}&=-\frac{1}{16\pi G_5}\int d^5x  \sqrt{-g}\left(\frac{g_{zz}}{2g_{xx}}(\partial h_{x}^{z})^2\right). \label{eq:Shxz}
\end{align} 
Notice also that the presence of a nontrivial dilaton field $\phi$ in the background does not alter the above expressions since its fluctuation $\delta\phi$ couples only to the diagonal part of the disturbed metric field (defining the so-called ``scalar channel''), which is needed to calculate the bulk viscosities but not the shear viscosities as aforementioned.

We obtain the shear viscosity transport coefficients, $\eta_{\perp}$ and $\eta_{\parallel}$, by employing the membrane paradigm directly to the quadratic fluctuated actions \eqref{eq:Shxy} and \eqref{eq:Shxz}, respectively,
\begin{align}
\eta_{\perp} &= \frac{1}{16\pi G_5}\sqrt{g_{xx}^{2}(r_H)g_{zz}(r_H)}, \\
\eta_{\parallel} &= \frac{1}{16\pi G_5}\sqrt{g_{zz}^{3}(r_H)}.
\end{align}

On the other hand, the entropy density $s$ obtained from the background \eqref{eq:genmetric} by using the Bekenstein-Hawking's relation \cite{bek1,bek2} is given by
\begin{equation}
s = \frac{\sqrt{g_{xx}^{2}(r_H)g_{zz}(r_H)}}{4G_5}.
\end{equation}
Therefore, the ratios $\eta_{\perp}/s$ and $\eta_{\parallel}/s$ become
\begin{align}
\frac{\eta_{\perp}}{s}&=\frac{1}{4\pi}, \label{eq:eta_perp_form}\\
\frac{\eta_{\parallel}}{s}&=\frac{1}{4\pi}\frac{g_{zz}(r_H)}{g_{xx}(r_H)}. \label{eq:eta_par_form} 
\end{align}
These formulas were first obtained in the context of the Einstein-axion-dilaton model \cite{Rebhan:2011vd}, and also in the particular cases of the EMD model given by the anisotropic Einstein-dilaton model \cite{Jain:2014vka} (see also \cite{Jain:2015txa})\footnote{We warn the reader that the notation for $\eta_\perp$ and $\eta_\parallel$ followed in Refs. \cite{Jain:2014vka,Jain:2015txa} are reversed compared to the notation adopted here and in Refs. \cite{Rebhan:2011vd,DK-applications2}. Moreover, we also remark that here and also in Ref. \cite{DK-applications2} we considered the fluctuation $h_{x}^{z}$ instead of the fluctuation $h_{z}^{x}$ considered in Refs. \cite{Rebhan:2011vd,Jain:2014vka,Jain:2015txa}.} and the Einstein-Maxwell model \cite{DK-applications2}.

The perpendicular shear viscosity to entropy density ratio $\eta_{\perp}/s$ associated with the fluctuation $h^{x}_{y}$ of the metric \eqref{eq:genmetric}, which has the residual $SO(2)$ rotational symmetry in the plane transverse to the magnetic field, does not deviate from the KSS result \eqref{eq:taxa}. Consequently, the goal of sections \ref{sec3.4} and \ref{sec4.4} will be to unveil how the parallel shear viscosity to entropy density ratio $\eta_{\perp}/s$ is modified relatively to the KSS formula in the presence of an external magnetic field for the magnetic brane and magnetic EMD models, respectively.

\section{Derivation of anisotropic Kubo formulas for viscosity from linear response theory}
\label{apb}

In this Appendix we investigate how the breaking of the $SO(3)$ rotation symmetry down to $SO(2)$ affects the dissipative properties of relativistic fluids, i.e., we shall discuss how one may generalize the viscosity coefficients (shear and bulk) in order to accommodate the anisotropic nature of the magnetized fluid in the presence of a magnetic field. Historically, calculations of anisotropic transport coefficients in Abelian plasmas were carried out in the 1950's, mainly by Braginskii \cite{Bragi,LandauKine} - see also the more recent Ref.\ \cite{Tuchin:2011jw}. Recently, there has been an increasing interest in the effects of strong fields on high energy relativistic systems, such as neutron stars \cite{Huang:2009ue,Huang:2011dc}, where the anisotropic nature of the plasma may play an important role. Although our discussion will be restricted to anisotropic viscosities in a plasma whose anisotropy is driven by an external magnetic field, we emphasize that this phenomenon occurs in various systems such as plastics and superfluids \cite{LandauEla}; see Refs.\ \cite{Natsuume:2010ky,Erdmenger:2010xm,Erdmenger:2012zu} for holographic approaches to the latter.

Ultimately, we are interested in relativistic viscous plasmas and, consequently, we want a causal and stable theory of relativistic magnetohydrodynamics. One approach to viscous magnetohydrodynamics corresponds to the Navier-Stokes-Fourier-Ohm theory \cite{Huang:2009ue}, which is an extension of the acausal and unstable \cite{Hiscock1,Hiscock2} relativistic Navier-Stokes theory - we shall not dwell into this approach here. Relativistic effects in magnetohydrodynamics for weakly collisional (Abelian) plasmas were studied in \cite{Chandra:2015iza}, which may be important to study black hole accretion flows where the magnetic field is intense. Recently, though, Ref.\ \cite{Molnar:2016vvu} extended the Israel-Stewart formalism \cite{Israel2} to derive the equations of motion of an anisotropic dissipative fluid obtained from the Boltzmann equation using the moments method developed in \cite{Denicol:2012cn}\footnote{The usefulness of the moments method in dealing with the relativistic Boltzmann equation goes beyond the context of heavy ion collision applications. In fact, the moments method may be used to obtain the equations of motion describing magnetohydrodynamics directly from the Boltzmann-Vlasov equations \cite{livro} and also the out-of-equilibrium dynamics of gases in an expanding universe \cite{Bazow:2015dha}.}.

The task now is to derive the form of the viscous stress tensor $\Pi_{\mu\nu}$ in order to obtain Kubo formulas for the anisotropic viscosities; for completeness we revisit Appendix A of Ref.\ \cite{DK-applications2} and some aspects of Ref.\ \cite{Huang:2009ue}. For instance, in isotropic theories, one has\footnote{We assume a 4D spacetime from now on.}
\begin{equation}\label{eq:IsoPmunu}
\Pi_{\mu\nu}=-2\eta\left(w_{\mu\nu}-\Delta_{\mu\nu}\frac{\theta}{3} \right)-\zeta\theta,
\end{equation}
where $w_{\mu\nu}=\frac{1}{2}\left(D_{\mu}u_{\nu}+D_{\nu}u_{\mu}\right)$, $u^\mu$ is the four-velocity with normalization $u^\mu u_\mu=-1$, $D_{\mu}=\Delta_{\mu\alpha}\partial^{\alpha}$, $\Delta_{\mu\nu}=g_{\mu\nu}+u_{\mu}u_{\nu}$ (orthogonal projector), and $\theta=\partial_\mu u^\mu$.

The expression \eqref{eq:IsoPmunu} cannot hold for a highly magnetized plasma since it has a reduced axial symmetry around the magnetic vector. From the gravity side of the holographic correspondence the anisotropic metric \eqref{eq:genmetric} tells us the same. Therefore, we need a rank-4 viscosity tensor $\eta^{\alpha\beta\mu\nu}$ obeying the following relation
\begin{equation}
\Pi^{\mu\nu}= \eta^{\mu\nu\alpha\beta}w_{\alpha\beta}.
\label{onsagercondition}
\end{equation}

The tensorial structure of the viscosity tensor $\eta^{\mu\nu\alpha\beta}$ depends solely on $\Delta^{\mu\nu}$, $b^{\mu}$ (a unit spacelike vector normal to the magnetic field), and $b^{\mu\nu}=\epsilon^{\mu\nu\alpha\beta}b^\alpha u^\beta$. Furthermore, the viscosity tensor must satisfy the following symmetry relations (where $B$ is the magnetic field)
\begin{align}
\eta^{\mu\nu\alpha\beta}(B)&= \eta^{\nu\mu\alpha\beta}(B) =\eta^{\mu\nu\beta\alpha}(B), \label{eq:Sym1} \\
\eta^{\mu\nu\alpha\beta}(B) &=  \eta^{\alpha\beta\mu\nu}(-B). \ \ \ \text{(Onsager principle)} \label{eq:Sym2}
\end{align}

The linearly independent structures which may be constructed using $\Delta^{\mu\nu}$, $b^{\mu}$, and $b^{\mu\nu}$, which respect the symmetries \eqref{eq:Sym1} and \eqref{eq:Sym2}, are given by
\begin{align}
\text{(i)}& \ \Delta^{\mu\nu}\Delta^{\alpha\beta}, \notag \\
\text{(ii)}& \ \Delta^{\mu\alpha}\Delta^{\nu\beta}+\Delta^{\mu\beta}\Delta^{\nu\alpha},  \notag \\
\text{(iii)}& \ \Delta^{\mu\nu}b^{\alpha}b^{\beta}+\Delta^{\alpha\beta}b^{\mu}b^{\nu}, \notag \\
\text{(iv)}& \ b^{\mu}b^{\nu}b^{\alpha}b^{\beta}, \notag \\
\text{(v)} & \ \Delta^{\mu\alpha}b^{\nu}b^{\beta}+\Delta^{\mu\beta}b^{\nu}b^{\alpha}+\Delta^{\nu\alpha}b^{\mu}b^{\beta}+\Delta^{\nu\beta}b^{\mu}b^{\alpha},  \notag \\
\text{(vi)}& \ \Delta^{\mu\alpha}b^{\nu\beta}+\Delta^{\mu\beta}b^{\nu\alpha}+\Delta^{\nu\alpha}b^{\mu\beta}+\Delta^{\nu\beta}b^{\mu\alpha},  \notag \\
\text{(vii)}& \ b^{\mu\alpha}b^{\nu}b^{\beta}+b^{\mu\beta}b^{\nu}b^{\alpha}+b^{\nu\alpha}b^{\mu}b^{\beta}+b^{\nu\beta}b^{\mu}b^{\alpha}.
\end{align}
The viscosity tensor will be composed by linear combinations of the above relations with the viscosity coefficients being the factors in front of each structure. Consequently, one concludes that there must be seven viscosity coefficients for this theory of viscous magnetohydrodynamics, divided into five shear viscosities and two bulk viscosities.

We adopt a similar convention of viscosity coefficients to the one followed in Ref.\ \cite{Huang:2011dc} (HSR)\footnote{This is a different convention from the one followed in Ref.\ \cite{Bragi} and in Chapter 13 of Ref.\ \cite{LandauKine}.}, except that in the present work (FCRN), $\eta_{2}^{\textrm{(FCRN)}}=-\eta_{2}^{\textrm{(HSR)}}$ and $\eta_{3}^{\textrm{(FCRN)}}=-2\eta_{3}^{\textrm{(HSR)}}$. In this case, the viscosity tensor assumes the form
\begin{align}\label{eq:r4ViscoTensMag}
\eta^{\mu\nu\alpha\beta} =& (-2/3\eta_0 +1/4\eta_1 +3/2\zeta_\perp)\text{(i)} + (\eta_0 )\text{(ii)} +(3/4\eta_1+3/2\zeta_\perp)\text{(iii)}\notag \\
  & +(9/4\eta_1 -4\eta_2 +3/2\zeta_\perp+3\zeta_\parallel )\text{(iv)} +(-\eta_2 )\text{(v)} +(-\eta_4)\text{(vi)} \notag \\
  & +(-\eta_3+\eta_4)\text{(vii)},
\end{align}
with the $\eta'$s being the shear viscosities and the $\zeta'$s being the bulk viscosities.

Substituting \eqref{eq:r4ViscoTensMag} into \eqref{onsagercondition} we find the following viscous tensor
\begin{align}\label{eq:ViscTensMag}
\Pi_{\mu\nu} &= -2\eta_0\left(w_{\mu\nu}-\Delta_{\mu\nu}\frac{\theta}{3} \right)-\eta_1\left(\Delta_{\mu\nu}-\frac{3}{2}\Xi_{\mu\nu} \right)\left(\theta-\frac{3}{2}\phi\right)+2\eta_2\left(b_\mu\Xi_{\nu\alpha}b_{\beta}+b_\nu\Xi_{\mu\alpha}b_{\beta}\right)w^{\alpha\beta} \notag \\
    & +\eta_3\left(\Xi_{\mu\alpha}b_{\nu\beta}+\Xi_{\nu\alpha}b_{\mu\beta}\right)w^{\alpha\beta}-2\eta_4\left(b_{\mu\alpha}b_{\nu}b_{\beta}+b_{\nu\alpha}b_{\mu}b_{\beta}\right)w^{\alpha\beta}-\frac{3}{2}\zeta_{\perp}\Xi_{\mu\nu}\phi - 3\zeta_{\parallel}b_{\mu}b_{\nu}\varphi,
\end{align}
where $w_{\mu\nu}=\frac{1}{2}\left(D_{\mu}u_{\nu}+D_{\nu}u_{\mu}\right)$, $D_{\mu}=\Delta_{\mu\alpha}\nabla^{\alpha}$, $\Xi_{\mu\nu}\equiv\Delta_{\mu\nu}-b_\mu b_\nu$ (orthogonal projector), $\theta=\nabla_\mu u^\mu$, $\phi\equiv\Xi_{\mu\nu}w^{\mu\nu}$ and $\varphi\equiv b_{\mu}b_{\nu}w^{\mu\nu}$. Note that the differential operator $D_\mu$ is given in terms of the covariant derivative, i.e., we are generalizing the viscous tensor to a curved spacetime; this will be essential to extract the Kubo formulas since they are obtained here by considering gravity fluctuations.

\subsection{Kubo formulas for viscous magnetohydrodynamics}

With the expression for the viscous tensor $\Pi_{\mu\nu}$ at hand, it is time to derive the Kubo formulas that relate the viscosity coefficients to the retarded Green's functions. We remark that Ref.\ \cite{Huang:2011dc} also derived the Kubo formulas although using the Zubarev formalism.

Let us summarize the usual procedure to obtain the Kubo formulas for the viscosity: adopting a Minkowski background, we perform small gravity perturbations in $\Pi_{\mu\nu}$ assuming that they are all homogeneous, which means that we can work only with the spatial indices, i.e.,  $g_{ij}= \eta_{ij}+h_{ij}(t)$, with $h_{00}=h_{0i}=0$. Also, we work in the rest frame of the fluid where $u^{\mu}=(1,0,0,0)$.\footnote{In other words, we will work in the Landau-Lifshitz frame where $u_\mu\Pi^{\mu\nu}=0$ and all the information about the viscosities are in the components $\lbrace i, j, k, l \rbrace$ of the retarded Green's function \eqref{eq:Green-Stress}.} Finally, we equate the fluctuated form of \eqref{eq:ViscTensMag} to $h_{kl}G^{R, \, kl}_{ij}(\omega)$ in order to extract the Kubo formulas. The novelty here is the presence of a magnetic field which is assumed to be constant and uniform along the $z$ direction, i.e., $b^{\mu}=(0,0,0,1)$.

Thus, we have the variation for the viscous tensor\footnote{Note that:
\begin{equation}
\delta\Xi_{\mu\nu}=h_{\mu\nu}, \ \ \ \delta\theta= \frac{1}{2}\partial_t h^{\lambda}_{\lambda}, \ \ \ \delta\varphi=\frac{1}{2}\partial_t h_{zz}. \notag
\end{equation}}
\begin{align}
\delta\Pi_{ij}=\delta(i)+\delta(ii)+\delta(iii)+\delta(iv)+\delta(v)+\delta(vi)+\delta(vii),
\end{align}
where
\begin{align}
\delta(i) &=  -\eta_0\left(\partial_t h_{ij}-\frac{1}{3}\delta_{ij}\partial_t h^{k}_{k}\right),
\end{align}
\begin{align}
\delta(ii) &=-\frac{1}{4} \eta_1 \left[ -(\delta_{ij}-3b_i b_j)\partial_t h^{k}_{k} +\frac{3}{2}(\delta^{kl}-b^{k}b^{l})(\delta_{ij}-3b_i b_j)\partial_t h_{kl}\right],
\end{align}
\begin{align}
\delta(iii) &=  \eta_2  \left[ b_i b^k\partial_t h_{jk}+b_j b^k\partial_t h_{ik} - 2b_i b_j b^k b^l\partial_t h_{kl} \right],
\end{align}
\begin{align}
\delta(iv) &=\frac{1}{2}\eta_3 \partial_t h_{kl} \left(\delta_{i}^{k}\epsilon_{j}^{\ lz}+\delta^{jl}\epsilon_{i}^{\ kz} - b_{i}b^{k}\epsilon_{j}^{\ lz} - b_{j}b^{k}\epsilon_{i}^{\ lz}  \right),
\end{align}
\begin{align}
\delta(v) &= -\eta_4 \left(b_{ik}b_{j}b_{l}+b_{jk}b_{i}b_{l} \right)\partial_t h^{kl},
\end{align}
\begin{align}
\delta(vi) &=-\frac{3}{4}\zeta_\perp\left(\delta_{ij}-b_ib_j \right) \left(\partial_t h^{k}_{k}+\partial_t h_{zz}\right),
\end{align}
\begin{align}
\delta(vii) &=  -\frac{3}{2}\zeta_\parallel b_ib_j\partial_t h_{zz}.
\end{align}

The next step is to write the variations above in Fourier space using a plane-wave Ansatz for the perturbations, which gives the following expressions
\begin{align}
\delta(i) &= \frac{i\omega}{2}h_{kl}(\omega)\left[ \eta_0 \left( \delta^{k}_{i}\delta^{l}_{j} +\delta^{l}_{i}\delta^{k}_{j}-\frac{2}{3}\delta_{ij}\delta^{kl}\right) \right],
\end{align}
\begin{align}
\delta(ii) &= \frac{i\omega}{2}h_{kl}(\omega)\frac{1}{4} \eta_1 \left[  -2\delta^{kl}(\delta_{ij}-3b_i b_j) +3(\delta^{kl}-b^{k}b^{l})(\delta_{ij}-3b_i b_j)\right],
\end{align}
\begin{align}
\delta(iii) &= -\frac{i\omega}{2}h_{kl}(\omega)\eta_2 \left( b_{i}b^{k}\delta^{l}_{j}+b_{i}b^{l}\delta^{k}_{j}+b_{j}b^{k}\delta^{l}_{i}+b_{j}b^{l}\delta^{k}_{i}-4b_{i}b_{j}b^{k}b^l \right),
\end{align}
\begin{align}
\delta(iv) &= -\frac{i\omega}{2}h_{kl}(\omega)\eta_3 \left(\delta_{i}^{k}\epsilon_{j}^{\ lz}+\delta^{jl}\epsilon_{i}^{\ kz} - b_{i}b^{k}\epsilon_{j}^{\ lz} - b_{j}b^{k}\epsilon_{i}^{\ lz} \right),
\end{align}
\begin{align}
\delta(iv) &= \frac{i\omega}{2}h_{kl}(\omega) \eta_4  \left(b_{i}^{\ k}b_{j}b^{l} +b_{i}^{\ l}b_{j}b^{k}+ b_{j}^{\ k}b_{i}b^{\ l}+b_{j}^{\ l}b_{i}b^{k} \right) ,
\end{align}
\begin{align}
\delta(v) &= \frac{i\omega}{2}h_{kl}(\omega)\left[ \frac{3}{2}\zeta_\perp\left(\delta_{ij}\delta^{kl}+\delta_{ij}\delta_{z}^{k}\delta_{z}^{l}-b_ib_j\delta^{kl}-b_ib_j\delta_{z}^{k}\delta_{z}^{l} \right)  \right],
\end{align}
\begin{align}
\delta(vii) &=\frac{i\omega}{2}h_{kl}(\omega)\left[ 3\zeta_\parallel b_ib_j\delta_{z}^{k}\delta_{z}^{l}  \right].
\end{align}

Collecting all the variations in Fourier space, we write 
\begin{align}
\delta\Pi_{ij}(\omega)&=\frac{i\omega}{2}h_{kl}(\omega)\left[ \eta_0\left(\delta^{k}_{i}\delta^{l}_{j}+\delta^{l}_{i}\delta^{k}_{j}-\frac{2}{3}\delta_{ij}\delta^{kl}\right) +\frac{1}{4} \eta_1 \left[  -2\delta^{kl}(\delta_{ij}-3b_i b_j) +3(\delta^{kl}-b^{k}b^{l})(\delta_{ij}-3b_i b_j)\right]  \right. \notag \\
 &\left.  - \eta_2 \left( b_{i}b^{k}\delta^{l}_{j}+b_{i}b^{l}\delta^{k}_{j}+b_{j}b^{k}\delta^{l}_{i}+b_{j}b^{l}\delta^{k}_{i}-4b_{i}b_{j}b^{k}b^l \right) -\eta_3 \left(\delta_{i}^{k}\epsilon_{j}^{\ lz}+\delta^{jl}\epsilon_{i}^{\ kz} - b_{i}b^{k}\epsilon_{j}^{\ lz} - b_{j}b^{k}\epsilon_{i}^{\ lz} \right) \right.\notag \\
   &\left. + \eta_4  \left(b_{i}^{\ k}b_{j}b^{l} +b_{i}^{\ l}b_{j}b^{k}+ b_{j}^{\ k}b_{i}b^{\ l}+b_{j}^{\ l}b_{i}b^{k} \right) +\frac{3}{2}\zeta_\perp\left(\delta_{ij}\delta^{kl}+\delta_{ij}\delta_{z}^{k}\delta_{z}^{l}-b_ib_j\delta^{kl}-b_ib_j\delta_{z}^{k}\delta_{z}^{l} \right)  \right.\notag \\
    &\left. +3\zeta_\parallel b_ib_j\delta_{z}^{k}\delta_{z}^{l}  \right],
\end{align}
which allows us to express the retarded Green's function in terms of the viscosities,
\begin{align}\label{eq:greenvisco}
 - \lim_{\omega\rightarrow 0}\frac{1}{\omega}\text{Im}\,G^{R, \, kl}_{ij}(\omega) = &   \eta_0\left(\delta^{k}_{i}\delta^{l}_{j}+\delta^{l}_{i}\delta^{k}_{j}-\frac{2}{3}\delta_{ij}\delta^{kl}\right) +\frac{1}{4} \eta_1 \left[ -2\delta^{kl}(\delta_{ij}-3b_i b_j) +3(\delta^{kl}-b^{k}b^{l})(\delta_{ij}-3b_i b_j)\right]   \notag \\
 & -\eta_2 \left( b_{i}b^{k}\delta^{l}_{j}+b_{i}b^{l}\delta^{k}_{j}+b_{j}b^{k}\delta^{l}_{i}+b_{j}b^{l}\delta^{k}_{i}-4b_{i}b_{j}b^{k}b^l \right)  \notag \\
   & -\eta_3 \left(\delta_{i}^{k}\epsilon_{j}^{\ lz}+\delta^{jl}\epsilon_{i}^{\ kz} - b_{i}b^{k}\epsilon_{j}^{\ lz} - b_{j}b^{k}\epsilon_{i}^{\ lz} \right) \notag \\
   & +  \eta_4  \left(b_{i}^{\ k}b_{j}b^{l} +b_{i}^{\ l}b_{j}b^{k}+ b_{j}^{\ k}b_{i}b^{\ l}+b_{j}^{\ l}b_{i}b^{k} \right)\notag \\
   & +\frac{3}{2}\zeta_\perp\left(\delta_{ij}\delta^{kl}+\delta_{ij}\delta_{z}^{k}\delta_{z}^{l}-b_ib_j\delta^{kl}-b_ib_j\delta_{z}^{k}\delta_{z}^{l} \right) +3\zeta_\parallel b_ib_j\delta_{z}^{k}\delta_{z}^{l}.
\end{align}

The final step is to isolate the viscosities and obtain the corresponding Kubo formulas. For such a task, we only need to select specific components of $G^{R}_{ij,kl}$. For instance, if we take $i=k=x$ and $j=l=y$ in \eqref{eq:greenvisco}, we have
\begin{equation}
\eta_0 = - \lim_{\omega\rightarrow 0}\frac{1}{\omega}G^{R}_{T_{xy}T_{xy}}(\omega),
\end{equation}
and so forth. Finally, we obtain the following Kubo formulas
\begin{align}
\eta_0 &= - \lim_{\omega \to 0}\frac{1}{\omega}\text{Im}\, G^{R}_{T_{xy}T_{xy}}(\omega)\label{eq:eta0} , \\
\eta_1 &= -\frac{4}{3}\eta_0+ 4\lim_{\omega \to 0}\frac{1}{\omega}\text{Im}\, G^{R}_{P_{\parallel}P_{\perp}}(\omega), \label{eq:eta1} \\
\eta_2 &= \eta_0+\lim_{\omega \to 0}\frac{1}{\omega}\text{Im}\, G^{R}_{T_{xz}T_{xz}}(\omega), \\
\eta_3 &=  \lim_{\omega\rightarrow 0}\frac{1}{\omega}G^{R}_{P_{\perp}T_{12}}(\omega), \label{eq:eta3} \\
\eta_4 &= - \lim_{\omega \to 0}\frac{1}{\omega}\text{Im}\, G^{R}_{T_{xz}T_{yz}}(\omega), \label{eq:eta4} \\
\zeta_\perp &= -\frac{2}{3}\lim_{\omega \to 0} \frac{1}{\omega}\left[ \text{Im}G^{R}_{P_\perp P_\perp}(\omega) + \text{Im}G^{R}_{P_\parallel, P_\perp}(\omega) \right], \label{eq:kubo1} \\
\zeta_\parallel &= -\frac{4}{3}\lim_{\omega \to 0}\frac{1}{\omega}\left[ \text{Im}G^{R}_{P_\perp  P_\parallel}(\omega) + \text{Im}G^{R}_{P_\parallel, P_\parallel}(\omega) \right],\label{eq:kubo2}
\end{align}
where
\begin{equation}
P_{\perp} \equiv\frac{1}{2}T^{a}_{\ a}=\frac{1}{2}(T^{x}_{\ x}+T^{y}_{\ y}), \ \ P_{\parallel} \equiv \frac{1}{2} T^{z}_{\ z}.
\end{equation}
The results in Eqs.\ \eqref{eq:eta0} --- \eqref{eq:kubo2} agree with the ones obtained in Ref.\ \cite{Hernandez:2017mch}.

At first sight, the Kubo formulas obtained here seem different from the ones obtained in Ref.\ \cite{Huang:2011dc}. The reason is that the formulas written in \cite{Huang:2011dc} are in a fully covariant form. However, if we use the following identity
\begin{equation}\label{eq:KuboTrick}
\left\langle \left[\int d^3x T^{00}, A \right]\right\rangle = \langle\left[H, A \right]\rangle = i\left\langle\frac{\partial A}{\partial t}\right\rangle = 0,
\end{equation}
where $A$ is a generic operator and $H$ is the Hamiltonian, we get rid of the term $\hat{\epsilon} \sim T^{00}$ - recall that the mean values $\langle \cdots \rangle$ for the Kubo formulas are always related to the equilibrium state. Furthermore, when we recover isotropy, i.e., when $B=0$, the formulas for both bulk viscosities, $\zeta_{\perp}$ and $\zeta_{\parallel}$, return to the well-known isotropic formula. Moreover, due to the structure of the Kubo formulas for the bulk viscosity, we have the relation
\begin{equation}\label{eq:zetas-relation}
\zeta =\frac{2}{3}\zeta_\parallel +\frac{1}{3}\zeta_\perp,
\end{equation}
where $\zeta$ is the isotropic bulk viscosity obtained by the Kubo formula,
\begin{align}
\zeta=-\frac{4}{9}\lim_{\omega\rightarrow 0}\frac{1}{\omega}\textrm{Im}\ G_{PP}^{R}(\omega,\vec{k}=\vec{0}),
\end{align}
where $P\equiv P_\perp+P_\parallel$. Following the usual convention, we define\footnote{Due to the different sign conventions for $\eta_2$, the definition for $\eta_\parallel$ differs from the one adopted in Ref.\ \cite{DK-applications2}.}
\begin{equation}
\eta_{\perp}\equiv \eta_0, \ \ \eta_{\parallel}\equiv \eta_0-\eta_2.
\end{equation}

Another common way to write the formulas for the shear viscosities is
\begin{equation}
\eta_{ijkl} = - \lim_{\omega \to 0} \frac{1}{\omega} \text{Im} \ G_{T_{ij}T_{kl}}^{R} (\omega, \vec{k} = \vec{0}) \ \ \text{with} \ i, j,k, l = x, y, z.
\end{equation}
For instance, in the above notation the isotropic shear viscosity $\eta_0$ reads
\begin{equation}
\eta_0=\eta_{xyxy}=\eta_{\perp}.
\end{equation}

We finish this appendix emphasizing that the Kubo formulas for $\eta_1$, $\eta_3$ and $\eta_4$ trivially vanish in the backgrounds considered here, which are of the form \eqref{eq:genmetric}. For instance, the Kubo formula for $\eta_3$ \eqref{eq:eta3} depends on the operators $P_\perp$ and $T^{xy}$; however, the dual bulk fields of these operators, $h_{xx}$ and $h_{xy}$, respectively, are decoupled in the disturbed on-shell action, which makes the corresponding 2-point Green's function vanish. Therefore, for the magnetic brane setup and the magnetic EMD model, one has to compute only two shear viscosities, $\eta_{\perp}$ and $\eta_{\parallel}$, as we have done in the previous sections.

\end{document}